\documentclass[12pt]{article}
\usepackage{amsmath,amsthm,amsfonts,amssymb,amscd,cite,graphicx,array,dsfont,color}
\RequirePackage{amsmath,amssymb,braket,geometry,sidecap,graphicx,tabularx,ragged2e,booktabs,caption,bm,relsize,empheq}

\newlength{\xtrawidth}
\newlength{\xtraheight}
\allowdisplaybreaks

\numberwithin{equation}{section}
\numberwithin{table}{section}

\def\tm{\mathfrak{m}}  
\def\adjs{\underline{\sigma}} 

\begin{document}
\begin{titlepage}
\begin{center}
\hfill BONN--TH--2018--03\\
\vskip 1in
{\Large\bf{The Geometry of Gauged Linear Sigma Model}}\\[2ex]
{\Large\bf{Correlation Functions}}
\vskip 0.6in
{\large{Andreas Gerhardus, Hans Jockers, Urmi Ninad}}\\
\vskip 0.4in
{\it Bethe Center for Theoretical Physics, Physikalisches Institut\\
der Universit\"at Bonn, Nussallee 12, D-53115 Bonn, Germany}
\vskip 0.2in
{\tt gerhardus@th.physik.uni-bonn.de}\\
{\tt jockers@uni-bonn.de}\\
{\tt urmi@th.physik.uni-bonn.de}
\end{center}
\vskip 0.1in
\begin{center} {\bf Abstract} \end{center}
Applying advances in exact computations of supersymmetric gauge theories, we study the structure of correlation functions in two-dimensional $\mathcal{N}=(2,2)$ Abelian and non-Abelian gauge theories. We determine universal relations among correlation functions, which yield differential equations governing the dependence of the gauge theory ground state on the Fayet--Iliopoulos parameters of the gauge theory. For gauge theories with a non-trivial infrared $\mathcal{N}=(2,2)$ superconformal fixed point, these differential equations become the Picard--Fuchs operators governing the moduli-dependent vacuum ground state in a Hilbert space interpretation. For gauge theories with geometric target spaces, a quadratic expression in the Givental $I$-function generates the analyzed correlators. This gives a geometric interpretation for the correlators, their relations, and the differential equations. For classes of Calabi--Yau target spaces, such as threefolds with up to two K\"ahler moduli and fourfolds with a single K\"ahler modulus, we give general and universally applicable expressions for Picard--Fuchs operators in terms of correlators. We illustrate our results with representative examples of two-dimensional $\mathcal{N}=(2,2)$ gauge theories.

\vfill
\noindent March, 2018

\end{titlepage}
\tableofcontents
\newpage

\section{Introduction}
With the seminal work \cite{Witten:1993yc} on two-dimensional $\mathcal{N}=(2,2)$ supersymmetric gauged linear sigma models, Witten offered a powerful machinery to study the geometry of the gauge theory target spaces together with their moduli spaces in terms of gauge theory techniques. For instance, Morrison and Plesser computed quantum-exact correlation functions as functions of the Fayet--Iliopoulos parameters and the theta angles in such gauged linear sigma models \cite{Morrison:1994fr}. Geometrically, such correlators become sections on the quantum K\"ahler moduli space of the target space geometry. The interplay between these two-dimensional gauge theories and the quantum geometry on the target space offers a far-reaching connection between two-dimensional gauge theories and their dualities on the one hand and Gromov--Witten theory and mirror symmetry on the other hand \cite{Morrison:1995yh,Hori:2000kt}.

The aim of this work is to systematically study the structure and the underlying geometry of a certain class of correlators of both Abelian and non-Abelian two-dimensional $\mathcal{N}=(2,2)$ supersymmetric gauged linear sigma models, which depend on the Fayet--Iliopoulos parameters and the theta angles of the gauge theory. The recent work \cite{Closset:2015rna} by Closset, Cremonesi and Park furnishes an important ingredient in our approach, as it offers techniques to exactly compute these correlators of gauged linear sigma models --- generalizing the methods of Morrison and Plesser \cite{Morrison:1994fr} to higher point correlators and to non-Abelian gauged linear sigma models.\footnote{See also ref.~\cite{Benini:2015noa} for correlators of two-dimensional $\mathcal{N}=(2,2)$ gauged linear sigma models.} Their approach is based upon modern localization techniques of supersymmetric gauge theories on curved spaces with a non-trivial (off-shell) supergravity background \cite{Festuccia:2011ws,Pestun:2007rz}, such that the quantum-exact correlators localize on a sum of non-trivial topological vortex sectors. As the performed localization calculation in the specified supergravity background directly relates to similar computations by Hori and Vafa in the context of A-twisted gauged linear sigma models on the symplectic side of mirror symmetry \cite{Hori:2000kt}, these correlators contain information about the quantum K\"ahler moduli space and the Gromov--Witten theory of the target space.

In this note --- starting from the residue integral of the localized gauge theory correlators provided in ref.~\cite{Closset:2015rna} --- we derive universal and non-trivial relations among the set of all gauge theory correlators, which are directly and easily obtained from the spectrum of the gauge theory. Giving a Hilbert space interpretation for the correlators, we map the obtained universal relations to differential operators that annihilate the ground state of the gauge theory. Realizing the gauge theory correlators as certain quadratic pairings of the Givental $I$-function --- as argued by Ueda and Yoshida \cite{Ueda:2016wfa} and proven for a particular class of target space geometries and conjectured more generally in ref.~\cite{Kim:2016uq} --- we argue that the obtained set of differential operators generates the GKZ system of differential equations governing the quantum cohomology of the target space.\footnote{For a connection between the Givental $I$-function of the target space and the vortex partition function of two-dimensional $\mathcal{N}=(2,2)$ quiver gauged linear sigma models see also ref.~\cite{Bonelli:2013mma}.} As a consequence, the obtained differential operators are in agreement with differential equations for the quantum periods of the A-twisted gauged linear sigma models studied in ref.~\cite{Hori:2000kt}.\footnote{For particular examples of gauge theories, Closset et. al. similarly deduce differential equations from the localized gauge theory correlators, which in our approach arise from universal correlator relations.}

For the important class of two-dimensional $\mathcal{N}=(2,2)$ supersymmetric gauged linear sigma models with a non-anomalous axial $U(1)_R$ R-symmetry, the gauge theories are known to flow in the IR to non-trivial families of two-dimensional $\mathcal{N}=(2,2)$ superconformal field theories \cite{Witten:1993yc}, where the Fayet--Iliopoulos parameters and the theta angles furnish the algebraic coordinates or UV~coordinates of the analyzed quantum K\"ahler moduli space. The moduli space of the IR family of conformal field theories is more suitably described in terms of IR coordinates or flat coordinates, which relate to the couplings of marginal operators. In a geometric phase with a non-linear sigma model description the correlation functions in the IR~coordinates yield generating functions of genus zero Gromov--Witten invariants of the Calabi--Yau target space. To arrive at the correlation functions in these flat coordinates the UV-IR map between these two coordinate systems is required. A standard technique to explicitly determine the UV-IR map is to compute the quantum periods of the family of conformal field theories \cite{Candelas:1990rm}, which in turn are obtained as solutions to the Picard--Fuchs differential operators $\mathcal{L}$ of the analyzed quantum K\"ahler moduli space.

For the aforementioned conformal class of gauge theories, we find that the universal correlator relations and their associated differential operators become the Picard--Fuchs operators $\mathcal{L}$ of the moduli spaces of these $\mathcal{N}=(2,2)$ superconformal field theories. Traditionally, the Picard--Fuchs operators and hence the quantum periods are often indirectly determined via mirror symmetry \cite{Candelas:1990rm}. Such computations are particularly powerful for setups that admit a systematic and known mirror construction, as it is available for compactifications on complete intersection Calabi--Yau manifolds in toric varieties \cite{Batyrev:1994hm,Batyrev:1994pg}. Our approach is complementary, as it offers an explicit algorithm to derive Picard--Fuchs differential equations for Calabi--Yau target spaces directly without the need to construct a mirror geometry, which for Calabi--Yau compactifications beyond complete intersection in toric varieties is not always known.\footnote{Advances in localization techniques have opened up a novel perspective on deriving IR gauge theory quantities directly, i.e., without employing mirror symmetry. Apart from the methodology introduced in this note, the sphere partition function computes the quantum-exact K\"ahler metric on the quantum K\"ahler moduli space \cite{Jockers:2012dk,Gomis:2012wy,Gomis:2015yaa}, while the hemisphere partition functions directly yields exact expressions for the quantum periods \cite{Honda:2013uca,Sugishita:2013jca,Hori:2013ika}.} Furthermore, we directly determine the Picard--Fuchs operators of Calabi--Yau manifolds without the need to further factor higher order differential operators, as is common to other approaches, see e.g., ref.~\cite{Hosono:1994ax}.

We further establish for Calabi--Yau geometries of a given dimension and with a given number of K\"ahler moduli universal expressions for Picard--Fuchs differential operators with coefficient functions in terms of gauge theory correlators. For instance, for Calabi--Yau threefolds with a single K\"ahler modulus we find for the Picard--Fuchs operator of the quantum K\"ahler moduli space the universal expression
\begin{multline}
 \mathcal{L}\,=\, \kappa _{0,3}^2 (\epsilon\Theta)^4 -\kappa _{0,3} \kappa _{0,4} (\epsilon\Theta)^3 +\left(\kappa _{0,4} \kappa _{1,3}-\kappa _{0,3} \kappa _{1,4}\right)(\epsilon\Theta)^2 \\
+\left(\kappa _{0,4} \kappa _{2,3}-\kappa _{0,3} \kappa _{2,4}\right)(\epsilon\Theta) +\left(\kappa _{1,4} \kappa _{2,3}-\kappa _{1,3} \kappa _{2,4}-\kappa _{0,3} \kappa _{3,4}\right) \ ,
\end{multline}
in terms of the algebraic coordinate $Q$, its logarithmic derivative $\Theta = Q \partial_Q$, and the gauge theory correlators $\kappa_{a,b}$. The coefficients of this Picard--Fuchs operator in terms of the correlators automatically fulfill a certain constraint, which has been established previously for Picard--Fuchs differential operators of Calabi--Yau threefolds with a single K\"ahler modulus in refs.~\cite{MR2282972,MR2282974}. While we find similar types of constraints for other classes of Picard--Fuchs operators as well, for this particular Picard--Fuchs operator the authors of refs.~\cite{MR2282972,MR2282974} show that it is a consequence $\mathcal{N}=2$ special geometry \cite{Strominger:1990pd}.

The organization of this paper is as follows. In Section~\ref{sec:CorRel}, we derive the universal gauge theory correlator relations from the localized residue integral of the gauge theory correlators presented in ref.~\cite{Closset:2015rna}. Furthermore, in a Hilbert space interpretation of these correlation functions, we obtain from the derived correlator relations differential operators that annihilate the gauge theory ground states. We analyze correlator relations both in Abelian and non-Abelian gauged linear sigma models. In Section~\ref{sec:CYtarget} we establish the systematics of deriving universal Picard--Fuchs operators for classes of Calabi--Yau target space geometries. We explicitly work out such universal expressions for elliptic curves, polarized K3 surfaces, Calabi--Yau threefolds, Calabi--Yau fourfolds with a single K\"ahler modulus, and for Calabi--Yau threefolds with two K\"ahler moduli. In Section~\ref{sec:Giv} we connect the correlators and their relations to the Givental $I$-function, offering a geometric interpretation of the results in terms of quantum cohomology established in the previous sections. In Section~\ref{sec:Examples} we illustrate our techniques with various explicit examples of both Abelian and non-Abelian $\mathcal{N}=(2,2)$ gauged linear sigma models, which give rise to Fano varieties, Calabi--Yau varieties, and varieties with ample canonical class as their target spaces. We conclude in Section~\ref{sec:Conc} with a short summary and an outlook. Finally, in Appendix~\ref{app:CY} we collect some technical aspects required in Section~\ref{sec:CYtarget} in order to derive the universal Picard--Fuchs differential operators for Calabi--Yau threefolds with two K\"ahler moduli and for Calabi--Yau fourfolds with an (non-minimal) order six Picard--Fuchs operator.

\section{Correlator relations in $\mathcal{N}=(2,2)$ gauge theories} \label{sec:CorRel}
In ref.~\cite{Benini:2015noa,Closset:2015rna} the authors perform an interesting localization computation of $\mathcal{N}=(2,2)$ supersymmetric two-dimensional gauge theories in a non-trivial off-shell supergravity background on the two-sphere $S^2$. In this background the gauge theory computes correlation functions of the A-twisted topological string at genus zero, to which are referred to as A-twisted correlators. Starting from the result of localization we here demonstrate that these correlators fulfill non-trivial relations. This is a consequence of the factorization of correlators into holomorphic blocks \cite{Cecotti:1991me}. We argue that these correlator relations become differential operators governing the parameter dependence of the gauge theory ground state. The relationship to the topological string does not come as a surprise, as the employed gauge theory formulation is closely related to the A-twisted gauged linear sigma model considered in the context of topological strings and mirror symmetry in ref.~\cite{Hori:2000kt}. 

\subsection{Correlator relations: Abelian gauge groups}  \label{sec:relcorr}
We consider two-dimensional $\mathcal{N}=(2,2)$ gauge theories with chiral matter multiplets. For simplicity, let us first focus on Abelian gauge groups $G=U(1)^r$ of rank $r$ together with $M$ charged matter multiplets $\phi_\ell$, $\ell=1,\ldots , M$, with gauge charges $\vec{\rho}_\ell=(\rho_{\ell,1}\ldots,\rho_{\ell,r})\in \mathbb{Z}^r$, $U(1)_R$ charges $\mathfrak{q}_\ell$, and twisted masses $\tm_\ell$. We further allow for a generic gauge-invariant superpotential $W$ of $R$-charge two that preserves the $U(1)_R$-symmetry.\footnote{Depending on the details of the chiral matter spectrum, the gauge theory may or may not admit a non-vanishing a $U(1)_R$- and gauge-invariant superpotential~$W$.} It is well-known that the target space geometries, given by the scalar fields in the chiral matter multiplets, (semi-classically) describe toric varieties or complete intersections therein \cite{Witten:1993yc}. 

\subsubsection{A-twisted correlators} \label{sec:Acorr}
For such gauge theories, Closset et. al. compute A-twisted correlators of the scalar components $\sigma_k$, $k=1,\ldots, r$, in the twisted chiral field strength $\Sigma_k$ associated to the $\mathcal{N}=(2,2)$ two-dimensional Abelian vector multiplets \cite{Closset:2015rna}.\footnote{In ref.~\cite{Benini:2015noa} A-twisted correlation functions are computed as well. In this work, however, we follow ref.~\cite{Closset:2015rna}, as it considers a more general class of correlators required for our forthcoming analysis.}  This is achieved by putting the gauge theory on a two-sphere with a suitable off-shell supergravity background that realizes a topological A-twist. As a consequence, in such a background the non-trivial twisted chiral correlators are of the form 
\begin{equation} \label{eq:AbCorr}
  \left\langle \vec\sigma_N^{\vec n} \vec\sigma_S^{\vec m} \right\rangle \,=\, \kappa_{\vec n,\vec m}(\vec Q, \tm_\ell, \epsilon) \ , \quad \vec n,\vec m \in \mathbb{Z}_{\ge 0}^r \ ,
\end{equation}
with the short-hand notation $\sigma^{\vec n} = \sigma^{n_1}_1 \cdots \sigma^{n_r}_r$. The correlation functions $\kappa_{\vec n,\vec m}$ are holomorphic in the variables $\vec Q=(Q_1,\ldots,Q_r)$ labelling the topological sectors of the gauge group $U(1)^r$, the twisted masses $\tm_\ell$, and the parameter $\epsilon$ for the off-shell supergravity background. The components $Q_k$ of $\vec Q$ can be interpreted as complex $\mathbb{C}^*$ variables
\begin{equation} \label{eq:Qdef}
  Q_k \,=\, e^{-2\pi \xi_k +  i \theta_k} \ ,
\end{equation}  
where $\xi_k$ and $\theta_k$ are the Fayet--Iliopoulos parameter and the $\theta$-angle of the Abelian $U(1)$-gauge group factors. The metric-dependence of the correlation functions $\kappa_{\vec n,\vec m}$ is encoded in the parameter $\epsilon$, see \cite{Closset:2015rna}, and their topological part is given by their value at $\epsilon = 0$, i.e., $\kappa_{\vec n,\vec m}(\vec Q,\tm_\ell,0)$. Since the latter is insensitive to the location of the insertion of the $\sigma$-fields one finds 
\begin{equation}
  \kappa_{ \vec n, \vec m}(\vec Q,\tm_\ell,0 ) \,=\, \kappa_{ \vec n', \vec m'}(\vec Q,\tm_\ell,0 )  \quad \text{for all} \quad \vec n + \vec m = \vec n' + \vec m' \in \mathbb{Z}^r_{\ge 0} \ .
\end{equation}
The (potentially anomalous) $U(1)_R$ axial symmetry yields a selection rule for correlation functions $\kappa_{\vec n,\vec m}$. It states that the coefficient in a series expansion with respect to the variables $\epsilon$, $\tm_\ell$, and $Q_k$ is only non-vanishing if the equality \cite{Morrison:1994fr,Closset:2015rna}
\begin{equation} \label{eq:srule}
  d + \#(\epsilon) + \# (\tm_\ell) + \sum_{k=1}^r \left(\sum_\ell \rho_{\ell,k}\right) \#(Q_k) \,=\, |\vec n|_1 + |\vec m|_1 \ .
\end{equation}  
holds. Here $\#(\, \cdot \, )$ denotes the exponent of the specified argument in the considered term of the series expansion, and $| \vec n |_1 \equiv \sum_k |n_k|$ is the taxicab norm of the vector $\vec n$. Upon coupling to gravity, $d$ is the gravitational contribution to the anomaly of the $U(1)_R$ axial symmetry, which for Abelian gauged linear sigma models reads
\begin{equation} \label{eq:ddef}
   d \,=\, \sum_\ell (1- \mathfrak{q}_\ell ) - r \ .
\end{equation}
If the analyzed gauge theory admits a geometric target space interpretation, it has $d$ complex dimensions. Furthermore, for gauged linear sigma models with a non-anomalous axial $U(1)_R$ symmetry, the central charge of the superconformal $\mathcal{N}=(2,2)$ theory at the infrared fixed point is $3 d$. 

We call a theory a gauged linear sigma model with the conformal property, if it has a non-anomalous axial $U(1)_R$ symmetry, i.e., if $\sum_\ell \vec \rho_\ell =0$ for all $\ell$. For such gauge theories the selection rule~\eqref{eq:srule} does not restrict the dependence on the parameters $\vec Q$. Nevertheless, the correlators are rational functions in the parameters $\vec Q$, as can be argued for as follows: Let $\Delta \subset (\mathbb{C}^*)^r$ be the discriminant of the two-dimensional gauge theory in the complexified Fayet--Iliopoulos moduli space \cite{Morrison:1994fr}, at which the gauged linear sigma model becomes singular due to non-compact Coulomb branches.\footnote{See ref.~\cite{Aspinwall:2017loy}, for a recent systematic analysis of the discriminant locus in Abelian $\mathcal{N}=(2,2)$ two-dimensional gauge theories.} The correlation functions $\kappa_{\vec n,\vec m}$ are then globally well-defined on the moduli space $\mathfrak{M}=(\mathbb{C}^*)^r \setminus \Delta$ of Fayet-Iliopoulos parameters. As a consequence, the correlation functions $\kappa_{\vec n,\vec m}$ become rational functions on the compactification $\overline{\mathfrak{M}}$, with poles only along the boundary components~$\Delta$ of the moduli space $\mathfrak{M}$. 

We denote a theory with an axial $U(1)_R$ anomaly arising from $\sum_\ell \vec \rho_\ell > 0$ for all $\ell$ as a gauged linear sigma model with the Fano property. The semi-classical Higgs branch vacuum then arises for large positive Fayet--Iliopoulos parameters, which is the limit $\vec Q \to 0$ \cite{Witten:1993yc}. As a consequence, we expect the correlators $\kappa_{\vec n,\vec m}$ to be finite in this limit. Moreover, assuming the correlators to be finite in the limit $\tm_\ell \to 0$ and $\epsilon \to 0$, the selection rules~\eqref{eq:srule} implies that the correlators $\kappa_{\vec n,\vec m}$ are (weighted homogeneous) polynomials in $\vec Q$ and $\epsilon$ with coefficient functions rational in $\tm_\ell$. For an axial $U(1)_R$ anomaly arising from $\sum_\ell \vec \rho_\ell < 0$ for all $\ell$ --- to be referred to as a gauged linear sigma model with the ample canonical bundle property --- we can argue analogously: Here, the semi-classical Higgs branch emerges in the limit $\vec Q \to \infty$ \cite{Witten:1993yc}. As long as the correlators are finite when $\tm_\ell \to 0$ and $\epsilon \to 0$, they are (weighted homogeneous) polynomials in $Q_1^{-1}, \ldots, Q_r^{-1}$.

Let us now come to the explicit form of the correlators $\kappa_{\vec n,\vec m}$. The starting point of our analysis is the localization formula for the correlation functions~\eqref{eq:AbCorr}, which reads \cite{Closset:2015rna}
\begin{equation}\label{eq:CorrAbGen}
   \kappa_{\vec n,\vec m}(Q,\tm_\ell,\epsilon) \,=\, \left\langle \vec{\sigma}_N^{\vec{n}}\, \vec{\sigma}_S^{\vec{m}}\right\rangle \,=\,
   \sum_{\vec{k}\,\in\,\gamma_m} \vec{Q}^{\vec{k}} 
   \operatorname{Res}_{\vec{\sigma},\vec{k}}^{\vec\xi}\left[\left(\vec{\sigma}-\frac{\epsilon}{2}\vec{k}\right)^{\vec{n}}
   \left(\vec{\sigma}+\frac{\epsilon}{2}\vec{k}\right)^{\vec{m}}Z_{\vec{k}}(\vec{\sigma},\tm_\ell,\epsilon)\right]\ .
\end{equation}
Here, the sum is taken over topological sectors $\vec k$ corresponding to the magnetic charge lattice $\gamma_m\simeq\mathbb{Z}^r$ of the $U(1)^r$ gauge theory, and
\begin{equation} \label{eq:1loopdet}
  Z_{\vec{k}}(\vec{\sigma},\tm_\ell,\epsilon) = \prod_{l=1}^M Z^{(\ell)}_{\vec{k}}(\vec{\sigma},\tm_\ell,\epsilon) \ , \quad
  Z^{(\ell)}_{\vec{k}}(\vec{\sigma},\tm_\ell,\epsilon) = 
  \epsilon^{\mathfrak{q}_l-\vec{\rho}_l\cdot\vec{k}-1} 
  \frac{\Gamma\left(\frac{\vec{\rho}_l\cdot \vec{\sigma}+\tm_\ell}{\epsilon}+\frac{\mathfrak{q}_l-\vec{\rho}_l\cdot\vec{k}}{2}\right)}
  {\Gamma\left(\frac{\vec{\rho}_l\cdot \vec{\sigma}+\tm_\ell}{\epsilon}-\frac{\mathfrak{q}_l-\vec{\rho}_l\cdot\vec{k}}{2}+1\right)}
\end{equation}
are the one loop determinants of the matter multiplets $\phi_\ell$, $\ell=1,\ldots,M$. Note that, due to the infinite sum over topological sectors $\vec k$, the algebraic properties of the correlators discussed above are not manifest in the localization formula~\eqref{eq:CorrAbGen}.

The residue symbol $\operatorname{Res}_{\vec \sigma, \vec{k}}^{\vec\xi}$ in formula~\eqref{eq:CorrAbGen} depends on the gauge theory phase as specified by the Fayet--Ilipoulos parameter $\vec\xi$ and deserves further explanations: Firstly, consider a set of $r$ chiral multiplets $\phi_{\ell_1}, \ldots, \phi_{\ell_r}$ with $1\leq \ell_1 < \ldots < \ell_r \leq M$ such that their charge vectors $\vec\rho_{\ell_1},\ldots,\vec\rho_{\ell_r}$ are linearly independent. These vectors span an $r$-dimensional cone in the electric charge lattice $\gamma_e \simeq \mathbb{Z}^r$ \cite{Closset:2015rna}, which we denote as $\sigma^{(r)}_{\ell_1,\ldots,\ell_r}$. Further, we define $\Sigma(r)$ as the set of all such cones. Secondly, let $\Pi^{(\ell_1,\ldots,\ell_r)}_{\vec{k}}$ be the (countable) set of poles in the variable of integration $\vec\sigma \in \mathbb{C}^r$ given by all simultaneous solutions to the equations $\left(Z^{(\ell_1)}_{\vec{k}}\right)^{-1}= 0, \ldots, \left(Z^{(\ell_r)}_{\vec{k}}\right)^{-1} = 0$. To each cone $\sigma^{(r)}_{\ell_1,\ldots,\ell_r}\in \Sigma(r)$ we then assign the residue symbol\footnote{For a rigorous definition of such higher-dimensional residues, see, for instance, ref.~\cite{MR1288523}.} 
\begin{equation}
    \operatorname{Res}^{\sigma^{(r)}_{\ell_1,\ldots,\ell_r}}_{\vec{\sigma},\vec{k}} \left( \strut\ldots\strut \right) \,=\,  \sum_{\vec{x}\, \in \, \Pi^{(\ell_1,\ldots,\ell_r)}_{\vec{k}}}  \operatorname{Res}_{\vec\sigma = \vec{x}} \left(\strut\ldots\strut \right) \ .
\end{equation}
Thirdly, we define the restricted set of $r$-dimensional cones $\Sigma^{\vec\xi}(r)$ constraint to contain the vector $\vec\xi$ of Fayet--Ilipoulos parameters, i.e.,
\begin{equation} \label{eq:ConeRestricted}
   \Sigma^{\vec\xi}(r) \,=\, \left\{ \, \sigma^{(r)}_{\ell_1,\ldots,\ell_r} \in \Sigma(r) \, \middle| \, \vec\xi \in \sigma^{(r)}_{\ell_1,\ldots,\ell_r} \, \right\} \ .
\end{equation}   
Then, the residue symbol $\operatorname{Res}_{\vec \sigma}^{\vec\xi}$ yields the sum of those poles attributed to cones in the restricted set $\Sigma^{\vec\xi}(r)$, namely
\begin{equation} \label{eq:ResSymbol}
    \operatorname{Res}_{\vec \sigma,\vec{k}}^{\vec\xi} \left( \strut\ldots\strut \right) \,=\, 
    \sum_{\sigma^{(r)}_{\ell_1,\ldots,\ell_r}\, \in \,\Sigma^{\vec\xi}(r)} \operatorname{Res}^{\sigma^{(r)}_{\ell_1,\ldots,\ell_r}}_{\vec{\sigma},\vec{k}} \left( \strut\ldots\strut \right) \ .
\end{equation}

It is important to stress that formula~\eqref{eq:CorrAbGen} is valid only for generic choices of the twisted masses $\tm_\ell$, such that the pole sets $\Pi^{(\ell_1,\ldots,\ell_r)}_{\vec{k}} \subset \mathbb{C}^r$ with the same $\vec{k}$ are mutually disjoint. Since the correlators $\kappa_{\vec n,\vec m}(Q,\tm_\ell,\epsilon)$ are continuous in the twisted masses $\tm_\ell$, we can extend their definition to non-generic values of $\tm_\ell$ with intersecting pole sets by taking the limit to these non-generic values --- as long as this limit exists. On the contrary, it is typically not allowed to take this limit in formula~\eqref{eq:CorrAbGen} already. If one nevertheless decides to set the twisted masses to zero right from the start, it is necessary to give a prescription for the evaluation of higher-dimensional residues with intersecting poles. This can, for instance, be achieved by introducing auxiliary parameters to separate the intersecting poles, which are eventually set to zero after evaluating the residue; see the Appendix of ref.~\cite{Gerhardus:2015sla} for a related discussion. Such auxiliary parameters play a similar role as a choice of generic twisted masses.

\subsubsection{Relations of correlators}\label{sec:RelofCor}
From the localization formula~\eqref{eq:CorrAbGen} we first deduce two basic properties of the correlators, namely
\begin{align}
  &\kappa_{\vec n+\vec e_i,\vec m}(\vec Q,\tm_\ell,\epsilon) - \kappa_{\vec n,\vec m+\vec e_i}(\vec Q,\tm_\ell,\epsilon) 
  \,=\, -\epsilon\, Q_i \partial_{Q_i} \kappa_{\vec n,\vec m}(\vec Q,\tm_\ell,\epsilon)
  \label{eq:RelDer} \ ,\\[1ex]
  &\kappa_{\vec n,\vec m}(\vec Q,\tm_\ell,\epsilon) \,=\, (-1)^{d+\sum_{i=1}^r (n_i+ m_i)} \kappa_{\vec m,\vec n}((-1)^{\sum_l \vec\rho_l}\vec Q,-\tm_\ell,\epsilon) \ .
  \label{eq:RelSym}
\end{align}
Here, we use the definition~\eqref{eq:ddef} and write $\vec e_i$ for the $i$-th unit vector in $\mathbb{Z}^r$ as well as $(-1)^{\vec \alpha} \vec Q = \left( (-1)^{\alpha_1} Q_1, \ldots, (-1)^{\alpha_r} Q_r \right)$. The former identity immediately follows from the localization formula~\eqref{eq:CorrAbGen}, and the latter is a consequence of the equality
\begin{equation}
  Z_{\vec{k}}(-\vec{\sigma},-\tm_\ell,\epsilon) =
 Z_{\vec{k}}(\vec{\sigma},\tm_\ell,\epsilon) \prod_{l=1}^M \frac{\sin\pi\left(1-\frac{(\vec{\rho}_l\cdot \vec{\sigma}+\tm_\ell)}{\epsilon}-\frac{\mathfrak{q}_l-\vec{\rho}_l\cdot \vec{k}}{2}\right)}{\sin\pi\left(-\frac{(\vec{\rho}_l\cdot \vec{\sigma}+\tm_\ell)}{\epsilon}+\frac{\mathfrak{q}_l-\vec{\rho}_l\cdot \vec{k}}{2}\right)}
 = (-1)^{d+r +\sum_{l} \vec{\rho}_l\cdot \vec{k}}  Z_{\vec{k}}(\vec{\sigma},\tm_\ell,\epsilon) \ .
\end{equation}

The primary objective of this section is to deduce linear relations among the correlators $\kappa_{\vec n,\vec m}$ of the form 
\begin{equation} \label{eq:Relansatz}
   R_S(\vec Q,\tm_\ell,\epsilon,\kappa_{\vec n,\,\cdot\,}) \,=\, \sum_{\vec m=0}^{\vec N} c_{\vec m}(\vec Q,\tm_\ell,\epsilon) \kappa_{\vec n,\vec m}(\vec Q,\tm_\ell,\epsilon)
   \,=\, \sum_{\vec m=0}^{\vec N} c_{\vec m}(\vec Q,\tm_\ell,\epsilon)\left\langle \sigma_N^{\vec n}\sigma_S^{\vec m} \right\rangle \,=\, 0 \ ,
\end{equation}
where we demand the coefficient functions $c_{\vec m}(\vec Q,\tm_\ell,\epsilon)$ to be polynomial in $\vec Q$. They can thus be expanded as
\begin{equation}
   c_{\vec m}(\vec Q,\tm_\ell,\epsilon) \,=\, \sum_{\vec p=0}^{\vec s} c_{\vec m,\vec p}(\tm_\ell,\epsilon){\vec Q}^{\vec p} \,=\,
   \sum_{p_1=0}^{s_1}\ldots  \sum_{p_r=0}^{s_r} c_{\vec m,p_1,\ldots,p_r}(\tm_\ell,\epsilon) Q^{p_1}_1 \cdots Q^{p_r}_r
\end{equation}
for some suitable finite vector $\vec s$. Note that the relations of the type~\eqref{eq:Relansatz} are non-trivally independent of the north pole insertions $\vec \sigma_N$ in that they hold for all $\vec n \in \mathbb{Z}^r_{\ge 0}$. We therefore refer to them as south pole relations. Analogously, we can define north pole relations $R_N(\vec Q,\tm_\ell,\epsilon,\kappa_{\,\cdot\,,\vec m})$, which do not depend on the south pole insertions $\vec \sigma_S$ and thus hold for all $\vec m \in\mathbb{Z}^r_{\ge 0}$. Due to the symmetry property~\eqref{eq:RelSym}, each south pole relation $R_S$ yields a north pole relation $R_N$ and vice versa by replacing the coefficient functions $c_{\vec m}$ according to $c_{\vec m}(\vec Q,\tm_\ell,\epsilon) \to (-1)^{\sum_i m_i} c_{\vec m}((-1)^{\sum_l \vec \rho_l}\vec Q,-\tm_\ell,\epsilon)$. South and north pole relations are thus in one to one correspondence.

Let us now determine the possible south pole relations $R_S$. First, we define a modified residue symbol that is independent of the topological sector $\vec{k}$. For this purpose we associate to every cone $\sigma^{(r)}_{\ell_1,\ldots,\ell_r} \in \Sigma(r)$ the pole lattice $P_{\sigma^{(r)}_{\ell_1,\ldots,\ell_r}}$ given by
\begin{equation} \label{eq:plattices}
  P_{\sigma^{(r)}_{\ell_1,\ldots,\ell_r}} \,=\, \left\{ \vec\sigma \in \mathbb{C}^r \, \middle| \, 
 2 \vec\rho_{\ell_i} \left( \frac{\vec\sigma}{\epsilon} \right)+ \left( \frac{2\mathfrak{m}_{\ell_i}}{\epsilon} + \mathfrak{q}_{\ell_i} \right) \in \mathbb{Z} \text{ for all $\ell_i$}\, \right\} \ .
\end{equation}
This is a discrete set due to the linear independence of the charge vectors $\vec\rho_\ell$, and it includes all poles associated to the cone $\sigma^{(r)}_{\ell_1,\ldots,\ell_r}$, namely $\Pi^{(\ell_1,\ldots,\ell_r)}_{\vec{k}}\subset  P_{\sigma^{(r)}_{\ell_1,\ldots,\ell_r}}$ for all $\vec{k}$. Further, under the assumption of generically chosen twisted masses $\tm_\ell$, the pole lattices are mutually disjoint. Introducing the modified residue symbol
\begin{equation} \label{eq:ModResSymbol}
   \widetilde{\operatorname{Res}}_{\vec\sigma}^{\vec\xi} \left( \strut\ldots\strut \right) \,=\,
   \sum_{\sigma^{(r)}\, \in \,\Sigma^{\vec\xi}(r)} \sum_{\vec{x}\, \in\, P_{\sigma^{(r)}}} \operatorname{Res}_{\vec\sigma = \vec{x}} \left(  \strut\ldots\strut \right) \ ,
\end{equation}
the localization formula~\eqref{eq:CorrAbGen} for the correlators can therefore be rewritten as
\begin{equation} \label{eq:Cormod}
\kappa_{\vec n,\vec m}(Q,\tm_\ell,\epsilon) \,=\, 
   \sum_{\vec{k}\,\in\,\gamma_m} \vec{Q}^{\vec{k}} \ 
    \widetilde{\operatorname{Res}}_{\vec\sigma}^{\vec\xi}
   \left[\left(\vec{\sigma}-\frac{\epsilon}{2}\vec{k}\right)^{\vec{n}}
   \left(\vec{\sigma}+\frac{\epsilon}{2}\vec{k}\right)^{\vec{m}}Z_{\vec{k}}(\vec{\sigma},\tm_\ell,\epsilon)\right]\ .
\end{equation}
Note, however, that this formula does not necessarily hold for non-generic twisted masses. Second, we insert eq.~\eqref{eq:Cormod} in the definition~\eqref{eq:Relansatz} and collect common powers of $\vec{Q}$. After the change of variables $\vec w = \vec \sigma + \epsilon \frac{\vec k+\vec p}2$, which maps the pole lattices to themselves, we arrive at
\begin{equation} \label{eq:AnsatzCor}
  0 = R_S(\vec Q,\kappa_{\vec n,\,\cdot\,}) = \sum_{\vec k \in \gamma_m} \vec Q^{\vec k} \sum_{\vec{m}=0}^{\vec{N}} \sum_{\vec{p}=0}^{\vec{s}}c_{\vec m,\vec p}\
  \widetilde{\operatorname{Res}}_{\vec w}^{\vec\xi}
  \left[  (\vec w - \epsilon \vec k)^{\vec n} (\vec w - \epsilon \vec p)^{\vec m} Z_{\vec k - \vec p}(\vec w - \epsilon \tfrac{\vec k+\vec p}2,\tm_\ell,\epsilon) \right] \ .
\end{equation}
This expression can only be zero if the coefficients of all powers of $\vec{Q}$ vanish separately. With the help of the Gamma function identity
\begin{equation}
    \Gamma(x-y) \,=\, \Gamma(x)\cdot \frac{\prod_{s=1+y}^{+\infty}(x-s)}{\prod_{s=1}^{+\infty}(x-s)}  \ ,
\end{equation}
we thus obtain the constraint
\begin{multline}
0 = \widetilde{\operatorname{Res}}_{\vec w}^{\vec\xi} \left[
(\vec{w}-\epsilon\,\vec{k})^{\vec{n}}Z_{\vec{k}}(\vec w - \epsilon \tfrac{\vec k}2,\tm_\ell,\epsilon)\right.\\
\left.\sum_{\vec{p} = 0}^{\vec{s}} \sum_{\vec{m}=0}^{\vec{N}}c_{\vec{m},\vec{p}} \left(\vec{w}-\epsilon \,\vec{p}\right)^{\vec{m}}\prod_{l=1}^M \frac{\prod\limits_{s=1}^{\infty}\big(\vec{w}\cdot\vec{\rho}_l+\tm_l+\epsilon\,(1-\frac{\mathfrak{q}_l}{2}-s)\big)}{\prod\limits_{s=1+\vec{\rho_l}\cdot\vec{p}}^{\infty}\big(\vec{w}\cdot\vec{\rho}_l+\tm_l+\epsilon\,(1-\frac{\mathfrak{q}_l}{2}-s)\big)}
 \right] \ .
\end{multline}
As this equation must hold for $\vec k$, it is necessary that the expression within the residue symbol vanishes itself. The constraint for a south pole relation thus takes the simple form
\begin{equation}\label{eq:RelFinalCond1}
  0 \,=\, \sum_{\vec{p}=0}^{\vec{s}} \alpha_{\vec{p}}(\vec{w},\tm_\ell,\epsilon) \cdot g_{\vec{p}}(\vec{w},\tm_\ell,\epsilon) \ ,
\end{equation}
in terms of the polynomials $\alpha_{\vec{p}}$ and rational functions $g_{\vec{p}}$ given by
\begin{equation} \label{eq:DefWandG}
\begin{aligned}
  \alpha_{\vec{p}}(\vec{w},\tm_\ell,\epsilon)  \,&=\, \sum_{\vec{n}=0}^{\vec{N}}c_{\vec{n},\vec{p}}(\tm_\ell,\epsilon) \left(\vec{w}-\epsilon \,\vec{p}\right)^{\vec{n}} \ , \\
  g_{\vec{p}}(\vec{w},\tm_\ell,\epsilon) \,&=\, 
  \prod_{l=1}^M \frac{\prod\limits_{s=1}^{\infty}\big(\vec{w}\cdot\vec{\rho}_l+\tm_\ell+\epsilon\,(1-\frac{\mathfrak{q}_l}{2}-s)\big)}{\prod\limits_{s=1+\vec{\rho_l}\cdot\vec{p}}^{\infty}\big(\vec{w}\cdot\vec{\rho}_l+\tm_\ell+\epsilon\,(1-\frac{\mathfrak{q}_l}{2}-s)\big)} \ .
\end{aligned}  
\end{equation}
Note that this expression is manifestly independent of the north pole insertions.

We observe that the rational functions $g_{\vec{p}}$ are entirely fixed by the spectrum of the gauge theory under consideration. Determining south pole relations $R_S$ of a given gauge theory thus amounts to finding polynomials $\alpha_{\vec{p}}$ satisfying the constraints~\eqref{eq:RelFinalCond1}. This is, in fact, a well-studied problem in commutative algebra: The set $M_S$ of polynomial solutions $\alpha_{\vec{p}}$ forms the syzygy module over the polynomial ring $\mathbb{C}(\tm_\ell)[\vec w,\epsilon]$ of the rational function $g_{\vec p}$.\footnote{The elements of the polynomial ring $\mathbb{C}(\tm_\ell)[\vec w,\epsilon]$ are polynomials in $\vec w$ and $\epsilon$ with coefficients in the field of complex rational functions in the twisted masses $\tm_\ell$, which is denoted by $\mathbb{C}(\tm_\ell)$.} From a given element $\alpha_{\vec p}$ in the south pole syzygy module $M_S$ we then readily reconstruct the south pole correlator relation as
\begin{equation} \label{eq:RSDef}
    0 \,=\, R_S(\vec Q,\tm_\ell,\epsilon,\kappa_{\vec n,\,\cdot\,}) \,=\, \sum_{\vec p=0}^{\vec s} \vec Q^{\vec p}
    \left\langle\vec\sigma_N^{\vec n} \, \alpha_{\vec p}(\vec\sigma_S + \epsilon \vec p,\tm_\ell,\epsilon) \right\rangle \ .
\end{equation}

We should reflect that the generators of the syzygy module $M_S$ do not necessarily yield independent south pole correlator relations. Namely, let us consider the quotient
\begin{equation}
  \frac{g_{\vec{p}+\vec{e}_i}(\vec{w}+\epsilon\,\vec{e}_i,\tm_\ell,\epsilon)}{g_{\vec{p}}(\vec{w},\tm_\ell,\epsilon)} 
  =\prod_{l=1}^M\frac{\prod\limits_{s=1-\rho_{l,i}}^\infty \big(\vec{w}\cdot\vec{\rho}_l+\tm_\ell+\epsilon\,(1-\frac{\mathfrak{q}_l}{2}-s)\big)}{\prod\limits_{s=1}^\infty \big(\vec{w}\cdot\vec{\rho}_l+\tm_\ell+\epsilon\,(1-\frac{\mathfrak{q}_l}{2}-s)\big)} \ ,
\end{equation}
in terms of the $i$-th unit vector $\vec e_i \in \mathbb{Z}^r$. We observe that this quotient is independent of $\vec{p}$. Therefore, given an element $\alpha_{\vec{p}}$ of the south pole syzygy module $M_S$, we arrive at another element $\tilde \alpha_{\vec p}$ of $M_S$ by setting
\begin{align}
\tilde{\alpha}_{\vec{p}}(\vec{w},\tm_\ell,\epsilon) \,=\,
\begin{cases}
\alpha_{\vec{p}-\vec{e}_i}(\vec{w}-\epsilon\,\vec{e}_i,\tm_\ell,\epsilon) & \text{if } p_i \geq 1 \\
0 & \text{otherwise \ .}
\end{cases}
\end{align}
On the level of south pole correlator relations these two module elements are trivially related as
\begin{equation} \label{eq:RSrels}
   \tilde{R}(\vec{Q},\tm_\ell,\epsilon,\kappa_{\vec n,\,\cdot\,}) \,=\, Q_i \cdot R(\vec{Q},\tm_\ell,\epsilon,\kappa_{\vec n,\,\cdot\,}) \ .
\end{equation}

Note that the module $M_S$ of relations is rather complicated as it is based upon an infinite (but countable) set of rational functions $g_{\vec p}$. However, due to redundancies in the definition of the module $M_S$ reflected in identities of the form~\eqref{eq:RSrels}, it nevertheless suffices to consider only a finite subset of functions~$g_{\vec p}$ to determine all south pole correlator relations. In practice, such a subset needs to be selected case by case.

\subsubsection{Reduced correlator relations for non-generic twisted masses} \label{sec:Reduction}
In the derivation of south pole relations we have so far assumed generically chosen twisted masses~$\tm_\ell$. Here, we consider non-generic values of the $\tm_\ell$ and argue that there may arise additional correlator relations for these cases.

Let us consider a non-generic choice of twisted masses $\tm_{0,\ell}$ such that the limit from generic twisted masses $\tm_\ell$ to $\tm_{0,\ell}$ yields finite and well-defined correlation functions $\kappa_{\vec n,\vec m}(\vec Q,\tm_{0,\ell},\epsilon)$. We can then take the same limit on the level of the syzygy module $M_S$,
\begin{equation}
  M_S^{\lim} \, =\, \lim_{\tm_\ell \to \tm_{0,\ell}} M_S \ ,
\end{equation}  
and the elements of $M_S^{\lim}$ become valid south pole relations for the non-generic twisted masses~$\tm_{0,\ell}$. Alternatively, we can first take the limit $\tm_\ell$ to $\tm_{0,\ell}$ on the level of rational functions~\eqref{eq:DefWandG}. These non-generic rational functions $g_{\vec p}(\vec w,\tm_{0,\ell},\epsilon)$ then define the syzygy module~$M_S^0$.

Since the limit to $\tm_{0,\ell}$ is well-defined for the defining equation~\eqref{eq:RelFinalCond1}, any relation $R_S^{\lim}$ in $M_S^{\lim}$ is also a relation in $M_S^0$. The converse, however, is not true in general: For specific values of the twisted masses $\tm_{0,\ell}$ the non-generic rational functions $g_{\vec p}(\vec w,\tm_{0,\ell},\epsilon)$ may give rise to additional relations $R_S$. For instance, a generically irreducible relation $R_S$ of $M_S$ may become reducible in the described limit, $R_S^{\lim}(\vec Q,\vec w,\tm_{0,\ell},\epsilon) = C(\vec Q,\vec w,\tm_{0,\ell},\epsilon) R_S^0(\vec Q,\vec w,\tm_{0,\ell},\epsilon)$, such that the factor $R_S^0$ is an element of $M_S^0$. On the level of the rational functions $g_{\vec{p}}$ this phenomenon appears if there are cancellations of factors originating from different fields in the considered limit. In summary, the limiting module $M_S^{\lim}$ is thus a submodule of the non-generic module $M_S^0$, namely
\begin{equation} \label{eq:SubMod}
   M_S^0 \, \supset \, M_S^{\lim} \ .
\end{equation}

Let us stress the following: The derivation presented in Section~\ref{sec:RelofCor} does not guarantee an element of $M_S^0 \setminus M_S^{\lim}$ to be a valid south pole correlator relation, because for non-generic twisted masses $\tm_{0,\ell}$ the pole lattices defined in eq.~\eqref{eq:plattices} may no longer be disjoint. If an intersection occurs between a pole lattice associated to a cone in $\Sigma^{\vec\xi}(r)$ and a cone in $\Sigma(r)\setminus\Sigma^{\vec\xi}(r)$, eq.~\eqref{eq:Cormod} is typically not correct. Demanding eq.~\eqref{eq:Cormod} to be applicable, is, however, too strong of a requirement. There is in fact a weaker condition ensuring validity of the relations in $M_S^0 \setminus M_S^{\lim}$: After a change of variable to $\vec{v} = \vec{\sigma} -\tfrac{\epsilon}{2}\vec{k}$, consider the union of all pole sets associated to cones in $\Sigma^{\vec\xi}(r)$ and similar the union of all pole sets associated to cones in $\Sigma(r) \setminus\Sigma^{\vec\xi}(r)$, namely
\begin{equation}
\begin{aligned}
\Theta(\vec \xi,\tm_\ell) &= \bigcup\limits_{\sigma^{(r)}_{\ell_1,\ldots,\ell_r}\,\in \,\Sigma^{\vec\xi}(r)} 
\left\{ \vec{v} \in \mathbb{C}^r \,\middle| \, Z^{(\ell_i)}_{\vec{k}}\left(\vec{v}+\tfrac{\epsilon}{2}\vec{k},\mathfrak{m}_\ell,\epsilon\right)^{-1}= 0 \,\text{ for all } 1\leq i \leq r\right\} \ , \\
\Omega(\vec \xi,\tm_\ell) &= \bigcup\limits_{\sigma^{(r)}_{\ell_1,\ldots,\ell_r}\,\in \,\Sigma(r)\setminus\Sigma^{\vec\xi}(r)}
\left\{ \vec{v} \in \mathbb{C}^r \,\middle| \, Z^{(\ell_i)}_{\vec{k}}\left(\vec{v}+\tfrac{\epsilon}{2}\vec{k},\mathfrak{m}_\ell,\epsilon\right)^{-1}= 0 \,\text{ for all } 1\leq i \leq r\right\} \ .
\end{aligned}
\end{equation}
For generic twisted masses $\tm_\ell$ the intersection $\Theta(\vec \xi,\tm_\ell) \cap \Omega(\vec\xi,\tm_\ell)$ is empty by construction. If the intersection is still empty for the non-generic twisted masses $\tm_{0,\ell}$, the non-generic syzygy module $M_S^0$ is guaranteed to correctly describe correlator relations in the limit $\tm_\ell \to \tm_{0,\ell}$.\footnote{Recall that we always assume a well-defined limit $\tm_\ell \to \tm_{0,\ell}$ on the level of correlators.} Intuitively, this condition ensures that there is no overlap between those poles that for a given Fayet--Ilipoulos parameter $\vec\xi$ contribute to the correlators and those poles that do not contribute. In summary:
\begin{equation} \label{eq:OverlapCond}
   \Theta(\vec \xi,\tm_{0,\ell}) \cap \Omega(\vec\xi,\tm_{0,\ell})=\emptyset \
   \Longrightarrow \
   \text{$M_S^0$ captures physically valid correlator relations $R_S^0$.} 
\end{equation}

In the discussion of explicit examples in Section~\ref{sec:Examples} we come back to the phenomenon of $M_S^{\lim}$ becoming a proper submodule of the non-generic module $M_S^0$ for certain non-generic choices of the twisted masses. Depending on the example under consideration, elements $R_S^0$ of $M_S^0$ may or may not be valid south pole correlator relations. 

\subsection{Differential operators from correlator relations} \label{sec:diffideal}
As discussed in refs.~\cite{Beem:2012mb,Benini:2015noa}, the localization formula~\eqref{eq:CorrAbGen} for the correlators decomposes into a quadratic form of suitable holomorphic blocks. Due to this decomposition property, the correlators enjoy an interpretation as matrix elements in a Hilbert space of states \cite{Cecotti:1991me}.

From this point of view we can interpret the south pole correlator relations~$R_S$ as operators~$\boldsymbol{R}_S$ annihilating the moduli-dependent ground state $\left| \Omega(\vec\xi,\vec\theta) \right\rangle$ of the gauge theory.  We thus explicitly have 
\begin{equation} \label{eq:RSOp}
  \boldsymbol{R}_S(\boldsymbol{\vec Q},\boldsymbol{\vec\sigma}_S,\tm_\ell,\epsilon)  \left| \Omega(\vec\xi,\vec\theta)\right\rangle \,=\, 0 \quad\text{with}\quad
  \boldsymbol{R}_S(\boldsymbol{\vec Q},\boldsymbol{\vec\sigma}_S,\tm_\ell,\epsilon) 
  \,=\, \sum_{\vec p=0}^{\vec s} \boldsymbol{\vec Q}^{\vec p} \alpha_{\vec p}(\boldsymbol{\vec\sigma}_S+\epsilon\vec p,\tm_\ell, \epsilon) \ ,
\end{equation}
where the boldface letters indicate the operator nature of $\boldsymbol{R}_S$ acting on the Hilbert space of states. 

For a given south pole relation $R_S$ arising from the polynomials $\alpha_{\vec p}(\vec w,\tm_\ell,\epsilon)$ we readily deduce other correlator relations $R'_S$, for instance by taking $\alpha'_{\vec p}(\vec w,\tm_\ell,\epsilon) = w_i \alpha_{\vec p}(\vec w,\tm_\ell,\epsilon)$ for all $\vec{p}$. These new polynomials correspond to an operator $\boldsymbol{R}'_S$, which is also obtained by multiplying  $\boldsymbol{\sigma}_{S,i}$ to $\boldsymbol{R}_S$ from the left together with the commutation relation
\begin{equation} \label{eq:com}
  \left[  \boldsymbol{\sigma}_{S,i} , \boldsymbol{Q}_j\right] \,=\, \delta_{ij} \epsilon \boldsymbol{Q}_j \ .
\end{equation} 
This can be seen from an explicit calculation:
\begin{equation}
\begin{aligned}
  \boldsymbol{R}'_S(\boldsymbol{\vec Q},\boldsymbol{\vec\sigma}_S,\tm_\ell,\epsilon) 
  \,&=\,\boldsymbol{\sigma}_{S,i} \sum_{\vec p=0}^{\vec s} \boldsymbol{\vec Q}^{\vec p} \alpha_{\vec p}(\boldsymbol{\vec\sigma}_S+\epsilon\vec p,\tm_\ell, \epsilon)
  \,=\, \sum_{\vec p=0}^{\vec s} \boldsymbol{\vec Q}^{\vec p} (\boldsymbol{\sigma}_{S,i}+\epsilon p_i)\alpha_{\vec p}(\boldsymbol{\vec\sigma}_S+\epsilon\vec p,\tm_\ell, \epsilon)\\
  \,&=\, \sum_{\vec p=0}^{\vec s} \boldsymbol{\vec Q}^{\vec p} \alpha'_{\vec p}(\boldsymbol{\vec\sigma}_S+\epsilon\vec p,\tm_\ell, \epsilon) \ .
\end{aligned}  
\end{equation}

Let us describe this in algebraic terms. The commutation relation~\eqref{eq:com} characterizes the non-commutative south pole ring
\begin{equation}
    \boldsymbol{\mathcal{R}}_S \,=\, \ \raisebox{1ex}{$\mathbb{C}(\tm_\ell)\langle \boldsymbol{\vec Q}, \boldsymbol{\vec\sigma}_S, \epsilon \rangle$}
         \Big/ 
         \raisebox{-1ex}{$\left[  \boldsymbol{\sigma}_{S,i} , \boldsymbol{Q}_j\right] = \delta_{ij} \epsilon \boldsymbol{Q}_j$} \ ,
\end{equation}
and the set of south pole operators $\boldsymbol{R}_S$ annihilating the ground state $\left| \Omega(\vec\xi,\vec\theta) \right\rangle$ forms a left ideal~$\boldsymbol{\mathcal{I}}_S$ in this ring $\boldsymbol{\mathcal{R}}_S$, which according to eq.~\eqref{eq:RelFinalCond1} is explicitly given by
\begin{equation}
   \boldsymbol{\mathcal{I}}_S \,=\, \left\{\, \sum_{\vec p} \boldsymbol{\vec Q}^{\vec p} \alpha_{\vec p}(\boldsymbol{\vec\sigma}_S+\epsilon\vec{p},\tm_\ell,\epsilon) \in \boldsymbol{\mathcal{R}}_S \, \middle|\,
   0=\sum_{\vec p} \alpha_{\vec p}(\vec\sigma_S,\tm_\ell,\epsilon) \cdot g_{\vec p}(\vec\sigma_S,\tm_\ell,\epsilon ) \right\} \ .
\end{equation} 

We note that an explicit representation of the commutation relation~\eqref{eq:com} and hence a representation of the non-commutative ring $\boldsymbol{\mathcal{R}}_S$ is given by
\begin{equation} \label{eq:OpRep}
\boldsymbol{Q}_i \,=\, Q_i \ , \qquad   \boldsymbol{\sigma}_{S,i} \,=\, \epsilon\,Q_i \frac{\partial}{\partial Q_i} \,=\, \epsilon\,\Theta_i \ ,
\end{equation}
which can be interpreted as a representation with respect to the eigenstates of the monopole operators $\boldsymbol{Q}_i$. Due to the relationship of $\vec Q$ with the complexified Fayet--Iliopoulos parameters as in eq.~\eqref{eq:Qdef}, we can then view the operators $\boldsymbol{R}_S \in  \boldsymbol{\mathcal{I}}_S$ as differential operators annihilating the $\vec Q$-dependent gauge theory ground state according to
\begin{equation}\label{eq:OperatorOnVac}
   \boldsymbol{R}_S(\vec Q,\epsilon \vec\Theta,\tm_\ell,\epsilon) \left| \Omega(\vec Q) \right\rangle \,=\, 0 \ .
\end{equation}
We subsequently refer to $\boldsymbol{\mathcal{I}}_S$ as the differential ideal, and the solutions to its differential equations $\boldsymbol{R}_S(\vec Q,\epsilon \vec\Theta,\tm_\ell,\epsilon) f(\vec{Q}) = 0$ capture the $\vec Q$-dependence of the gauge theory ground state $\left| \Omega(\vec Q) \right\rangle$. As we will see in the explicit examples discussed in Section~\ref{sec:Examples}, for gauge theories with a geometric target space interpretation $\boldsymbol{\mathcal{I}}_S$ becomes the differential ideal governing the Gromov--Witten theory of the target space.

For particular target space geometries Closset et al.~also derive certain subclasses of differential operators from different but related considerations \cite{Closset:2015rna}. Their findings are in agreement with our general formulas for the differential operators obtained from the south pole correlator relations.

Analogously, we can derive the north pole differential ideal from the north pole operators~$\boldsymbol{R}_N$. This does not give novel information since north pole operators are in one-to-one correspondence with south pole operators.

\subsection{Correlator relations: Non-Abelian gauge groups} \label{sec:nonAbCor}
In this section we extend the study of correlators and their relations to gauged linear sigma models with non-Abelian gauge groups $G$. In doing so, our first task is to introduce the non-Abelian gauge theory correlators. To derive their relations and the corresponding differential operators, we analyze the non-Abelian gauge theory in its Coulomb branch, where the gauge group $G$ is spontaneously broken to its maximal torus $T\simeq U(1)^{\operatorname{rk} G}$. As explained in ref.~\cite{Closset:2015rna}, the $W$-bosons of the twisted chiral fields $\Sigma$ in the adjoint representation of $G$ then contribute to the one-loop determinants as chiral multiplets with their respective charge under the unbroken Abelian group $T\simeq U(1)^{\operatorname{rk} G}$ and with R-charge two. This essentially reduces the derivation of non-Abelian correlator relations to the previously discussed Abelian case, see Sections~\ref{sec:relcorr} and \ref{sec:diffideal}.

\subsubsection{Non-Abelian correlation functions}
Let us now consider correlation functions of gauged linear sigma models with non-Abelian compact gauge groups~$G$ of the form
\begin{equation} \label{eq:nonAbG}
   G \,=\, (U(1)^{r'} \times G') / \Gamma \ .
 \end{equation}
Here, the non-Abelian factor $G'$ (with $\operatorname{rk} G'=\operatorname{rk} G - r'$) is a product of semi-simple Lie groups and $\Gamma$ is a discrete and normal subgroup of $U(1)^{k'}\times G'$.\footnote{The discrete subgroup $\Gamma$ can certainly be trivial.} Further, the non-Abelian gauge theory spectrum consists of the twisted chiral multiplet $\underline{\Sigma}$ in the adjoint representation of $G$, as well as the chiral multiplets $\phi_\alpha$ in irreducible representations $\rho_\alpha$, $\alpha=1,\ldots,A$, with R-charge $\mathfrak{q}_\alpha$ and twisted mass $\tm_\alpha$. This is summarized in the left column of Table~\ref{tab:SpecCoulombBranch}. 

The topological sectors of this non-Abelian gauge theory are characterized by the rank $\operatorname{rk} G$ magnetic charge lattice of the gauge group $G$ \cite{Goddard:1976qe}. These sectors are labelled by the formal parameters 
\begin{equation}
   Q_k \,=\, e^{-2\pi\xi_k +  i \theta_k} \ , \qquad k=1,\ldots, \operatorname{rk} G \ .
\end{equation}   
Here the first $r'$ parameters $Q_1,\ldots,Q_{r'}$ are associated to the Abelian factor $U(1)^{r'}$ in eq.~\eqref{eq:nonAbG}, while the remaining parameters are auxiliary and vanish on the level of physical quantities. The correlation functions therefore depend only on the parameters $\vec Q' = (Q_1,\ldots, Q_{r'})$ associated to $U(1)^{r'}$ in $G$, whereas the remaining parameters $Q_k$ with $k=r'+1,\ldots,\operatorname{rk} G$ are set to one, i.e., $\vec Q = ( \vec Q',1,\ldots,1)$.\footnote{Note that for non-trivial discrete subgroup $\Gamma$ the chosen parameters $\vec Q = ( Q_1, \ldots, Q_{\operatorname{rk} G})$ do not coincide with the canonical Fayet--Iliopoulos parameters of the maximal torus $T\simeq U(1)^{\operatorname{rk} G}$.}

As correlation functions correspond to physical measurements, they must be independent of gauge choices. That is to say, given a polynomial $f(\adjs_N,\adjs_S)$ in terms of the adjoint-valued twisted chiral operator insertions $\adjs_S$ and $\adjs_N$ at the south and north pole, the associated correlators $\kappa_{f}$ obey
\begin{equation} \label{eq:NonAbCorr}
  \kappa_{f}(\vec Q', \tm_\alpha, \epsilon) \,=\, \left\langle f(\adjs_N,\adjs_S) \right\rangle \,=\, \left\langle f(g^{-1}\adjs_Ng,g^{-1}\adjs_Sg) \right\rangle 
  \quad \text{for all $g\in G$} \ .
\end{equation}  
For these correlators the selection rule~\eqref{eq:srule} generalizes to
\begin{equation} \label{eq:sruleNonAb}
   d + \#(\epsilon) + \# (\tm_\ell) + \sum_{k=1}^r \left(\sum_\ell \rho_{\ell,k}\right) \#(Q_k) \,\in\, \operatorname{pow}(f) \ ,
\end{equation}
where $\operatorname{pow}(f)$ denotes set of degrees of all monomials in $\adjs_N$ and $\adjs_S$ that appear in $f(\adjs_N,\adjs_S)$, and the gravitational anomaly \eqref{eq:ddef} generalizes to
\begin{equation}\label{eq:NonAbCorrdim}
  d\,=\, \sum_\alpha (1- \mathfrak{q}_\alpha) \dim \rho_\alpha - \dim \mathfrak{g} \ ,
\end{equation}
in terms of the dimensions of the representations $\rho_\alpha$ and of the Lie algebra $\mathfrak{g}$ of $G$.

Let us put the non-Abelian correlators $\kappa_{f}$ in the context of invariant theory, see e.g. refs. \cite{MR0255525,MR2004511}. As already used in eq.~\eqref{eq:NonAbCorr}, the $G$-action of the gauge group on the fields $\adjs_N$ and $\adjs_S$ canonically extends to a $G$-action on the polynomial ring $\mathbb{C}[\mathfrak{g}\times\mathfrak{g}]$ in terms of the Lie algebra $\mathfrak{g}$ of $G$. Due to the gauge invariance we can average the correlator $\kappa_{f}$ over the entire compact gauge group $G$ according to
\begin{equation}
    \kappa_{f}(\vec Q', \tm_\alpha, \epsilon) \,=\, \int_G d\mu(g)\,\left\langle f(g^{-1}\adjs_Ng,g^{-1}\adjs_Sg) \right\rangle \ ,
\end{equation}
where $d\mu(g)$ is the probability Haar measure of the compact group $G$ with $\int_G  d\mu(g) = 1$. Due to linearity of the correlator and $G$-invariance of the Haar measure we further find
\begin{equation} \label{eq:ReynoldsOp}
  \kappa_{f}(\vec Q', \tm_\alpha, \epsilon) \,=\, \kappa_{f^*}(\vec Q', \tm_\alpha, \epsilon) \ , \qquad
  f^*(\adjs_N,\adjs_S) \,=\,  \int_G d\mu(g)f(g^{-1}\adjs_Ng,g^{-1}\adjs_Sg) \ ,
\end{equation}
where $f^*(\adjs_N,\adjs_S)$ is an element in the ring of gauge invariant polynomials $\mathbb{C}[\mathfrak{g}\times\mathfrak{g}]^G$. The averaging map $*: \mathbb{C}[\mathfrak{g}\times\mathfrak{g}] \to \mathbb{C}[\mathfrak{g}\times\mathfrak{g}]^G, f \mapsto f^*$ defined by eq.~\eqref{eq:ReynoldsOp} is known as the Reynolds operator \cite{MR0255525,MR2004511}. We can thus compute the correlator of any polynomial $f$ upon projecting to the gauge invariant polynomial $f^*$ with the Reynolds operator, and it suffices to study non-Abelian correlators on the polynomial ring of invariant polynomials $\mathbb{C}[\mathfrak{g}\times\mathfrak{g}]^G$. According to the Hilbert--Nagata theorem this ring $\mathbb{C}[\mathfrak{g}\times\mathfrak{g}]^G$ is finitely generated for the compact Lie group~$G$ \cite{MR0179268}. In practice this simply means that any gauge invariant combination of $\adjs_N$ and $\adjs_S$ can be expressed in terms of a finite generating set of gauge invariant expressions. As an example, for the gauge group $U(2)$ all gauge invariant correlator insertions are functions of the gauge invariant combinations $\operatorname{tr}(\adjs_{N/S})$, $\operatorname{tr}(\adjs_{N/S}^2)$, and $\operatorname{tr}(\adjs_{N}\adjs_{S})$ only. 

In this work we mainly focus on the subclass of correlators $\kappa_{f_N,f_S}$ given by
\begin{equation} \label{eq:NonAbCorrSpecial}
  \left\langle f_N(\adjs_N) f_S(\adjs_S) \right\rangle \,=\, \kappa_{f_N,f_S}(\vec Q', \tm_\alpha, \epsilon) 
  \quad \text{with} \quad 
  f_N,f_S \in \mathbb{C}[\mathfrak{g}]^G \ .
\end{equation}
The product $f_N(\adjs_N)f_S(\adjs_S)$ obviously yields a polynomial in the invariant ring $\mathbb{C}[\mathfrak{g}\times\mathfrak{g}]^G$, while the converse --- namely that an invariant polynomial in $\mathbb{C}[\mathfrak{g}\times\mathfrak{g}]^G$ decomposes into a sum of products of invariant north and south pole polynomials --- does not hold in general.

\subsubsection{Abelianization of the gauge group and the correlation functions}
We now consider the non-Abelian gauged linear sigma model in its Coulomb branch, where the gauge group $G$ is spontaneously broken to $T\simeq U(1)^{\operatorname{rk} G}$ by a non-vanishing generic expectation value of the adjoint-valued scalar field $\adjs$ in the twisted chiral multiplet $\underline{\Sigma}$. This process is also known as Abelianization of the gauge group.

Let $\vec\omega_{\Sigma_i}$ with $i=1,\ldots,\dim \operatorname{adj}(G)$ be the roots of $G$ ---  that is to say the weights of the adjoint representation of $G$. Further, let $\vec\omega_{\alpha_i}$ with $i=1,\ldots,\dim \rho_\alpha$ be the weights of the irreducible representation of dimension $\dim \rho_\alpha$ of the chiral field $\phi_\alpha$. These vectors $\vec\omega_{\Sigma_i}$ and $\vec\omega_{\alpha_i}$ correspond to gauge charges under the unbroken Abelian gauge group $T=U(1)^{\operatorname{rk} G}$. The charged components of $\underline{\Sigma}$ correspond to chiral multiplets with $U(1)_R$-charge two, whose scalar component fields are the $W$-bosons of the spontaneous breaking to $U(1)^{\operatorname{rk} G}$ \cite{Closset:2015rna}. Due to eq.~\eqref{eq:1loopdet} and the Gamma function identity $\Gamma(z)\Gamma(1-z) = \frac{\pi}{\sin \pi z}$, they contribute to the partition function with the factor \cite{Closset:2015rna}
\begin{equation}\label{eq:ZVec}
   Z^{(\Sigma)}_{\vec k}(\vec\sigma,\mathfrak{q}_\ell,\epsilon) \,=\, \prod_{\vec{\omega}_\Sigma>0} 
   (-1)^{\vec{\omega}_\Sigma \cdot \vec k+1}
   \left(\vec{\omega}_\Sigma \cdot \vec\sigma +\frac{\vec{\omega}_\Sigma \cdot \vec k}2\epsilon\right)
   \left(\vec{\omega}_\Sigma \cdot \vec\sigma - \frac{\vec{\omega}_\Sigma \cdot \vec k}2\epsilon\right)\ .
\end{equation}
Here, the product is taken over the positive roots $\vec\omega_\Sigma$, and we used that the non-zero roots come in pairs $(\vec{\omega}_\Sigma,-\vec{\omega}_\Sigma)$. This shows that the $W$-bosons do not give rise to any poles in the one-loop determinant~\eqref{eq:1loopdet}. Further, we denote the twisted chiral multiplets and their scalar field components of the unbroken Abelian subgroup $U(1)^{\operatorname{rk} G}$ by $\vec\Sigma$ and $\vec\sigma$. The non-Abelian spectrum together with its Abelianization is summarized in Table~\ref{tab:SpecCoulombBranch}. We also note that the factor~\eqref{eq:ZVec} could be reinterpreted as a polynomial in the operator insertions $\vec{\sigma}_S$ and $\vec{\sigma}_N$.

\begin{table}[t]
\hfil
\hbox{
\vbox{
\offinterlineskip
\halign{\strut\vrule width1.2pt height15pt#&\hfil~#~\hfil\vrule width0.5pt&\hfil~#~\hfil&#\vrule width1.2pt&\hfil~#~\hfil&\vrule width0.5pt\hfil~#~\hfil&\vrule width1.2pt\hfil~#~\hfil\vrule width1.2pt&\hfil~#~\hfil\vrule width1.2pt\cr 
\noalign{\hrule height 1.2pt}
&\multispan2 \hfil non-Abelian spectrum \hfil && \multispan2 \hfil Coulomb branch spectrum \hfil & $U(1)_R$ & twisted \cr
&non-Abelian mult. & $G$-rep. && Abelian mult. & $U(1)^{\operatorname{rk} G}$-rep. &  charge & mass \cr
\noalign{\hrule height 1.2pt}
&twisted chiral & $\operatorname{adj}(G)$ && chiral mult. $W_1$ & $\vec\omega_{\Sigma_1}$ & $2$ & 0 \cr
&mult. $\underline{\Sigma}$ &&& $\vdots$ & $\vdots$ & $\vdots$ & $\vdots$ \cr
&&&& chiral mult. $W_{\dim \operatorname{adj}(G)}$ & $\vec\omega_{\Sigma_{\dim \operatorname{adj}(G)}}$ & $2$ &0 \cr
\noalign{\hrule height 0.5pt}
&chiral mult. $\phi_\alpha$ & $\rho_\alpha$ && chiral mult. $\phi_{\alpha_1}$ & $\vec\omega_{\alpha_1}$ & $\mathfrak{q}_\alpha$ & $\tm_\alpha$ \cr
&$\alpha=1,\ldots,A$ &&& $\vdots$ & $\vdots$ & $\vdots$ & $\vdots$ \cr
&&&&chiral mult. $\phi_{\alpha_{\dim \rho_\alpha}}$ & $\vec\omega_{\Sigma_{\dim \rho_\alpha}}$ & $\mathfrak{q}_\alpha$ & $\tm_\alpha$ \cr
\noalign{\hrule height 1.2pt}
}
}}
\hfil
\caption{The table shows the decomposition of the non-Abelian gauge theory spectrum into the Abelian spectrum of the Coulomb branch of the gauge theory, where the non-Abelian gauge group $G$ is spontaneously broken to the maximal torus $U(1)^{\operatorname{rk} G}$. The Abelian charge vectors $\vec\omega_{\Sigma_i}$ and $\vec\omega_{\alpha_i}$ are the weights of the non-Abelian representations of the multiplets.}  
\label{tab:SpecCoulombBranch}
\end{table}

On the Coulomb branch the non-Abelian remnant of the gauge transformations acting on the functions $f({\vec{\sigma}}_{N},{\vec{\sigma}}_{S})$, $\dim {\vec{\sigma}}_{N/S}=\operatorname{rk} G$, is given by the Weyl group $\mathcal{W}_G$ of $G$. By definition this is the normal subgroup of $G$ that preserves the maximal torus $\mathfrak{t} \subset \mathfrak{g}$ of dimension ${\operatorname{rk} G}$ modulo the maximal torus $\mathfrak{t}$, namely
\begin{equation}
  \mathcal{W}_G \, =\, 
   \left.\raisebox{0.25ex}{$\left\{ \, g\in G \, \middle| \, g^{-1} \,\mathfrak{t}\, g = \mathfrak{t} \, \right\}$} \middle/ \raisebox{-0.25ex}{$\mathfrak{t}$} \right.  \ .
\end{equation}   

Since the non-Abelian parameters $\vec Q'$ are by construction invariant with respect to the full gauge group $G$, they are in particular invariant with respect to the action of the Weyl group $\mathcal{W}_G$. The Abelianized correlator associated to any polynomial $f(\vec\sigma_N,\vec\sigma_S) \in \mathbb{C}[\mathfrak{t}\times\mathfrak{t}]$ becomes therefore invariant with respect to the action of the Weyl group $\mathcal{W}_G$ upon setting $\vec Q=(\vec Q',1,\ldots,1)$, i.e., 
\begin{equation}
  \kappa_f(\vec Q',\tm_\alpha,\epsilon)\,=\, \left\langle f(\vec\sigma_N,\vec\sigma_S) \right\rangle
  \,=\, \left\langle f(w(\vec\sigma_N),w(\vec\sigma_S)) \right\rangle
   \quad \text{for all $w \in \mathcal{W}_G$} \ .
\end{equation}
Here, $w(\vec\sigma_{N/S})$ denotes to the action of the Weyl group element $w$ on the fields $\vec\sigma_{N/S}$. 

In analogy to the non-Abelian group correlators, we can on the Coulomb branch define the Reynolds operator $*: \mathbb{C}[\mathfrak{t}\times\mathfrak{t}] \to  \mathbb{C}[\mathfrak{t}\times\mathfrak{t}]^{\mathcal{W}_G}$ by projection to the Weyl invariant part of the polynomials $f(\sigma_N,\sigma_S)$, namely
\begin{equation} \label{eq:ReynoldsOp2}
  \kappa_{f}(\vec Q', \tm_\alpha, \epsilon) \,=\, \kappa_{f^*}(\vec Q', \tm_\alpha, \epsilon) \ , \qquad
  f^*(\vec\sigma_N,\vec\sigma_S) \,=\,  \frac{1}{| \mathcal{W}_G |} 
  \sum_{w\in\mathcal{W}_G} f(w(\vec\sigma_N),w(\vec\sigma_S)) \ .
\end{equation}
Here ${| \mathcal{W}_G |}$ is the order of the Weyl group $\mathcal{W}_G$.

Moreover, Coulomb branch correlators based on the $\mathcal{W}_G$-invariant polynomials $f_N(\vec\sigma_N)$ and  $f_S(\vec\sigma_S)$ are of particular interest in this  work and take the specialized form
\begin{equation} \label{eq:AbCorrSpecial}
  \left\langle f_N(\vec\sigma_N) f_S(\vec\sigma_S) \right\rangle \,=\, \kappa_{f_N,f_S}(\vec Q', \tm_\alpha, \epsilon) 
  \quad \text{with} \quad 
  f_N, f_S \in \mathbb{C}[\mathfrak{t}]^{\mathcal{W}_G}  \ .
\end{equation}

The connection between the non-Abelian correlators and the Coulomb branch correlators is established through the Luna--Richardson theorem \cite{MR544240}. Applied to our situation it asserts that the restriction map from G-invariant to $\mathcal{W}_G$-invariant polynomials,
\begin{equation} \label{eq:Restriction}
   R: \mathbb{C}[\mathfrak{g}]^G \stackrel{\simeq}{\longrightarrow} \mathbb{C}[\mathfrak{t}]^{\mathcal{W}_G} \ ,
 \end{equation}  
is an isomorphism. We can thus ambiguously reconstruct the non-Abelian correlators \eqref{eq:NonAbCorrSpecial} in terms of the adjoint-valued fields $\adjs_N$ and $\adjs_S$ from the Coulomb branch correlators \eqref{eq:AbCorrSpecial} by lifting the Weyl-invariant polynomials $f_N(\vec\sigma_N)$ and $f_S(\vec\sigma_S)$ to $G$-invariant polynomials $f_N(\adjs_N)$ and $f_S(\adjs_S)$ with the inverse map $R^{-1}$.  

\subsubsection{Non-Abelian correlator relations}
A non-Abelian south pole correlator relation $R_S^G(\vec Q',\kappa_{f_N,\, \cdot})$ is defined to be a universal linear relation among correlators of the form $\kappa_{f_N,\, \cdot\,}$, where `$\,\cdot\,$' now refers to a finite collection of $G$-invariant polynomials $f_S$ in $\mathbb{C}[\mathfrak{g}]^G$. Similarly as for Abelian correlator relations, the non-trivial universality property states that $R_S^G(\vec Q',\kappa_{f_N,\, \cdot}) = 0$ holds for any choice of north pole polynomial $f_N$.\footnote{As we require the polynomials $f_S$ in $\kappa_{f_N,\, \cdot\,}$ to be $G$-invariant, eq.~\eqref{eq:ReynoldsOp} asserts that a south pole correlator relation $R_S^G$ holds for any --- not necessarily $G$-invariant --- north pole polynomial $f_N \in \mathbb{C}[\mathfrak{g}]$.}

We now determine such non-Abelian relations $R_S^G$ from the Abelian correlator relations $R_S^\text{Ab}$ of the Coulomb branch theory in two steps:

Firstly, we compute the rational functions~\eqref{eq:DefWandG} for the Abelian Coulomb branch spectrum listed in Table~\ref{tab:SpecCoulombBranch} and use the constraint~\eqref{eq:RelFinalCond1} to determine the Abelian syzygy module $M_S^{\text{Ab}}$. When then set the auxilary $Q$-parameters to one, namely $\vec Q=(\vec Q',1,\ldots,1)$, and employ the Reynolds operator~\eqref{eq:ReynoldsOp2} to project $R_S^\text{Ab} \in M_S^{\text{Ab}}$ to the $\mathcal{W}_G$-invariant correlator relation $R_S^{\mathcal{W}_G}$ given by
\begin{equation}
   R_S^{\mathcal{W}_G}(\vec Q',\mathfrak{m}_\alpha, \epsilon, \kappa_{f_N,\cdot}) \,=\, 
   \frac{1}{| \mathcal{W}_G |}  \sum_{w\in\mathcal{W}_G} R_S^\text{Ab}(\vec Q',\mathfrak{m}_\alpha, \epsilon, \kappa^\text{Ab}_{f_N,w(\cdot)}) \ .
\end{equation}
By construction of the projection, the Abelian correlators $\kappa^\text{Ab}_{f_N,\cdot}$ in $R_S^\text{Ab}$ arrange themselves to $\mathcal{W}_G$-invariant correlators $\kappa_{f_N,\cdot}$ in $R_S^{\mathcal{W}_G}$. These $\mathcal{W}_G$-invariant correlator relations $R_S^{\mathcal{W}_G}$ define the $\mathcal{W}_G$-invariant syzygy module $M_S^{\mathcal{W}_G}$. Note that, in order to construct a set generators for $M_S^{\mathcal{W}_G}$, it is in general not sufficient to project a set of generators of the syzygy module $M_S^\text{Ab}$ to $M_S^{\mathcal{W}_G}$. The projection may remove non-$\mathcal{W}_G$-invariant parts, which nevertheless --- by multiplication with suitable non-$\mathcal{W}_G$-invariant factors before projection --- may give rise to additional $\mathcal{W}_G$-invariants that are required to form a generating set for $M_S^{\mathcal{W}_G}$. 

Secondly, we obtain non-Abelian correlator relations $R_S^G$ from the Abelianized $\mathcal{W}_G$-invariant correlator relations $R_S^{\mathcal{W}_G}$ by application of the (inverse) Luna--Richardson restriction isomorphism \eqref{eq:Restriction} to the correspondence
\begin{equation}
   R_S^G (\vec Q',\mathfrak{m}_\alpha,\epsilon,\kappa_{f_N,\cdot}) \,=\,
   R_S^{\mathcal{W}_G}(\vec Q',\mathfrak{m}_\alpha,\epsilon,\kappa_{R(f_N),R(\,\cdot\,)}) \ .
\end{equation}  
This maps a set of generators of $M_S^{\mathcal{W}_G}$ to a set of generators of the non-Abelian $G$-invariant syzygy module $M_S^{G}$.

Note that the constructed $G$-invariant syzygy module $M_S^{G}$ is based on the special subclass of generators \eqref{eq:NonAbCorrSpecial}. If the placeholder `$\,\cdot\,$' in the set of correlators~$\kappa_{f_N,\cdot}$ admits non-$G$-invariant polynomials in $\mathbb{C}[\mathfrak{g}]$, there typically are further non-Abelian relations. Due to the gauge invariance~\eqref{eq:ReynoldsOp}, such correlator relations project for $G$-invariant insertions $f_N$ again to $G$-invariant correlator relations. For non-$G$-invariant insertions $f_N$, however, they encode novel relations among $G$-invariant correlators of the general type~$\kappa_f$ with $f \in \mathbb{C}[\mathfrak{g}\times\mathfrak{g}]^G$.

\subsubsection{The non-Abelian differential ideal}
In this work we restrict ourselves to the subset of differential operators obtained from non-Abelian south pole correlator relations $R_S^{G,\text{lin}}$, which can be written as functions of those $G$-invariants that are only linear in $\adjs_S$. As an example, for the gauge group $G=U(N)$ the south pole insertion $\operatorname{tr}(\adjs_S)$ is the only $G$-invariant linear in $\adjs_S$. Since there are no linear group invariants in the adjoint representation of semi-simple Lie groups, the linear invariants in the gauge group \eqref{eq:nonAbG} are in one-to-one correspondence with its $U(1)$ factors. We denote the associated linear $G$-invariant fields simply by $\sigma_k$, $k=1,\ldots,r'$, and in analogy to the case of Abelian gauge theories introduce the short-hand notation $\vec\sigma^{\vec n}=\sigma_1^{n_1}\cdots\sigma_{r'}^{n_{r'}}$, $\vec n \in \mathbb{Z}^{r'}_{\ge 0}$, for the product of such $G$-invariant fields. The subclass of correlator relations $R_S^{G,\text{lin}}$ then becomes a function of the correlators~$\kappa_{f_N,\cdot}$ where `$\,\cdot\,$' now refers to south pole insertions of the type $\vec\sigma^{\vec n}_S$. Furthermore, analogously to the Abelian gauge theories, we also denote the correlator of linear $G$-invariant operators by $\kappa_{\vec n,\vec m}$. 

By following the same steps as in Section~\ref{sec:diffideal}, this type of relations can be interpreted as differential operators in the differential ideal $\boldsymbol{\mathcal{I}}_S^{G,\text{lin}}$ with the commutation relation
\begin{equation} \label{eq:comGlin}
  \left[  \boldsymbol{\sigma}_{S,i} , \boldsymbol{Q}'_j\right] \,=\, \delta_{ij} \epsilon \boldsymbol{Q}'_j \ , \quad i,j=1,\ldots,r' \ .
\end{equation} 
Here, the operators $\boldsymbol{\sigma}_{S,i}$ and $\boldsymbol{Q}'_i$ are deduced from the linear $G$-invariant south pole insertions $\vec\sigma_{S}$ and the parameters $\vec Q'$, respectively.

The differential ideal $\boldsymbol{\mathcal{I}}_S^{G,\text{lin}}$ is by construction certainly not sensitive the non-linear $G$-invariant operators corresponding to $G$-invariant insertions of the non-Abelian semi-simple group factors in \eqref{eq:nonAbG}. Nevertheless, the ideal $\boldsymbol{\mathcal{I}}_S^{G,\text{lin}}$ is in principle sufficient to determine the entire quantum cohomology ring of the target space, due to its connection to the Givental $I$-function discussed in Section~\ref{sec:Giv} and Givental's reconstruction theorem \cite{MR2276766}. In order to extend the discussion of Section~\ref{sec:diffideal} to the fullfledged differential ideal acting on a Hilbert space of states of the non-Abelian gauge theory, it is necessary to include non-gauge invariant operators in the ring of operators as well as to define a Hilbert space of physical states in terms of an adequate BRST cohomology. The relevance of non-gauge invariant operators and the appearance of physical states in terms of BRST cohomology elements are both well-known phenomena in non-Abelian Yang--Mills theories; see for example refs.~\cite{KlubergStern:1974rs,Henneaux:1992ig}. We hope to get back to these aspects in the future.

\section{Calabi--Yau target spaces} \label{sec:CYtarget}
In this section we consider gauged linear sigma models whose target space is a Calabi--Yau manifold. We argue in Section~\ref{sec:CYTargetspaces} that the differential ideal $\boldsymbol{\mathcal{I}}_S$ then corresponds to the system of Picard--Fuchs differential operators governing the quantum cohomology of the Calabi--Yau manifold. In Section~\ref{sec:SolveForPF} we explain how this leads to expressions for Picard--Fuchs operators in terms of A-twisted correlators. It should be stressed that this equally works even for non-Abelian gauge theories, in case of which the methods of Section~\ref{sec:CorRel} may become rather tedious. Sections~\ref{sec:Curve} to \ref{sec:CY4} are dedicated to the discussion of several examples for this procedure, including the derivation of explicit formulas. In Sections~\ref{sec:Recurs} and \ref{sec:FixedModels} we further comment on recursion relations amongst the correlators and an alternative approach for fixed models.

\subsection{The differential ideal for Calabi--Yau target spaces}\label{sec:CYTargetspaces}
A geometrically interesting class of two-dimensional gauge theories realize Calabi--Yau target space geometries. First we start from a $d$-dimensional compact weak Fano toric variety $\mathbb{P}^d_\Delta$, which in our context is obtained from a $U(1)^r$ gauge theory with chiral multiplets $X_\ell$, $\ell=1,\ldots,d+r$, of gauge charge $\vec{\rho}^{\,x}_\ell$ and vanishing $U(1)_R$ charge $\mathfrak{q}^x_\ell=0$. The weak Fano condition implies that $\sum_\ell \rho^x_{\ell,s} \ge 0$ for all $s=1,\ldots,r$. The gauge theory now realizes the target space geometry $\mathbb{P}^d_\Delta$ as the vector space $\mathbb{C}^{d+r}$ spanned by the chiral multiplets $X_\ell$ modulo the gauge transformations acting on the fields $X_\ell$. More geometrically the weak Fano toric variety $\mathbb{P}^d_\Delta$ appears in the gauge theory as the symplectic quotient \cite{Witten:1993yc}
\begin{equation} \label{eq:SympQuotient}
   \mathbb{P}^d_\Delta \,=\, \left. \mathbb{C}^{d+r} \middle/ \mu^{-1}(\vec \xi)\right. \ ,
\end{equation}   
where $\mu: \mathbb{C}^{d+r} \to \mathfrak{u}(1)^r$ is the moment map for the gauge group action $U(1)^r$ on the chiral fields $X_\ell$ and $\vec\xi$ is the Fayet--Iliopoulos parameter of the analyzed gauge theory phase.

We can arrive at a Calabi--Yau target space by adding additional chiral multiplets $P_i$, $i=1,\ldots,n$ with gauge charge $-\vec{\rho}^{\,p}_i$ and $U(1)_R$ charges $\mathfrak{q}^p_i$ such that the axial anomaly of the classical axial $U(1)_R$-symmetry is cancelled. On the level of the gauge theory charges this corresponds to the well-known condition \cite{Witten:1993yc}
\begin{equation}
   \sum_\ell \vec{\rho}^{\,x}_\ell \,=\, \sum_i \vec{\rho}^{\,p}_i  \ .
\end{equation}    

For our first class of Calabi--Yau geometries we set the $U(1)_R$ charges of the $P$-fields to zero, i.e., $\mathfrak{q}^p_i = 0$ for $i=1,\ldots,n$. Then the target space of the gauge theory describes the non-compact toric Calabi--Yau variety $X_\text{nc}$ of dimension $d+n$, which is the total space of the vector bundle
\begin{equation} \label{eq:CYTot}
   X_\text{nc} \,=\, \operatorname{Tot}\left[\begin{CD} \bigoplus_{i=1}^n \mathcal{O}_\Delta(-\vec \rho^{\,p}_i) \\ @VVV\\   \mathbb{P}^d_\Delta \end{CD} \right] \ .
\end{equation}   
Due to the non-compactness of the target space $X_\text{nc}$, the correlation functions $\kappa_{\vec n,\vec m}$ of this gauge theory require a regularization with twisted mass $\tm_i^p$ for the chiral fields $P_i$ associated to the non-compact directions of $X_\text{nc}$.

The differential ideal $\boldsymbol{\mathcal{I}}_S$ of the gauge theory annihilates the moduli space of ground states, which --- due to the interpretation of the $\epsilon$-deformation as the topological $A$-twist \cite{Closset:2014pda,Closset:2015rna} --- is identified with the quantum K\"ahler moduli space of the two-dimensional $\sigma$-model with the non-compact Calabi--Yau geometry $X_\text{nc}$. As a result the ideal $\boldsymbol{\mathcal{I}}_S$ is identified with the GKZ system of differential operators governing the quantum periods of the described non-compact Calabi--Yau geometry $\mathcal{X}_\text{nc}$, where the twisted masses~$\tm_i^p$ become the equivariant parameters for the $(\mathbb{C}^*)^n$ action on the non-compact directions of $X_\text{nc}$.

Note that in the limit $\tm_i^p\to 0$ the differential ideal $\boldsymbol{\mathcal{I}}_S^{\lim}$ deduced from $M_S^{\lim}$ becomes generically a proper subideal of the non-generic differential ideal $\boldsymbol{\mathcal{I}}_S^0$ associated to the the non-generic module $M_S^0$ (with proper submodule $M_S^{\lim}$). However, for the non-compact Calabi--Yau geometry $X_\text{nc}$ the additional generators in the non-generic differential ideal $\boldsymbol{\mathcal{I}}_S^0$ do not give rise to physical differential operators, which is a consequence of the violation of the pole condition discussed at the end of Section~\ref{sec:Reduction}.

For our second class of Calabi--Yau geometries we specialize to gauge theories with $n<d$ and with the $U(1)_R$ charges $\mathfrak{q}^p_i = 2$ for $i=1,\ldots,n$. Due to the assigned $R$-charges of the $P$-fields the gauge theory typically admits a non-vanishing superpotential $W$ of the form
\begin{equation}
  W \,=\, \sum_i P_i G_i(X) \ ,
\end{equation}
in terms of generic polynomials $G_i(X)$ homogenous with respect to the weights of the $U(1)^r$ gauge symmetry of multi-degree $\vec{\rho}^{\,p}_i$. Assuming that the $(d+r)\times n$-dimensional Jacobian matrix $\partial_\ell G_i(x)$ has maximal rank $n$ for all $x \in \mathbb{P}^d_\Delta$, we arrive at the compact Calabi--Yau variety $X$ of dimension $d-n$ of the form
\begin{equation}
   X \,=\, \mathbb{P}_\Delta^d( \vec{\rho}^{\,p}_1,\ldots,\vec{\rho}^{\,p}_n) \, \subset \, \mathbb{P}_\Delta^d \ .
\end{equation}   
Here $\mathbb{P}_\Delta^d( \vec{\rho}^{\,p}_1,\ldots,\vec{\rho}^{\,p}_n)$ refers to the common zero locus of the homogeneous polynomials $G_i(X)$, $i=1,\ldots,n$. Note that --- due to the compactness of the Calabi--Yau geometry $X$ --- the correlation functions $\kappa_{\vec n,\vec m}$ are finite even for vanishing twisted masses. Thus on the level of correlators the limit $\tm_\ell \to 0$ is well-defined. 

Let us now turn to the structure of the differential ideal $\boldsymbol{\mathcal{I}}_S$ of the discussed gauge theory. For the non-generic case of non-vanishing twisted masses, we get a non-generic extended differential ideal $\boldsymbol{\mathcal{I}}_S^0$, for which the limiting ideal $\boldsymbol{\mathcal{I}}_S^{\lim}$ is a proper subideal. However, in the case of the compact Calabi--Yau geometry $X$ the non-generic differential ideal $\boldsymbol{\mathcal{I}}_S^0$ is of physical significance as the pole condition of Section~\ref{sec:Reduction} is satisfied, which is a consequence of the $R$-charge assignment to the $P$-fields. Thus, the non-generic differential ideal $\boldsymbol{\mathcal{I}}_S^0$ captures the moduli dependence of the ground state in the K\"ahler moduli space. As a consequence we can identify the non-generic differential ideal $\boldsymbol{\mathcal{I}}_S^0$ with the ideal of Picard--Fuchs operators governing the quantum periods of the compact Calabi--Yau geometry. Compared to $\boldsymbol{\mathcal{I}}_S^0$ the limiting ideal $\boldsymbol{\mathcal{I}}_S^{\lim}$ is generated by higher order differential operators than $\boldsymbol{\mathcal{I}}_S^0$, and geometrically the generators of $\boldsymbol{\mathcal{I}}_S^{\lim}$ are identified with the GKZ system of differential operators.

Thus in summary, we observe that for the compact Calabi--Yau geometry $X$ the proper inclusion of the differential ideal $\boldsymbol{\mathcal{I}}_S^{\lim}$ in the differential ideal $\boldsymbol{\mathcal{I}}_S^0$ for $\tm_\ell \to 0$ describes on the level of differential operator the reduction of the system of GKZ operators~$\mathcal{L}_\alpha^\text{GKZ}$ to the system of Picard--Fuchs operators $\mathcal{L}^\text{PF}_\beta$, i.e., 
\begin{equation}\label{eq:PFIdeal}
  \left\langle\!\!\left\langle \mathcal{L}_\alpha^\text{GKZ} \right\rangle\!\!\right\rangle \,=\, \boldsymbol{\mathcal{I}}_S^{\lim}  \, \subset \,
  \boldsymbol{\mathcal{I}}_S^{0}   \,=\,  \left\langle\!\!\left\langle \mathcal{L}_\beta^\text{PF} \right\rangle\!\!\right\rangle \ . 
\end{equation} 

\subsection{Picard--Fuchs operators from A-twisted correlators}\label{sec:SolveForPF}
In the previous section we have presented a method to rapidly determine correlator relations by inspection of the gauge theory spectrum. Here, we take on a different perspective: Assume we know by some independent argument that a linear relation of the type introduced in eq.~\eqref{eq:Relansatz},
\begin{equation} \label{eq:RelansatzRepeat}
   R_S(\vec Q,\tm_\ell,\epsilon,\kappa_{\vec n,\,\cdot\,}) \,=\, \sum_{\vec m=0}^{\vec N} c_{\vec m}(\vec Q,\tm_\ell,\epsilon) \kappa_{\vec n,\vec m}(\vec Q,\tm_\ell,\epsilon)
    \,=\, 0 \ ,
\end{equation}
exists. Without any reference to a gauge theory realization, we can then view eq.~\eqref{eq:RelansatzRepeat} as an infinite family of homogeneous linear equations. By solving these for sufficiently many values of $\vec{n}$ we are thus able to express the finite set of coefficient functions $c_{\vec m}(\vec Q,\tm_\ell,\epsilon)$ appearing in eq.~\eqref{eq:RelansatzRepeat} in terms of the correlators $\kappa_{\vec n,\vec m}(\vec Q,\tm_\ell,\epsilon)$. Since by the arguments of Section~\ref{sec:CYTargetspaces} these $c_{\vec m}(\vec Q,\tm_\ell,\epsilon)$  determine a Picard--Fuchs differential operator as
\begin{equation}\label{eq:LFromR}
\mathcal{L}(\vec{Q},\tm_\ell,\epsilon)\,=\, \boldsymbol{R}_S(\vec Q,\epsilon\, \vec\Theta,\tm_\ell,\epsilon)\,=\, \sum_{\vec m=0}^{\vec N} c_{\vec m}(\vec Q,\tm_\ell,\epsilon) \, \left(\epsilon\,\vec{\Theta}\right)^{\vec{m}}\ ,
\end{equation}
this leads to expressions for Picard--Fuchs operators in terms of A-twisted correlators. As we shall see below, knowledge of the vector $\vec{N}$ is essential for this procedure. Henceforth we assume $\vec{N}$ to be chosen such that none of its components $N_i$ (recall $1\leq i\leq r$) can be decreased without changing the relation~\eqref{eq:RelansatzRepeat}. We further denote $|\vec{a}|_1 = \sum_{i=1}^r |a_i|$, write $\vec{a}\leq \vec{b}$ if $a_i\leq b_i$ for all $i$, and for $r=1$ drop the vector notation.

Let us now specialize to vanishing twisted masses, $\tm_\ell = 0$, and for this case discuss the outlined procedure in more detail: We aim to express elements $\mathcal{L}$ of the differential ideal $\boldsymbol{\mathcal{I}}_S^{0}$ of Picard--Fuchs operators in terms of correlators. Any such operator $\mathcal{L}$ can be expanded as
\begin{equation}\label{eq:LFromRWithoutM}
\mathcal{L}(\vec{Q},\epsilon)\,=\, \sum_{\vec m=0}^{\vec N} c_{\vec m}(\vec Q,\epsilon) \, \left(\epsilon\,\vec{\Theta}\right)^{\vec{m}}\quad \text{with}  \quad
\sum_{\vec m=0}^{\vec N} c_{\vec m}(\vec Q,\epsilon)\, \kappa_{\vec n,\vec m}(\vec Q,0,\epsilon) = 0
\end{equation}
for all $\vec{n}\in\mathbb{Z}^r_{\geq 0}$. For the remainder of this section all correlators are evaluated for vanishing twisted masses and we abbreviate
\begin{equation}
\kappa_{\vec{n},\vec{m}} =\kappa_{\vec n,\vec m}(\vec Q,0,\epsilon) \ , \qquad c_{\vec{m}} = c_{\vec m}(\vec Q,\epsilon)\ , \qquad \mathcal{L} = \mathcal{L}(\vec{Q},\epsilon) \ .
\end{equation}
To concisely write the constraints imposed by the right hand part of eq.~\eqref{eq:LFromRWithoutM} we introduce an ordered list $I$ and a vector $c(I)$ as
\begin{equation}\label{eq:DefCVector}
I = \left\{\, \vec{m} \, \in \, \mathbb{Z}^r_{\geq 0} \, \middle| \, \vec{m} \leq \vec{N}\, \right\} \ , \qquad c(I)_{k} = c_{I(k)} \ ,
\end{equation}
where $I(k)$ denotes the $k^{\text{th}}$ entry of $I$. For every other ordered list $J$ of vectors $\vec{n}\in\mathbb{Z}^r_{\geq 0}$ we further introduce the $|J|\times|I|$ matrix $M(J,I)$ by
\begin{equation}\label{eq:MGeneral}
M(J,I)_{k,l} =(-1)^{|J(k)|_1} \, \kappa_{J(k),I(l)} \ .
\end{equation}
Here, the power of $(-1)$ is conventional, its advantage will become clear soon. Equation~\eqref{eq:LFromRWithoutM} is then equivalent to
\begin{equation}\label{eq:KernelCondition}
\mathcal{L} = \sum_{k=1}^{|I|} c(I)_k \left(\epsilon \vec{\Theta}\right)^{I(k)} \quad \text{with} \quad M(J,I) \cdot c(I)\, = \, 0 \qquad \forall \,J \ ,
\end{equation}
i.e. the vector $c(I)$ of coefficient functions is in the kernel of $M(J,I)$ for all index lists $J$. To understand how this fact can be used to determine $c(I)$, we here list several of its aspects and implications:

\begin{enumerate}
\item The fact that $M(J,I) \cdot c(I) = 0$ for some $J$ is not sufficient for existence of a correlator relation, it needs to be true for all $J$. Therefore, existence of a relation needs to be guaranteed by some independent argument.
\item Assume $M(J,I) \cdot c(I) = 0$ for some non-zero $c(I)$. Item (i) then raises the question of how to decide, whether this $c(I)$ corresponds to a relation. For answering this in generality, we need to know how many relations are expected for the given $I$. Thus, we require knowledge of how $\boldsymbol{\mathcal{I}}_S^{0}$ is generated:

Say $\boldsymbol{\mathcal{I}}_S^{0}$ is generated by $M$ operators $\mathcal{L}^{(\alpha)}$ where $1\leq \alpha \leq M$ and $\vec{N} = \vec{N}^{(\alpha)}$. Then, for all $\alpha$ and all $\vec{k}\in\mathbb{Z}^r_{\geq0}$ the operator $\Theta^{\vec{k}}\mathcal{L}^{(\alpha)}$ is an element of $\boldsymbol{\mathcal{I}}_S^{0}$ too. For every linearly independent operator $\Theta^{\vec{k}}\mathcal{L}^{(\alpha)}$, which is such that $\{\, \vec{m} \, \in \, \mathbb{Z}^r_{\geq 0} \, | \, \vec{m} \leq \vec{N}^{(\alpha)}+\vec{k}\, \} \subset I$, we expect one linearly independent kernel element.
\item Assume we know how $\boldsymbol{\mathcal{I}}_S^{0}$ is generated and thus know that for some $J$ the kernel of $M(J,I)$ needs to be $d$-dimensional. From the definition of $M(J,I)$ in eq.~\eqref{eq:MGeneral} this might, however, not be manifest. Then:
\begin{enumerate}
\item If the kernel of $M(J,I)$ appears to be of smaller dimension, the requirement $\operatorname{dim} \operatorname{ker} M(J,I) = d$ imposes constraints on the $\kappa_{\vec{n},\vec{m}}$.
\item If the kernel of $M(J,I)$ is of higher dimension, there needs to be $J^\prime$ such that the excessive kernel elements of $M(J,I)$ are not in the kernel of $M(J^\prime,I)$.
\end{enumerate}
The combination of items (a) and (b) complicates the analysis: If it is possible to impose conditions on the $\kappa_{\vec{n},\vec{m}}$, for which the dimension of the kernel becomes too big, we are not able to derive universal expressions for the Picard--Fuchs operators.
\end{enumerate}

For the most part of this section our aim is to derive general expressions for Picard--Fuchs operators in terms of correlators, i.e., without specializing to fixed models. The latter is only commented on in Section~\ref{sec:FixedModels}. For the derivation general formulae it turns out necessary to employ the basic properties of the $\kappa_{\vec{n},\vec{m}}$, these are:
\begin{enumerate}
\item The derivative rule~\eqref{eq:RelDer} for $\tm_\ell=0$,
\begin{equation}\label{eq:RelDerGen}
 \epsilon\Theta_i\, \kappa_{\vec n,\vec m} =  \kappa_{\vec n,\vec m+\vec e_i} -\kappa_{\vec n+\vec e_i,\vec m} \ ,
\end{equation}
where $\Theta_i = Q_i \partial_{Q_i}$. This property does not depend on the dimension of the target space.
\item Due to the Calabi--Yau condition $\sum \vec{\rho}_{\ell} = 0$ and for vanishing twisted masses $\tm_\ell=0$, the behavior under the exchange of north and south poles as given in eq.~\eqref{eq:RelSym} simplifies to 
\begin{equation}\label{eq:SymEasy}
\kappa_{\vec{n},\vec{m}} = (-1)^{\operatorname{dim}_\mathbb{C} X + |\vec{n}|_1 + |\vec{m}|_1} \kappa_{\vec{m},\vec{n}} \ ,
\end{equation}
with the (complex) dimension of the target space $X$. It can be verified with eq.~\eqref{eq:NonAbCorrdim} that this continues to hold in the Cartan theory of the non-Abelian gauge theory. Equation~\eqref{eq:SymEasy} implies $M(I,I)$ to be symmetric for even-dimensional target spaces and anti-symmetric in case of an odd dimension. This is the reason for the conventional factor of $(-1)$ in eq.~\eqref{eq:MGeneral}. In particular, $\kappa_{\vec n,\vec n} = 0$ in the odd-dimensional case.
\item In the discussed Calabi--Yau setting, the anomalous contribution~\eqref{eq:ddef} of the selection rule~\eqref{eq:srule} is identified with the (complex) dimension of the target space $X$ (or alternatively with one third of the central charge). Therefore, the selection rule~\eqref{eq:srule} implies
\begin{equation}\label{eq:SelecRule}
  \kappa_{\vec n,\vec m} = 0 \quad \text{for} \quad |\vec{n}|_1 + |\vec{m}|_1 < \dim_\mathbb{C} X \ .
\end{equation}
Here we assume that the correlators are well-behaved in the limit $\epsilon \to 0$. 
\end{enumerate}

This sets the stage for the explicit discussion of several classes of examples in the subsequent parts of this section. We will use the properties~\eqref{eq:SymEasy} and \eqref{eq:SelecRule} without explicit mentioning.

\subsection{Elliptic curves}\label{sec:Curve}
As a first example we consider the easiest class of Calabi--Yau target space geometries: Elliptic curves. These are always parameterized by a single K\"ahler parameter, therefore $\operatorname{dim}X=1$ and $r=1$. The correlators thus satisfy
\begin{align}
\epsilon\Theta\, \kappa_{n,m} &=  \kappa_{n,m+1} -\kappa_{n+1,m} \ , \label{eq:RelDerEllipticCurve} \\
\kappa_{n,m}  &= (-1)^{1+n+m} \kappa_{m,n}  \ .\label{eq:SymEllipticCurve}
\end{align}
Note that the selection rule $\kappa_{0,0}=0$ does not carry additional information.

From variation of Hodge structures the single Picard--Fuchs operator $\mathcal{L}$ generating $\boldsymbol{\mathcal{I}}_S^{0}$ is known to be of order $N=2$. We thus choose $I=\left\{0,1,2\right\}$ and consider the matrix
\begin{equation}
M = M(I,I) =  \left(
\begin{array}{ccc}
 0 & \kappa _{0,1} & \kappa _{0,2} \\
 -\kappa _{0,1} & 0 & -\kappa _{1,2} \\
 -\kappa _{0,2} & \kappa _{1,2} & 0 \\
\end{array}
\right) \ .
\end{equation}
Due to its antisymmetry $M$ has a kernel automatically, we therefore do not need to impose an additional constraint on the $\kappa_{n,m}$. Moreover, assuming that $\kappa_{0,1}$ does not vanish, the kernel is exactly one-dimensional, and the (up to an overall normalization) unique kernel element $c(I)$ is given by
\begin{equation}
 c_0 =-\kappa _{1,2} \ , \qquad
 c_1 =-\kappa _{0,2} = - \epsilon \Theta \, \kappa_{0,1} \ , \qquad
 c_2 = +\kappa _{0,1} \ .
\end{equation}
Here, use has been made of the derivative rule~\eqref{eq:RelDerEllipticCurve} to reduce the number of correlators entering the formula. The corresponding Picard--Fuchs operator reads
\begin{equation}\label{eq:LCY1}
\mathcal{L} = \kappa_{0,1} \left(\epsilon \Theta\right)^2 - \kappa_{0,2} \left(\epsilon \Theta\right)-\kappa_{1,2} 
= \kappa_{0,1} \left(\epsilon \Theta\right)^2 - \left(\epsilon \Theta\,\kappa_{0,1}\right) \left(\epsilon \Theta\right)-\kappa_{1,2} \ .
\end{equation}

Now reconsider the assumption that $\kappa_{0,1}$ does not vanish. Since $\kappa_{0,1}$ is the one-dimensional analog of the Yukawa coupling of a Calabi--Yau threefold, it is not identically zero. For a generic point in moduli space the assumption is thus fulfilled and formula~\eqref{eq:LCY1} is universally valid for all elliptic curves.

\subsection{One parameter polarized K3 surfaces}\label{sec:K3}
Our second example are gauge theories in which the target space is a one parameter polarized K3 surface. Then $\operatorname{dim} X =2 $ and $r =1 $, such that the correlators satisfy
\begin{align}
\epsilon\Theta\, \kappa_{n,m} &=  \kappa_{n,m+1} -\kappa_{n+1,m} \ , \label{eq:K3RelDer}\\
\kappa_{n,m}  &= (-1)^{n+m} \kappa_{m,n}  \ ,\label{eq:K3Sym} \\
\kappa_{n,m} &=0 \quad \text{for} \quad  n+m <2 \ . \label{eq:K3SelecRule}
\end{align}
Note that there is one independent selection rule, $\kappa_{0,0} = 0$.

Variation of Hodge structure shows that the single Picard--Fuchs operator generating $\boldsymbol{\mathcal{I}}_S^{0}$ is of order $N=3$. We therefore take $I=\left\{0,1,2,3\right\}$ and consider the matrix $M = M(I,I)$:
\begin{equation}\label{eq:K3cEq}
\begin{aligned}
M &= \left(
\begin{array}{cccc}
 0 & 0 & \kappa _{0,2} & \kappa _{0,3} \\
 0 & -\kappa _{1,1} & -\kappa _{1,2} & -\kappa _{1,3} \\
 \kappa _{0,2} & -\kappa _{1,2} & \kappa _{2,2} & \kappa _{2,3} \\
 \kappa _{0,3} & -\kappa _{1,3} & \kappa _{2,3} & -\kappa _{3,3} \\
\end{array}
\right) \\[0.1em]
&= \left(
\begin{array}{cccc}
 0 & 0 & \kappa _{0,2} & \frac{3}{2} \epsilon  \Theta \, \kappa _{0,2} \\
 0 & -\kappa _{0,2} & -\frac{1}{2} \epsilon  \Theta \, \kappa _{0,2} & -\frac{1}{2} \epsilon ^2 \Theta ^2\,\kappa _{0,2}-\kappa _{2,2} \\
 \kappa _{0,2} & -\frac{1}{2} \epsilon  \Theta \, \kappa _{0,2} & \kappa _{2,2} & \frac{1}{2} \epsilon  \Theta \, \kappa _{2,2} \\
 \frac{3}{2} \epsilon  \Theta \, \kappa _{0,2} & -\frac{1}{2} \epsilon ^2 \Theta ^2\, \kappa _{0,2}-\kappa _{2,2} & \frac{1}{2} \epsilon  \Theta \, \kappa _{2,2} & -\kappa _{3,3} \\
\end{array}
\right) \ .
\end{aligned}
\end{equation}
Due to the different symmetry property~\eqref{eq:K3Sym}, here $M$ is symmetric rather than anti-symmetric and the kernel condition $\operatorname{det}M=0$ is no longer trivial. Since $\kappa_{0,2}=\kappa_{1,1}$ corresponds to the Yukawa coupling, it can vanish at special points in moduli space only. For generic points the second line of eq.~\eqref{eq:K3cEq} thus shows $\operatorname{rank} M \geq 3$, and $\operatorname{rank} M = 3$ (or $\operatorname{det}M=0$) is equivalent to
\begin{equation}
\begin{aligned}
\kappa_{3,3} =&\left(16\, \kappa _{0,2}^3\right)^{-1}\Big\{
9 \left(\epsilon \Theta\, \kappa _{0,2}\right)^4
-12\kappa _{0,2} \left(\epsilon \Theta \,\kappa _{0,2}\right)^2 \left(\epsilon ^2 \Theta^2\,\kappa_{0,2}-\kappa _{2,2}\right) \\
&\,\,4 \kappa _{0,2}^2 \left[4 \kappa_{2,2}^2+ \left(\epsilon ^2 \Theta^2\,\kappa_{0,2}\right)^2+4\kappa _{2,2}\left(\epsilon^2 \Theta^2\,\kappa_{0,2}\right)-6 \left(\epsilon\Theta\, \kappa_{0,2}\right)\left(\epsilon \Theta\, \kappa _{2,2}\right)\right]\Big\} \ .
\end{aligned}   
\end{equation}
Upon inserting this constraint into $M$ we immediately find the kernel element $c(I)$ and arrive at the Picard--Fuchs operator
\begin{equation}\label{eq:K3Op}
\begin{aligned}
\mathcal{L} =
&+8 \kappa _{0,2}^3 \left(\epsilon\Theta\right)^3
-12\kappa _{0,2}^2 \left(\epsilon \Theta\kappa _{0,2}\right) \left(\epsilon\Theta\right)^2\\
&+ \left[6\kappa _{0,2} \left(\epsilon \Theta \kappa _{0,2}\right)^2 -4 \kappa_{0,2}^2 \left(2 \kappa _{2,2}+\Theta ^2 \epsilon ^2 \kappa _{0,2}\right)\right]\left(\epsilon \Theta\right)\\
&+3 \left(\epsilon \Theta \kappa _{0,2}\right)^3 -4 \kappa _{0,2}^2\left(\epsilon  \Theta
   \kappa _{2,2}\right) -2\kappa _{0,2} \left(\epsilon\Theta \kappa _{0,2} \right) \left(\Theta ^2 \epsilon ^2 \kappa _{0,2}-4 \kappa _{2,2}\right) \ ,
\end{aligned}
\end{equation}
This formula is valid for all one-parameter polarized K3 surfaces.

The Picard--Fuchs operator is constituted by the coefficents functions $c_k$ multiplying $(\epsilon\Theta)^k$, here $k=1,\ \ldots, 3$. Due to the freedom of rescaling all $c_k$ with a common function, $c_k \mapsto g \cdot c_k$, the $c_k$ are not significant themselves. Rather, the invariant information is encoded in three independent ratios, e.g. $c_k/c_0$ for $k=1,2,3$. The right hand side of eq.~\eqref{eq:K3Op} is, however, entirely expressed in terms of the two correlators $\kappa_{0,2}$ and $\kappa_{2,2}$. We therefore expect one differential-algebraic relation between the four $c_k$. Indeed, we find
\begin{equation}\label{eq:cRelK3}
\begin{aligned}
54 c_0 c_3^2 =
&+9 c_2 c_3\left( \epsilon^2\Theta^2 \,c_3\right) -9 c_3^2\left( \epsilon^2\Theta^2\, c_2\right)-18 c_2 \left(\epsilon\Theta\, c_3\right)^2+18 c_3 \left(\epsilon\Theta\, c_2\right)\left(\epsilon\Theta\, c_3\right)+18 c_2^2\left( \epsilon\Theta\, c_3\right)\\
&-18 c_2 c_3\left( \epsilon\Theta\, c_2\right) +27 c_3^2 \left(\epsilon\Theta\, c_1 \right)-27 c_1 c_3\left( \epsilon\Theta\, c_3\right)-4c_2^3+18 c_1 c_2 c_3 \ ,
\end{aligned}
\end{equation}
which unambiguously fixes $c_0$. Equation~\eqref{eq:cRelK3} is invariant under a common rescaling of the $c_k$ and thus not an artifact of the specific representation chosen in eq.~\eqref{eq:K3Op}.

\subsection{One parameter Calabi--Yau threefolds}
Let us now turn to the important case in which the target space is a three-dimensional Calabi--Yau manifold with a single K\"ahler parameter as target space, i.e., $\operatorname{dim}X=3$ and $r=1$. The correlators therefore obey
\begin{align}
\epsilon\Theta\, \kappa_{n,m} &=  \kappa_{n,m+1} -\kappa_{n+1,m} \ , \label{eq:RelDerCY3} \\
\kappa_{n,m}  &= (-1)^{1+n+m} \kappa_{m,n}  \ , \\
\kappa_{n,m} &=0 \quad \text{for} \quad  n+m <3 \ . \label{eq:SelecRuleCY3}
\end{align}
Note that $\kappa_{0,1} = 0$ is the single non-trivial, independent selection rule.

Variation of Hodge structures shows that the single Picard--Fuchs operator generating $\boldsymbol{\mathcal{I}}_S^{0}$ is of order $N=4$. We therefore take $I=\left\{0,1,2,3,4\right\}$ and consider the matrix
\begin{equation}
M = M(I,I) = \left(
\begin{array}{ccccc}
 0 & 0 & 0 & \kappa _{0,3} & \kappa _{0,4} \\
 0 & 0 & -\kappa _{1,2} & -\kappa _{1,3} & -\kappa _{1,4} \\
 0 & \kappa _{1,2} & 0 & \kappa _{2,3} & \kappa _{2,4} \\
 -\kappa _{0,3} & \kappa _{1,3} & -\kappa _{2,3} & 0 & -\kappa _{3,4} \\
 -\kappa _{0,4} & \kappa _{1,4} & -\kappa _{2,4} & \kappa _{3,4} & 0 \\
\end{array}
\right) \ ,
\end{equation}
which due to its antisymmetry has a kernel automatically. Since the geometric Yukawa coupling $\kappa_{0,3}$ can vanish at special points in moduli space only and $\kappa_{1,2}=\kappa_{0,3}$ as a result of eqs.~\eqref{eq:RelDerCY3} and \eqref{eq:SelecRuleCY3}, we find $\operatorname{rank} M = 4$ at generic points in moduli space. The solution to the equation $M \cdot c(I) = 0$ is thus unique up to rescaling and leads to the Picard--Fuchs operator
\begin{equation}\label{eq:LCY3v0}
\begin{aligned}
\mathcal{L}\,=\, &\kappa _{0,3}^2 (\epsilon\Theta)^4 -\kappa _{0,3} \kappa _{0,4} (\epsilon\Theta)^3 +\left(\kappa _{0,4} \kappa _{1,3}-\kappa _{0,3} \kappa _{1,4}\right)(\epsilon\Theta)^2 \\
&+\left(\kappa _{0,4} \kappa _{2,3}-\kappa _{0,3} \kappa _{2,4}\right)(\epsilon\Theta) +\left(\kappa _{1,4} \kappa _{2,3}-\kappa _{1,3} \kappa _{2,4}-\kappa _{0,3} \kappa _{3,4}\right) \ .
\end{aligned}
\end{equation}
By fully exploiting eq.~\eqref{eq:RelDerCY3} in order to reduce the number of required correlators, this can be rewritten as
\begin{equation}\label{eq:LCY3}
\begin{aligned}
\mathcal{L}\,=\, &+\kappa_{0,3}^2 (\epsilon\Theta)^4-2\kappa_{0,3}\left(\epsilon\Theta\,\kappa_{0,3}\right)(\epsilon\Theta)^3+\left[2 \left(\epsilon\Theta\, \kappa _{0,3}\right)^2-\kappa _{0,3} \left(\epsilon^2 \Theta^2\,\kappa _{0,3}+\kappa _{2,3}\right)\right](\epsilon\Theta^2)\\
&+ \left[2 \kappa _{2,3} \left(\epsilon\Theta\, \kappa _{0,3}\right)- \kappa _{0,3}\left(\epsilon \Theta\,\kappa _{2,3}\right)\right](\epsilon \Theta)\\
&+\left[\kappa_{2,3}^2-\kappa _{0,3} \kappa _{3,4}-\left(\epsilon \Theta\, \kappa_{0,3}\right)\left(\epsilon \Theta\, \kappa _{2,3}\right)+\kappa _{2,3} \left(\epsilon^2\Theta^2\,\kappa _{0,3}\right)\right]\ .
\end{aligned}
\end{equation}
These formulae are valid for all three-dimensional Calabi--Yau manifolds with a single K\"ahler parameter.

The five coefficient functions $c_k$ of $\mathcal{L}$ in eq.~\eqref{eq:LCY3} are expressed in terms of the three correlators $\kappa_{0,3}$, $\kappa_{2,3}$ and $\kappa_{3,4}$. Consequently, there needs to be one differential-algebraic relation between the $c_k$. In fact,
\begin{equation}\label{eq:cRelCY3}
\begin{aligned}
8 c_1 c_4^2 = &-8 c_3 \left(\epsilon \Theta c_4\right)^2+8 c_4 \left(\epsilon\Theta c_3\right)\left(\epsilon\Theta c_4\right)+4 c_3 c_4 \left(\epsilon ^2 \Theta^2\, c_4\right)-4 c_4^2 \left( \epsilon ^2 \Theta^2\, c_3\right)\\
&+6 c_3^2\left( \epsilon\Theta\,c_4\right)-6 c_3 c_4\left( \epsilon \Theta\, c_3\right)+8 c_4^2\left( \epsilon \Theta\, c_2\right)-8 c_2 c_4\left( \epsilon \Theta\, c_4\right)-c_3^3+4 c_2 c_4 c_3 \ ,
\end{aligned}
\end{equation}
which is invariant under a common rescaling of all $c_k$ and unambiguously fixes $c_1$. Equation~\eqref{eq:cRelCY3} does not contain $c_0$, since all the $c_k$ for $k=1,\ldots, 4$ are independent of $\kappa_{3,4}$.

Note that the obtained differential-algebraic relation~\eqref{eq:cRelCY3} is a consequence of the underlying $\mathcal{N}=2$ special geometry of Calabi--Yau threefolds \cite{Strominger:1990pd}. This has been shown in refs.~\cite{MR2282972,MR2282974}, and it plays an important role in the classification of Picard--Fuchs operators for Calabi--Yau threefolds with a single K\"ahler modulus. It is gratifying to see that we recover this condition form the gauge theory considerations. More generally, we also find differential-algebraic equations for Calabi--Yau target spaces in other dimensions; see, e.g., eq.~\eqref{eq:CY4algdiff} for Calabi--Yau fourfolds. It would be interesting to understand their geometric origin as well.

\subsection{Two parameter Calabi--Yau threefolds}\label{sec:CY32Param}
The techniques of Section~\ref{sec:SolveForPF} can also be applied to target spaces with more than a single K\"ahler parameter. Here, we consider three-dimensional Calabi--Yau manifolds with two parameters, in case of which the correlators satisfy
\begin{align}
\epsilon\Theta_i\, \kappa_{\vec{n},\vec{m}} &=  \kappa_{\vec{n},\vec{m}+\vec{e}_i} -\kappa_{\vec{n}+\vec{e}_i,\vec{m}} \ , \label{eq:RelDerCY32Param} \\
\kappa_{\vec{n},\vec{m}}  &= (-1)^{1+|\vec{n}|_1+|\vec{m}|_1} \kappa_{\vec{m},\vec{n}}  \ , \label{eq:SymCY32Param}\\
\kappa_{\vec{n},\vec{m}} &=0 \quad \text{for} \quad  |\vec{n}|_1+|\vec{m}|_1 <3 \ . \label{eq:SelecRuleCY32Param}
\end{align}
These properties allow to express all correlators in terms of those $\kappa_{\vec{n},\vec{m}}$ with $\vec{n} = 0$ and $|\vec{m}|_1\in 2\mathbb{N}+1$. As an example, for $|\vec{n}|_1+|\vec{m}|_1=3$ the derivative relations are solved by
\begin{equation}\label{eq:3Fold2Param3PointRel}
\begin{aligned}
 \kappa _{(1,0),(2,0)} &= \kappa _{(0,0),(3,0)} \ , \quad &&\kappa _{(1,0),(1,1)} = \kappa _{(0,0),(2,1)} \ , \quad  &&\kappa _{(0,2),(1,0)} = \kappa _{(0,0),(1,2)} \ ,  \\
\kappa _{(0,1),(2,0)} &= \kappa _{(0,0),(2,1)} \ , \quad &&\kappa _{(0,1),(1,1)} = \kappa _{(0,0),(1,2)} \ , \quad &&\kappa _{(0,1),(0,2)} = \kappa _{(0,0),(0,3)} \ ,
\end{aligned}
\end{equation}
All other three-point correlators are related to these by the symmetry~\eqref{eq:SymCY32Param}. As a consequence, at least one of the four correlators $\kappa_{(0,0),(a,3-a)}$ with $0 \leq a \leq 3$ must not vanish identically. Further relevant implications of the derivative rule are listed in Table~\ref{tab:RepList}.

Variation of Hodge structure shows that $\boldsymbol{\mathcal{I}}_S^{0}$ is generated by two independent operators, $\mathcal{L}^{(2)}$ and $\mathcal{L}^{(3)}$. These are of order two and three, respectively, and are expanded as
\begin{equation}
\begin{aligned}\label{eq:LExpandCY32}
\mathcal{L}^{(2)} &= \sum_{k=1}^6 c^{(2)}(I^{(2)})_k \left(\epsilon\vec{\Theta}\right)^{I^{(2)}(k)} = \sum_{k_1 = 0}^2\sum_{k_2 = 0}^{2-k_1} c^{(2)}_{(k_1,k_2)} \left(\epsilon\Theta_1\right)^{k_1} \left(\epsilon\Theta_2\right)^{k_2} \ , \\
\mathcal{L}^{(3)} &= \sum_{k=1}^{10} c^{(3)}(I^{(3)})_k \left(\epsilon\vec{\Theta}\right)^{I^{(3)}(k)} = \sum_{k_1 = 0}^3\sum_{k_2 = 0}^{3-k_1} c^{(3)}_{(k_1,k_2)} \left(\epsilon\Theta_1\right)^{k_1} \left(\epsilon\Theta_2\right)^{k_2} \ . \\
\end{aligned}
\end{equation}
Here, the vectors $c^{(i)}=c^{(i)}(I^{(i)})$ with $i=1,2$ are defined as in eq.~\eqref{eq:DefCVector} in terms of the index sets
\begin{align}
I^{(2)} &= \{(0, 0), (0, 1), (1, 0), (0, 2), (1, 1), (2, 0) \} \ , \\
I^{(3)} &= \{(0, 0), (0, 1), (1, 0), (0, 2), (1, 1), (2, 0), (0, 3), (1, 2), (2,1), (3, 0)\} \ .
\end{align}
The derivation of explicit formulas for $\mathcal{L}^{(2)}$ and $\mathcal{L}^{(3)}$ is rather tedious and technical. We therefore relegate the detailed discussion to appendix~\ref{App:3Fold2Param} and here give a short summary.

\paragraph{Convention and assumption:}
First, recall that not all four correlators $\kappa_{(0,0),(a,3-a)}$ with $0 \leq a \leq 3$ can vanish identically. Second, we make the assumption that
\begin{equation}
\begin{aligned}
\kappa _{(0,0),(0,3)}\kappa _{(0,0),(2,1)} &\neq\kappa _{(0,0),(1,2)}^2  \\ \text{or} \quad \kappa _{(0,0),(3,0)}\kappa _{(0,0),(1,2)} &\neq \kappa_{(0,0),(2,1)}^2 \\ \text{or} \quad \kappa _{(0,0),(3,0)}\kappa _{(0,0),(0,3)} &\neq \kappa _{(0,0),(1,2)} \kappa_{(0,0),(2,1)} \ .
\end{aligned}
\end{equation}
Given these two points, it is possible to rotate the generators of the $U(1)^2$ factor in the gauge group such that
\begin{equation}\label{eq:CY32Assumption}
\begin{aligned}
\kappa_{(0,0),(0,3)} &\neq 0  \\
\text{and} \quad \kappa _{(0,0),(0,3)}\kappa _{(0,0),(2,1)} &\neq \kappa _{(0,0),(1,2)}^2 \ .
\end{aligned}
\end{equation}
In the formulas presented below we assume eq.~\eqref{eq:CY32Assumption}.

\paragraph{Formula for $\mathcal{L}^{(2)}$:} 
Equation~\eqref{eq:CY32Assumption} guarantees that $\mathcal{L}^{(2)}$ is represented by the following coefficient functions:
\begin{equation}\label{eq:3Fold2ParamL2SolveMainText}
\begin{aligned}
 c^{(2)}_{(0,0)} = &+\kappa_{(0,0),(1,2)}\Big(\kappa _{(0,0),(3,0)} \kappa _{(0,2),(0,3)}+\kappa _{(0,2),(2,0)} \kappa _{(0,3),(1,0)}-\kappa _{(0,1),(0,3)} \kappa _{(1,1),(2,0)}\Big) \\
 &+\kappa _{(0,0),(2,1)}
   \Big(\kappa _{(0,1),(0,3)} \kappa _{(0,2),(2,0)}-\kappa _{(0,2),(1,1)} \kappa _{(0,3),(1,0)}+\kappa _{(0,0),(1,2)} \kappa _{(0,3),(1,1)} \\
 &+\kappa _{(0,0),(0,3)} \kappa _{(0,3),(2,0)}\Big)-\kappa _{(0,3),(2,0)} \kappa _{(0,0),(1,2)}^2-\kappa _{(0,0),(2,1)}^2 \kappa _{(0,2),(0,3)}\\ 
  &-\kappa _{(0,0),(3,0)} \kappa _{(0,1),(0,3)} \kappa _{(0,2),(1,1)}-\kappa _{(0,0),(0,3)} \kappa _{(0,0),(3,0)} \kappa _{(0,3),(1,1)} \\
 &-\kappa _{(0,0),(0,3)}
   \kappa _{(0,3),(1,0)} \kappa _{(1,1),(2,0)}\\
 c^{(2)}_{(0,1)} = &-\kappa _{(0,0),(0,3)} \Big(\kappa _{(0,0),(3,0)} \kappa _{(0,2),(1,1)}-\kappa _{(0,0),(2,1)} \kappa _{(0,2),(2,0)}+\kappa _{(0,0),(1,2)} \kappa _{(1,1),(2,0)}\Big) \\
 c^{(2)}_{(1,0)} = &+ \kappa _{(0,0),(0,3)} \Big(\kappa _{(0,0),(2,1)} \kappa _{(0,2),(1,1)}-\kappa _{(0,0),(1,2)} \kappa _{(0,2),(2,0)}+\kappa _{(0,0),(0,3)} \kappa _{(1,1),(2,0)}\Big) \\
  c^{(2)}_{(0,2)} = &+\kappa _{(0,0),(0,3)} \Big(\kappa _{(0,0),(2,1)}^2-\kappa _{(0,0),(1,2)} \kappa _{(0,0),(3,0)}\Big)\\
 c^{(2)}_{(1,1)} = &+\kappa _{(0,0),(0,3)} \Big(\kappa _{(0,0),(0,3)} \kappa _{(0,0),(3,0)}-\kappa _{(0,0),(1,2)} \kappa _{(0,0),(2,1)}\Big) \\
 c^{(2)}_{(2,0)} = &+\kappa _{(0,0),(0,3)} \Big(\kappa _{(0,0),(1,2)}^2-\kappa _{(0,0),(0,3)} \kappa _{(0,0),(2,1)}\Big)\ .
\end{aligned}
\end{equation}
Note that $c^{(2)}_{(2,0)}$ can never be constantly zero. Hence, these coefficient functions indeed correspond to an order two operator. With the formulas listed in Table~\ref{tab:RepList} they can be expressed in terms of the seven correlators $\kappa_{(0,0),(a,3-a)}$ with $0 \leq a \leq 3$ and $\kappa_{(0,0),(b,5-b)}$ with $0 \leq b \leq 2$ only.

\paragraph{Formula for $\mathcal{L}^{(3)}$:}
Equation~\eqref{eq:CY32Assumption} guarantees that $\mathcal{L}^{(3)}$ is represented by the following coefficient functions:
\begin{equation}\label{eq:CY32L3MainText}
\begin{aligned}
c^{(3)}_{(0,0)} = 
&+\kappa _{(0,3),(1,2)} \kappa _{(0,0),(1,2)}^2-\kappa _{(0,2),(1,2)} \kappa _{(0,3),(1,0)} \kappa _{(0,0),(1,2)}-\kappa _{(0,2),(0,3)} \kappa _{(1,0),(1,2)} \kappa _{(0,0),(1,2)}\\
&+\kappa _{(0,1),(0,3)} \kappa
   _{(1,1),(1,2)} \kappa _{(0,0),(1,2)}+\kappa _{(0,1),(1,2)} \left(\kappa _{(0,2),(1,1)} \kappa _{(0,3),(1,0)}-\kappa _{(0,0),(1,2)} \kappa _{(0,3),(1,1)}\right)\\
&+\kappa _{(0,0),(2,1)} \left(\kappa
   _{(0,1),(1,2)} \kappa _{(0,2),(0,3)}-\kappa _{(0,1),(0,3)} \kappa _{(0,2),(1,2)}-\kappa _{(0,0),(0,3)} \kappa _{(0,3),(1,2)}\right)\\
&+\kappa _{(0,1),(0,3)} \kappa _{(0,2),(1,1)} \kappa _{(1,0),(1,2)}+\kappa
   _{(0,0),(0,3)} \kappa _{(0,3),(1,1)} \kappa _{(1,0),(1,2)}\\
&+\kappa _{(0,0),(0,3)} \kappa _{(0,3),(1,0)} \kappa _{(1,1),(1,2)}\\
c^{(3)}_{(0,1)}=&+\kappa_{(0,0),(1,2)}\left(\kappa _{(0,0),(2,1)} \kappa _{(0,2),(0,3)}+\kappa _{(0,2),(1,1)} \kappa _{(0,3),(1,0)}+\kappa _{(0,0),(0,3)} \kappa _{(1,1),(1,2)}\right) \\
&+\kappa _{(0,0),(0,3)} \left(\kappa _{(0,2),(1,1)} \kappa _{(1,0),(1,2)}-\kappa _{(0,0),(2,1)} \kappa _{(0,2),(1,2)}\right)-\kappa _{(0,3),(1,1)} \kappa _{(0,0),(1,2)}^2\\
c^{(3)}_{(1,0)}=&+\kappa_{(0,0),(1,2)}\left(\kappa _{(0,1),(0,3)} \kappa _{(0,2),(1,1)}+\kappa _{(0,0),(0,3)} \left(\kappa _{(0,2),(1,2)}+\kappa _{(0,3),(1,1)}\right)\right)\\
&-\kappa _{(0,0),(0,3)} \left(\kappa _{(0,1),(1,2)} \kappa _{(0,2),(1,1)}+\kappa _{(0,0),(0,3)} \kappa _{(1,1),(1,2)}\right)-\kappa _{(0,2),(0,3)} \kappa _{(0,0),(1,2)}^2\\
c^{(3)}_{(0,2)}=&+\kappa _{(0,3),(1,0)} \kappa _{(0,0),(1,2)}^2+\kappa _{(0,0),(1,2)}\left(\kappa _{(0,0),(2,1)} \kappa _{(0,1),(0,3)}+\kappa _{(0,0),(0,3)} \kappa _{(1,0),(1,2)}\right)\\
&-\kappa _{(0,0),(0,3)} \kappa
   _{(0,0),(2,1)} \kappa _{(0,1),(1,2)}\\
c^{(3)}_{(1,1)}=&-\kappa _{(1,0),(1,2)} \kappa _{(0,0),(0,3)}^2+\kappa _{(0,0),(1,2)}\kappa _{(0,0),(0,3)} \left(\kappa _{(0,1),(1,2)}-\kappa _{(0,3),(1,0)}\right) \\
&-\kappa _{(0,0),(1,2)}^2 \kappa _{(0,1),(0,3)}\\
c^{(3)}_{(2,0)}= &\,\, 0\\
c^{(3)}_{(0,3)}=&+\kappa _{(0,0),(1,2)}\left(\kappa _{(0,0),(1,2)}^2-\kappa _{(0,0),(0,3)}\kappa _{(0,0),(2,1)}\right)\\
c^{(3)}_{(1,2)}=&+\kappa _{(0,0),(0,3)} \left(\kappa _{(0,0),(0,3)} \kappa _{(0,0),(2,1)}-\kappa _{(0,0),(1,2)}^2\right)\\
c^{(3)}_{(2,1)} = &\,\,0\\
c^{(3)}_{(3,0)}= &\,\,0
\end{aligned}
\end{equation}
Note that $c^{(3)}_{(1,2)}$ can never be constantly zero. Hence, these coefficient functions indeed correspond to an order three operator. Moreover, this operator can not be a linear combination of $\mathcal{L}^{(2)}$ and its derivatives. With the formulas listed in Table~\ref{tab:RepList} the above coefficient functions can be expressed in terms of the same seven correlators required to determine $c^{(2)}$, namely $\kappa_{(0,0),(a,3-a)}$ with $0 \leq a \leq 3$ and $\kappa_{(0,0),(b,5-b)}$ with $0 \leq b \leq 2$.

\subsection{One parameter Calabi--Yau fourfolds}\label{sec:CY4}
As a last example we consider four-dimensional Calabi--Yau manifolds with a single K\"ahler parameter as target spaces. Thus $\operatorname{dim}X = 4$ and $r=1$, such that the correlators satisfy
\begin{align}
\epsilon\Theta\, \kappa_{n,m} &=  \kappa_{n,m+1} -\kappa_{n+1,m} \ , \label{eq:RelDerCY4}\\
\kappa_{n,m}  &= (-1)^{n+m} \kappa_{m,n}  \ ,\label{eq:CY4Sym} \\
\kappa_{n,m} &=0 \quad \text{for} \quad  n+m <4 \ .\label{eq:CY4SelecRule}
\end{align}
These properties imply that there is only one independent correlator $\kappa_{n,m}$ for each even value of $n+m\geq 4$, which can be chosen to $\kappa_{n,n}$ with $n\geq 2$. In particular, we find the equality $\kappa_{0,4} = \kappa_{1,3} = \kappa_{2,2}$. To keep formulas short, we will not make all of these implications explicit until the final result in eq.~\eqref{eq:CY4N5cs}. Note that there are two independent selection rules, $\kappa_{0,0}=0$ and $\kappa_{1,1}=0$.

Variation of Hodge structures shows that there is a single Picard--Fuchs operator $\mathcal{L}$ generating the ideal $\boldsymbol{\mathcal{I}}_S^{0}$.  Its order $N$ is, however, not universally fixed. While $N$ is at least five, it can be bigger in general \cite{Gerhardus:2016iot}. We can therefore not derive a universal formula for $\mathcal{L}$, but need to analyze cases of different order seperately. We here consider the most frequently encountered situation of minimal order, i.e., we know specialize to $N=5$. A discussion of the next-to-minimal case with $N=6$ is relegated to the appendix~\ref{App:4FoldN6}.

We thus define the index set $I =\{0, 1, 2, 3, 4, 5\}$ and consider the vector of coefficient functions $c^{(5)} = c(I)$, which determines the order five Picard--Fuchs operator $\mathcal{L}^{(5)}$ as
\begin{equation}\label{eq:LCY4Expand}
\mathcal{L}^{(5)} \, = \, \sum_{k=0}^5 c^{(5)}_k \left(\epsilon \Theta\right)^k \ .
\end{equation} 
By assumption there is such a vector $c^{(5)}$ and it needs to be in the kernel of the $6 \times 6$ matrix $M^{(5)}$ with
\begin{equation}
M^{(5)}=M(I,I) = \left(
\begin{array}{cccccc}
 0 & 0 & 0 & 0 & \kappa _{0,4} & \kappa _{0,5} \\
 0 & 0 & 0 & -\kappa _{0,4} & -\kappa _{1,4} & -\kappa _{1,5} \\
 0 & 0 & \kappa _{0,4} & \kappa _{2,3} & \kappa _{2,4} & \kappa _{2,5} \\
 0 & -\kappa _{0,4} & \kappa _{2,3} & -\kappa _{3,3} & -\kappa _{3,4} & -\kappa _{3,5} \\
 \kappa _{0,4} & -\kappa _{1,4} & \kappa _{2,4} & -\kappa _{3,4} & \kappa _{4,4} & \kappa _{4,5} \\
 \kappa _{0,5} & -\kappa _{1,5} & \kappa _{2,5} & -\kappa _{3,5} & \kappa _{4,5} & -\kappa _{5,5} \\
\end{array}
\right) \ .
\end{equation}
The analysis now goes similar as in Section~\ref{sec:K3}: At a generic point in moduli space the geometric Yukawa coupling $\kappa_{0,4}$ can not vanish. Thus $\operatorname{rank} M^{(5)} \geq 5$, and the non-trivial kernel condition $\operatorname{det} M^{(5)}= 0$ is equivalent to $\kappa_{5,5} = \kappa_{5,5}^{N=5}$ given by
\begin{equation}\label{eq:k55N5CY4}
\begin{aligned}
\kappa_{5,5}^{N=5} \cdot \kappa_{0,4}^5 = &-\kappa _{2,5}^2 \kappa _{0,4}^4+2 \kappa _{1,5} \kappa _{3,5} \kappa _{0,4}^4-2 \kappa _{0,5} \kappa _{4,5} \kappa _{0,4}^4+2 \kappa _{1,5} \kappa _{2,3} \kappa _{2,5} \kappa_{0,4}^3-\kappa _{0,5}^2 \kappa _{1,4}^2 \kappa _{2,3}^2\\
&+2 \kappa _{0,5} \kappa _{2,4} \kappa _{2,5} \kappa _{0,4}^3-\kappa _{1,5}^2 \kappa _{3,3} \kappa _{0,4}^3-2 \kappa _{0,5} \kappa _{1,5} \kappa _{3,4} \kappa_{0,4}^3-2 \kappa _{0,5} \kappa _{1,4} \kappa _{3,5} \kappa _{0,4}^3\\
&+\kappa _{0,5}^2 \kappa _{4,4} \kappa _{0,4}^3-\kappa _{1,5}^2 \kappa _{2,3}^2 \kappa _{0,4}^2-\kappa_{0,5}^2 \kappa _{2,4}^2 \kappa _{0,4}^2-2 \kappa _{0,5} \kappa _{1,5} \kappa _{2,3} \kappa _{2,4} \kappa _{0,4}^2\\
&-2 \kappa _{0,5} \kappa _{1,4} \kappa _{2,3} \kappa _{2,5}
   \kappa _{0,4}^2+2 \kappa _{0,5} \kappa _{1,4} \kappa _{1,5} \kappa _{3,3} \kappa _{0,4}^2+2 \kappa _{0,5}^2 \kappa _{1,4} \kappa _{3,4} \kappa _{0,4}^2\\
&+2 \kappa _{0,5} \kappa_{1,4} \kappa _{1,5} \kappa _{2,3}^2 \kappa _{0,4}+2 \kappa _{0,5}^2 \kappa _{1,4} \kappa _{2,3} \kappa _{2,4} \kappa _{0,4}-\kappa _{0,5}^2 \kappa _{1,4}^2 \kappa _{3,3}
   \kappa _{0,4}\ .
\end{aligned}
\end{equation}
Upon inserting this constraint into $M^{(5)}$, the vector $c^{(5)}$ is found as the up to scaling unique kernel element and reads
\begin{equation}\label{eq:CY4N5cs}
\begin{aligned}
c^{(5)}_0 = &
-\tfrac{1}{2}\kappa _{0,4}^4 \left(\epsilon\Theta\,\kappa_{4,4}\right)-\tfrac{45}{32}\left(\epsilon\Theta\,\kappa_{0,4}\right)^5-\tfrac{3}{8} \kappa _{0,4} \left(\epsilon\Theta\,\kappa_{0,4}\right)^3 \left[4 \kappa _{3,3}-7 \left(\epsilon^2\Theta^2\,\kappa_{0,4}\right)\right]\\
&-\tfrac{3}{8}\kappa _{0,4}^2 \left(\epsilon\Theta\,\kappa_{0,4}\right) \big\{4 \kappa _{3,3}^2+2 \kappa _{3,3} \left(\epsilon^2\Theta^2\,\kappa_{0,4}\right)+3 \left(\epsilon^2\Theta^2\,\kappa_{0,4}\right)^2\\
&+\left(\epsilon\Theta\,\kappa_{0,4}\right)
   \left[\left(\epsilon^3\Theta^3\,\kappa_{0,4}\right)-7 \left(\epsilon\Theta\,\kappa_{3,3}\right)\right]\big\}+\tfrac{1}{4} \kappa _{0,4}^3 \big\{4 \kappa _{4,4} \left(\epsilon\Theta\,\kappa_{0,4}\right)-\left(\epsilon\Theta\,\kappa_{3,3}\right) \left(\epsilon^2\Theta^2\,\kappa_{0,4}\right)\\
&-3 \left(\epsilon\Theta\,\kappa_{0,4}\right) \left(\epsilon^2\Theta^2\,\kappa_{3,3}\right)+2 \kappa _{3,3} \left[2 \left(\epsilon\Theta\,\kappa_{3,3}\right)+\left(\epsilon^3\Theta^3\,\kappa_{0,4}\right)\right]+\left(\epsilon^2\Theta^2\,\kappa_{0,4}\right)
   \left(\epsilon^3\Theta^3\,\kappa_{0,4}\right)\big\}
\\[0.2em]
c^{(5)}_1 = &
-\tfrac{15}{16}\kappa_{0,4} \left(\epsilon\Theta\,\kappa_{0,4}\right)^4-\tfrac{1}{2} \kappa _{0,4}^4 \left[2 \kappa _{4,4}+\left(\epsilon^2\Theta^2\,\kappa_{3,3}\right)\right]+\tfrac{9}{8} \kappa _{0,4}^2 \left(\epsilon\Theta\,\kappa_{0,4}\right)^2 \left[\left(\epsilon^2\Theta^2\,\kappa_{0,4}\right)-2
   \kappa _{3,3}\right]\\
  & +\tfrac{1}{4} \kappa _{0,4}^3 \left\{4 \kappa _{3,3}^2+8 \kappa _{3,3} \left(\epsilon^2\Theta^2\,\kappa_{0,4}\right)+\left(\epsilon\Theta\,\kappa_{0,4}\right) \left[2 \left(\epsilon\Theta\,\kappa_{3,3}\right)-\left(\epsilon^3\Theta^3\,\kappa_{0,4}\right)\right]\right\}
\\[0.2em]
c^{(5)}_2 = & -\tfrac{15}{8} \kappa _{0,4}^2\left(\epsilon\Theta\,\kappa_{0,4}\right)^3 +\tfrac{3}{4}\kappa _{0,4}^3 \left(\epsilon\Theta\,\kappa_{0,4}\right) \left[4 \kappa _{3,3}+3 \left(\epsilon^2\Theta^2\,\kappa_{0,4}\right)\right]\\
&-\tfrac{1}{2} \kappa _{0,4}^4
   \left[3 \left(\epsilon\Theta\,\kappa_{3,3}\right)+\left(\epsilon^3\Theta^3\,\kappa_{0,4}\right)\right]
\\[0.2em]
c^{(5)}_3 = & -\kappa _{0,4}^4 \left[\kappa _{3,3}+2 \left(\epsilon^2\Theta^2\,\kappa_{0,4}\right)\right]+\tfrac{15}{4}\kappa_{0,4}^3 \left(\epsilon\Theta\,\kappa_{0,4}\right)^2\\[0.2em]
c^{(5)}_4 = & -\tfrac{5}{2} \kappa _{0,4}^4 \left(\epsilon\Theta\, \kappa_{0,4}\right) \\[0.2em]
c^{(5)}_5 = & +\kappa _{0,4}^5 \ .
\end{aligned}
\end{equation}
With eq.~\eqref{eq:LCY4Expand} this determines the Picard--Fuchs operator $\mathcal{L}^{(5)}$.

The above formulae express $\mathcal{L}^{(5)}$ in terms of the three correlators $\kappa_{0,4} =\kappa_{2,2}$, $\kappa_{3,3}$ and $\kappa_{4,4}$. We thus expect two differential-algebraic relations among the coefficient functions. Indeed,
\begin{equation} \label{eq:CY4algdiff}
\begin{aligned}
12500 c_0 c_5^4 &= 64 c_4^5-1600 c_{5,1} c_4^4+14000 c_{5,1}^2 c_4^3-200 c_3 c_5 c_4^3+1600 c_5 c_{4,1} c_4^3-4000 c_5 c_{5,2} c_4^3\\
&\quad-50000 c_{5,1}^3 c_4^2-1500 c_5^2 c_{3,1} c_4^2+4000 c_5^2
   c_{4,2} c_4^2+3000 c_3 c_5 c_{5,1} c_4^2-20000 c_5 c_{4,1} c_{5,1} c_4^2\\
&\quad+40000 c_5 c_{5,1} c_{5,2} c_4^2-5000 c_5^2 c_{5,3} c_4^2+60000 c_{5,1}^4 c_4+2500 c_1 c_5^3 c_4+6000
   c_5^2 c_{4,1}^2 c_4\\
&\quad-13750 c_3 c_5 c_{5,1}^2 c_4+70000 c_5 c_{4,1} c_{5,1}^2 c_4+15000 c_5^2 c_{5,2}^2 c_4-3750 c_5^3 c_{3,2} c_4\\
&\quad+5000 c_5^3 c_{4,3}
   c_4+11250 c_5^2 c_{3,1} c_{5,1} c_4-25000 c_5^2 c_{4,2} c_{5,1} c_4+5000 c_3 c_5^2 c_{5,2} c_4\\
&\quad-90000 c_5 c_{5,1}^2 c_{5,2} c_4-25000 c_5^2 c_{4,1} c_{5,2} c_4+20000 c_5^2
   c_{5,1} c_{5,3} c_4-2500 c_5^3 c_{5,4} c_4\\
&\quad+18750 c_3 c_5 c_{5,1}^3-60000 c_5 c_{4,1} c_{5,1}^3-18750 c_5^2 c_{3,1} c_{5,1}^2+30000 c_5^2 c_{4,2} c_{5,1}^2+6250 c_5^4
   c_{1,1}\\
&\quad-3125 c_5^4 c_{3,3}-3750 c_5^3 c_{3,1} c_{4,1}-1250 c_3 c_5^3 c_{4,2}+10000 c_5^3 c_{4,1} c_{4,2}+2500 c_5^4 c_{4,4}\\
&\quad-6250 c_1 c_5^3 c_{5,1}-20000 c_5^2 c_{4,1}^2
   c_{5,1}+9375 c_5^3 c_{3,2} c_{5,1}+6250 c_3 c_5^2 c_{4,1} c_{5,1}\\
&\quad+9375 c_5^3 c_{3,1} c_{5,2}-15000 c_5^3 c_{4,2} c_{5,2}-18750 c_3 c_5^2 c_{5,1}
   c_{5,2}+60000 c_5^2 c_{4,1} c_{5,1} c_{5,2}\\
&\quad+3125 c_3 c_5^3 c_{5,3}-10000 c_5^3 c_{4,1} c_{5,3}-1500 c_3 c_5^2 c_{4,1} c_4-10000 c_5^3 c_{4,3} c_{5,1} \ ,\\[0.2em]
50 c_2 c_5^2 &= -8 c_4^3+60 c_{5,1} c_4^2-100 c_{5,1}^2 c_4+30 c_3 c_5 c_4-60 c_5 c_{4,1} c_4+50 c_5 c_{5,2} c_4\\
&\quad+75 c_5^2 c_{3,1}-50 c_5^2 c_{4,2}-75 c_3 c_5 c_{5,1}+100 c_5 c_{4,1} c_{5,1}\ ,
\end{aligned}
\end{equation}
where we abbreviate $c_{k,n} = \epsilon^n\Theta^n\, c^{(5)}_k$. Both relations are invariant under a commong rescaling of all the $c^{(5)}_k$, and they unambiguously fix $c^{(5)}_0$ and $c^{(5)}_2$.

\subsection{Descendant relations and recursion}\label{sec:Recurs}
The existence of a relation as defined in eq.~\eqref{eq:RelansatzRepeat} or eq.~\eqref{eq:LFromRWithoutM} implies a whole tower of descendent relations and thereby puts severe constraints on the set of correlators. In the case of one K\"ahler parameter  a single relation recursively determines almost all correlators in terms of a finite subset needed to start the recursion. Here, we derive these properties explicitly and comment on their implications for the examples of the previous subsections.

Let us start with a relation and its corresponding Picard--Fuchs operator $\mathcal{L}$ as defined in eq.~\eqref{eq:LFromRWithoutM}. By differentiating the relation and taking an appropriate linear combination with a different number of north pole insertions we obtain the descendant relation
\begin{equation}\label{eq:DescendantRelation1}
0 \, = \, \sum_{\vec{p}=0}^{\vec{m}} {\vec{m} \choose \vec{p}} \left(\epsilon\vec{\Theta}\right)^{\vec{p}} \, \sum_{\vec{\ell}=0}^{\vec{N}} c_{\vec{\ell}} \,\kappa_{\vec{m}-\vec{p}+\vec{n},\vec{l}} 
 \,= \, \sum_{\vec{q}=0}^{\vec{m}} \sum_{\vec{\ell}=0}^{\vec{N}} {\vec{m} \choose \vec{q}} \left[\left(\epsilon\vec{\Theta}\right)^{\vec{m}-\vec{q}}c_{\vec{\ell}}\, \right] \kappa_{\vec{n},\vec{l}+\vec{q}} \ ,
\end{equation}
which holds for all $\vec{n},\vec{m}\in \mathbb{Z}^r_{\geq 0}$. Here, the binomial of two vectors is understood as the product of binomials of their components. The second equal sign hides a tedious but elementary calculation that uses eq.~\eqref{eq:LFromRWithoutM} and the derivative rule~\eqref{eq:RelDerGen} only. To make the above descendant relation more transparent we recast it into the form
\begin{equation}\label{eq:DescendantRelation2}
0 = \sum_{\vec{\ell} = 0}^{\vec{N}+\vec{m}} \tilde{c}^{(\vec{m})}_{\vec{\ell}}\,  \kappa_{\vec{n},\vec{l}} \qquad \text{with} \qquad \tilde{c}^{(\vec{m})}_{\vec{\ell}} = \sum_{\vec{q}=\operatorname{Max}(0,\vec{\ell}-\vec{N})}^{\operatorname{Min}(\vec{m},\vec{\ell)}} {\vec{m}\choose \vec{q}}\left(\epsilon \vec{\Theta}\right)^{\vec{m}-\vec{q}} c_{\vec{\ell}-\vec{q}} \ ,
\end{equation}
and note that the coefficient functions $\tilde{c}_{\vec{\ell}}$ correspond to the operator $(\epsilon\vec{\Theta})^{\vec{m}} \mathcal{L}$. Equation~\eqref{eq:DescendantRelation1} thus says that if $\mathcal{L}$ is a valid operator, so are all its derivatives. So far the discussion is not restricted to the Calabi--Yau case, it equally works for non-vanishing twisted masses.

We now specialize to the case of a single K\"ahler parameter. Using the fact that $\tilde{c}^{(m)}_{N+m} = c_N$, equation~\eqref{eq:DescendantRelation2} then gives
\begin{equation}\label{eq:Recurse1Param}
\kappa_{n,N+m} = -\frac{1}{c_N} \sum_{\ell = 0}^{N+m-1} \tilde{c}^{(m)}_{\ell}\,  \kappa_{n,l}
\end{equation}
for all $n,m\geq 0$. Without loss of generality we can assume $c_N \neq 0$, otherwise we replace $N$ with $(N-1)$ from the beginning. Formula~\eqref{eq:Recurse1Param} is thus well defined and expresses the $(n+N+m)$-point correlator $\kappa_{n,N+m}$ in terms of lower-point correlators. This leaves only those $\kappa_{n,m}$ with $m<N$ undetermined. In the case of vanishing twisted masses we can moreover employ the symmetry property~\eqref{eq:SymEasy} to determine all $\kappa_{n,m}$ with $n>m$ in terms of $\kappa_{m,n}$. Once a single relation is given, it thus fixes all but finitely many correlators. If further this relation is expressed in terms of a finite set of $\kappa_{n,m}$, only finitely many correlators determine all the others. Let us make this explicit for the one-parameter cases discussed in the previous subsections:
\begin{itemize}
\item \textbf{Elliptic curve:}\\
Given the order $N=2$ operator, all correlators can be recursed back to $\kappa_{0,1}$. Thus, $\kappa_{0,1}$ and $\kappa_{1,2}$ determine all correlators.
\item \textbf{One-parameter polarized K3 surface:}\\
Given the order $N=3$ operator, all correlators can be recursed back to $\kappa_{0,2}$. Thus, $\kappa_{0,2}$ and $\kappa_{2,2}$ determine all correlators.
\item \textbf{One-parameter Calabi--Yau threefold:}\\
Given the order $N=4$ operator, all correlators can be recursed back to $\kappa_{0,3}$. Thus, $\kappa_{0,3}$, $\kappa_{2,3}$ and $\kappa_{3,4}$ determine all correlators.
\item \textbf{One-parameter Calabi--Yau fourfold:}\\
Given an order $N=5$ operator, all correlators can be recursed back to $\kappa_{0,4}$. Then, $\kappa_{0,4}$, $\kappa_{3,3}$, and $\kappa_{4,4}$ determine all correlators.\\
Given an order $N=6$ operator, all correlators can be recursed back to $\kappa_{0,4}$ and $\kappa_{3,3}$. Then, $\kappa_{0,4}$, $\kappa_{3,3}$, $\kappa_{4,4}$ and $\kappa_{5,5}$ determine all correlators.\\
\end{itemize}
In the case of several K\"ahler parameters the recursion is not as simple. This is due to there generically not being a unique $\kappa_{\vec{n},\vec{m}}$ with the maximal number of north plus south pole insertions in eq.~\eqref{eq:DescendantRelation2}. Equivalently, a Picard--Fuchs operator depending on several K\"ahler parameter generically does not have a unique term with the maximal number of derivatives.

\subsection{Picard--Fuchs operators for specific models}\label{sec:FixedModels}
So far we have derived general formulae for Picard--Fuchs operators that apply to certain classes of examples. For cases with several K\"ahler parameters or high-dimensional target spaces their derivation may be rather tedious, see for example the case of two-parameter Calabi--Yau threefolds in Section~\ref{sec:CY32Param}.

An alternative approach is to calculate several correlators so as to determine the appropriate matrix $M(I,J)$ and its kernel explicitly. All non-trivial conditions amongst the correlators are then met automatically. It can not be excluded, however, that for certain cases some kernel elements may not correspond to a proper relation. Therefore, knowledge of the number of independent relations and their respective orders is still essential. If the explicit calculation determines exactly as many relations as are expected, they necessarily correspond to valid relations.

Moreover, this alternative approach equally works for non Calabi--Yau cases and the case of non-vanishing twisted masses. Due to the more complicated symmetry relation~\eqref{eq:RelSym}, it would be harder to derive general formulae in these situations.

\section{Correlators and the Givental $I$-function} \label{sec:Giv}
In this section we spell out the relationship between the analyzed gauge theory correlators and the Givental $I$-function, as discussed by Ueda and Yoshida in ref.~\cite{Ueda:2016wfa} and established by Kim et. al. in ref.~\cite{Kim:2016uq}.\footnote{See also ref.~\cite{Bonelli:2013mma} for a related analysis, realizing the Givental $I$-function in the context of target spaces of $\mathcal{N}=(2,2)$ quiver gauged linear sigma models.} This connection allows us then to directly interpret the gauged linear sigma model correlators geometrically in the context of Gromov--Witten theory \cite{MR1653024,MR2276766}. 

In ref.~\cite{MR1653024} Givental introduces the $I$-function for complete intersections $X=\mathbb{P}^d_\Delta(\vec\rho_1^{\,p},\ldots,\vec\rho_n^{\,p})$ of codimension $n$ in a (non-singular) compact weak Fano toric variety $\mathbb{P}^d_\Delta$ given in terms of the symplectic quotient~\eqref{eq:SympQuotient}. The $I$-function is a formal function in terms of the input $\vec t=(t_1,\ldots,t_r)$ --- the coordinates on $H^2(X)$ with respect to the basis $\vec p=(p_1,\ldots,p_r)$ of $H^2(X)$ --- and in the parameter $\hbar$. It maps to the (even) cohomology ring $H^\text{ev}(X)$ and reads
\begin{equation} \label{eq:GivI}
    I_X( \vec t,\hbar) 
    \,=\, e^{{\vec t \cdot \vec p}/\hbar}\sum_{\vec k} e^{\vec t \cdot \vec k} 
    \frac{\prod_{i=1}^n \prod_{s=-\infty}^{\vec k(v_i)} (v_i + s \hbar) \prod_{\ell=1}^{d+r}\prod_{s=-\infty}^0 (u_\ell + s \hbar)}
    {\prod_{i=1}^n \prod_{s=-\infty}^{0} (v_i + s \hbar) \prod_{\ell=1}^{d+r}\prod_{s=-\infty}^{\vec k(u_\ell)} (u_\ell + s \hbar)}
    \,\in\, H^\text{ev}(X) \ .
\end{equation}
Here $u_\ell$, $\ell=1,\ldots,d+r$, are the toric hyperplane classes of $\mathbb{P}^d_\Delta$ generating the ring $H^{\text{ev}}(\mathbb{P}^d_\Delta)$, and $v_i$, $i=1,\ldots, n$, are the first Chern classes of the non-negative line bundles $\mathcal{O}_\Delta(\vec\rho_i^{\,p})$ associated with the complete intersection $X$. The sum runs over the semi-group of compact holomorphic curves $\vec k$ in the variety $X$, and $\vec k(\,\cdot\,)$ abbreviates the intersection pairing $\int_{\vec k}\,\cdot\,$.

Kim, Oh, Ueda and Yoshida conjecture and prove for certain classes of examples a direct relationship between the Givental $I$-function and the gauged linear sigma model correlators \cite{Kim:2016uq}. Starting from a pairing of Givental $I$-function, defined in ref.~\cite{MR1653024,Kim:2016uq} as
\begin{equation}
    \Phi(\vec{t},\vec{t}',\hbar) \,=\, \int_X I_X(\vec t,-\hbar) \cup I_X(\vec{t}',\hbar) \ ,
\end{equation} 
the authors establish that $\Phi$ is the generating function of the discussed gauge linear sigma model correlators. 

\begin{table}[th]
\hfil
\hbox{
\vbox{
\offinterlineskip
\halign{\strut\vrule width1.2pt\hfil~#~\hfil\vrule width0.5pt&\hfil~#~\hfil\vrule width0.5pt&\hfil~#~\hfil\vrule width0.5pt&\hfil~#~\hfil\vrule width1.2pt\cr
\noalign{\hrule height 1.2pt}
Chiral multiplets & $U(1)^r$~charge & $U(1)_R$~charge & twisted masses\cr
\noalign{\hrule height 1.2pt}
$X_\ell$, $\ell=1,\ldots,d+r$ & ${\vec\rho}^{\,x}_\ell$ & $0$ & $\tm_\ell$ \cr
\noalign{\hrule height 0.5pt}
$P^i$, $i=1,\ldots,n$ & $-\vec\rho_i^{\,p}$ & $2$ & $\tm^i_P$ \cr
\noalign{\hrule height 1.2pt}
}}}
\hfil
\caption{This table shows the matter spectrum of the $U(1)^r$ gauged linear sigma model with the semi-classical large volume target space $X=\mathbb{P}^d_\Delta(\vec\rho_1^{\,p},\ldots,\vec\rho_n^{\,p})$. The $U(1)^r$ charge vectors of the chiral fields $X_\ell$ correspond to the one-dimensional cones in the fan $\Delta$, realizing the toric variety $\mathbb{P}^d_\Delta$ as the ambient space of $X$. Furthermore, the chiral fields $P^i$ are responsible for the complete intersection locus $X \subset \mathbb{P}^d_\Delta$, which arises in the gauge theory from the F-terms of the superpotential.} \label{tab:IFuncSpec}
\end{table}
That is to say, let us consider the Abelian gauged linear sigma model with the chiral matter spectrum displayed in Table~\ref{tab:IFuncSpec}. It realizes the complete intersection $X$ as its semi-classical target space in the large volume phase. Then, upon identifying the arguments of the Givental $I$-function \eqref{eq:GivI} with the gauge theory parameters according to
\begin{equation} \label{eq:Idic}
   \epsilon \,=\, \hbar \ , \qquad e^{\vec t \cdot \vec k} \,=\, (-1)^{\sum_{i=1}^n \vec\rho_i^p} \vec Q^{\vec k} \ ,
\end{equation}
the correlators~\eqref{eq:AbCorr} of the gauged linear sigma model in Table~\ref{tab:IFuncSpec} are given by \cite{Kim:2016uq}
\begin{equation} \label{eq:CorrCorrespondence}
    \kappa_{\vec n,\vec m}(\vec Q,0,\epsilon) \,=\, 
    \epsilon^{|\vec m|_1 }(-\epsilon)^{|\vec n|_1 }
    \left. 
    \frac{ \partial^{|\vec n|_1 + |\vec m|_1}  \Phi(\vec{t},\vec{t}',\epsilon)}
    { \partial t_1^{n_1} \cdots \partial t_r^{n_r} \partial {t'_1}^{m_1} \cdots \partial {t'_r}^{m_r}}
    \right|_{\vec t = \vec{t}' = \log\left( \pm {\vec Q} \right)} \ .
\end{equation}
Here, we explicitly spell out the correlator correspondence for vanishing twisted masses $\tm_\ell$ and $\tm^i_P$. By matching the equivariant parameters $\Lambda_\ell$, $\ell=1,\ldots,d+r$, of the toric $\mathbb{C}^*$-symmetries and $\Lambda'_i$, $i=1,\ldots,n$, of the $\mathbb{C}^*$-symmetries of the line bundles $\mathcal{O}_\Delta(\vec \rho_i^{\,p})$ with the twisted masses $\tm_\ell$ and $\tm^i_P$, respectively, it is straight-forward to restore the twisted masses in order to obtain the generalized correlator correspondence in the equivariant setting. 

Note that in the stated correlator correspondence~\eqref{eq:CorrCorrespondence} the parameter~$\epsilon$ of the gauge theory on $S^2$ is identified with the parameter $\hbar$ in the Givental $I$-function according to the dictionary~\eqref{eq:Idic}. As a consequence the dependence on $\epsilon$ in the correlation functions captures Gromov--Witten invariants with insertions of the classes $\psi^k$, $k=0,1,2,\ldots$, at their marked points \cite{MR1653024,MR2276766}. Here the class $\psi$ denotes the first Chern class of the universal cotangent line bundle over the moduli space of stable maps. 

Inserting the geometric definition~\eqref{eq:CorrCorrespondence} of the gauge theory correlators into a south pole correlator relation~\eqref{eq:Relansatz} and using eqs.~\eqref{eq:RSOp} and \eqref{eq:OpRep}, we find
\begin{equation}
\begin{aligned}
    0 \,=\, R_S(\vec Q,\epsilon,\kappa_{\vec n, \cdot }) 
    \,&=\,  (-1)^{|n|_1}\int_X (\epsilon \vec\Theta)^{\vec n} I_X(\vec Q,-\epsilon)\cup\left(\sum_{\vec m} c_{\vec m}(\vec Q,\epsilon) (\epsilon\vec\Theta)^{\vec m} I_X(\vec Q,\epsilon) \right) \\
    \,&=\,  (-1)^{|n|_1} \int_X (\epsilon \vec\Theta)^{\vec n} I_X(\vec Q,-\epsilon)\cup\left(\boldsymbol{R}_S(\vec Q,\epsilon \vec\Theta,\epsilon) I_X(\vec Q,\epsilon) \right) \ .
\end{aligned}    
\end{equation}
Here we express the Givental $I$-function in terms of the gauge theory parameters~$\vec Q$ instead of the parameters $\vec t$, and we suppress the twisted masses for simplicity. As the above relation holds for general $\vec n$, we conclude that the differential operators $\boldsymbol{R}_S$ of the south pole correlator relations annihilate the Givental $I$-function, i.e.,
\begin{equation}
    \boldsymbol{R}_S(\vec Q,\epsilon \vec\Theta,\epsilon) I_X(\vec Q,\epsilon) \,=\, 0 \ .
\end{equation}
This result explicitly connects the differential operators obtained from the gauge theory correlator relations with the quantum cohomology of the target space geometry. The established relationship of the correlator relations to the Givental $I$-function also reflects the close relationship between the analyzed correlators and the quantum A-periods of the A-twisted gauged linear sigma model considered in ref.~\cite{Hori:2000kt}.

In ref.~\cite{MR3126932} Ciocan-Fontanine, Kim and Maulik generalize the Givental $I$-function to more general GIT~quotients than toric varieties. These geometries relate to gauged linear sigma model target spaces of non-Abelian gauge groups. We expect that the stated correlator conjecture~\eqref{eq:CorrCorrespondence} holds beyond the class of Abelian gauged linear sigma models. As a matter of fact in ref.~\cite{Kim:2016uq} the authors prove the correspondence for Grassmannian target spaces. But it would interesting to examine this connection further in the context of non-Abelian gauged linear sigma models and more general GIT~quotients.

\section{Examples} \label{sec:Examples}
In this section we study explicit examples of various gauged linear sigma models with focus on the properties of their target space geometries. This is meant to illustrate the general concepts and methods introduced in the previous sections. For the examples we determine correlator relations directly from their gauge theory spectrum as derived in Section~\ref{sec:CorRel}, we compute differential operators and for Calabi--Yau geometries analyze their coefficients in terms of gauge theory correlators illustrating the analysis in Section~\ref{sec:CYtarget}, and we exhibit the connection to the Givental $I$-function as covered in Section~\ref{sec:Giv}.

\subsection{Fano varieties as target spaces}
Our first three examples are gauged linear sigma models with Fano varieties as their semi-classical target spaces: Namely the projective line $\mathbb{P}^1$, the quartic threefold $\mathbb{P}^4[4]$, and the Grassmannian fourfold $\operatorname{Gr}(2,4)$.

\subsubsection{The complex projective line $\mathbb{P}^1$}
We start by considering a gauged linear sigma model with Abelian gauge group $U(1)$ and charged matter spectrum as listed in Table~\ref{tab:P1}.
\begin{table}[h]
\hfil
\hbox{
\vbox{
\offinterlineskip
\halign{\strut\vrule width1.2pt\hfil~#~\hfil\vrule width0.5pt&\hfil~#~\hfil\vrule width0.5pt&\hfil~#~\hfil\vrule width0.5pt&\hfil~#~\hfil\vrule width1.2pt\cr
\noalign{\hrule height 1.2pt}
Chiral multiplets & $U(1)$~charge & $U(1)_R$~charge & twisted masses\cr
\noalign{\hrule height 1.2pt}
$\phi_i$, $i=1,2$ & $+1$ & $0$ & $\tm_i$ \cr
\noalign{\hrule height 1.2pt}
}}}
\hfil
\caption{Matter spectrum of the $U(1)$ gauged linear sigma model of the projective line $\mathbb{P}^1$.} \label{tab:P1}
\end{table}
It is well known that by minimizing the associated classical scalar potential for a positive Fayet--Iliopoulos parameter, the symplectic quotient~\eqref{eq:SympQuotient} yields the complex projective line $\mathbb{P}^1$ as classical target space geometry.

From the matter spectrum in Table~\ref{tab:P1} we readily determine the functions defined in eq.~\eqref{eq:DefWandG} as
\begin{equation} \label{eq:gsP1}
  g_p(w,\tm_i,\epsilon) \,=\, \prod_{s=0}^{p-1}(w + \tm_1 - \epsilon s)(w + \tm_2 - \epsilon s) \ , \quad p=0,1,2,\ldots \ .
\end{equation}
Thus $g_0=1$ and $g_1=(w +\tm_1)(w+\tm_2)$, for which the constraint~\eqref{eq:RelFinalCond1} is easily solved by the polynomials 
\begin{equation}
  \alpha_0(w,\tm_i,\epsilon)\,=\,(w+\tm_1)(w+\tm_2)\ , \quad \alpha_1(w,\tm_i,\epsilon)\,=\,-1\ .
\end{equation}
Together with eq.~\eqref{eq:RSDef} these determine the south pole correlator relation
\begin{equation}
  R_S(Q,\tm_i,\epsilon,\kappa_{n,\,\cdot\,}) 
  \,=\, \left\langle \sigma_N^n  (\sigma_S+\tm_1)(\sigma_S+\tm_2) \right\rangle -Q  \left\langle \sigma_N^n \right\rangle \ ,
\end{equation}
which according to eqs.~\eqref{eq:RSOp} and \eqref{eq:OpRep} further corresponds to the differential operator 
\begin{equation} \label{eq:LP1}
  \mathcal{L}(Q,\epsilon,\tm_i) = (\epsilon \Theta +\tm_1)(\epsilon \Theta + \tm_2)  - Q \ ,
\end{equation}
given in terms of the logarithmic derivative  $\Theta = Q \partial_Q$. It can be checked that this operator generates the entire differential ideal $\boldsymbol{\mathcal{I}}_S$ of the gauge theory. That is to say, other south pole correlator relations obtained from the higher degree polynomials~\eqref{eq:gsP1} yield differential operators in the differential ideal generated by the above operator~\eqref{eq:LP1}.

For the projective line the Givental $I$-function takes the form \cite{MR1653024,MR2276766}
\begin{equation}
  I_{\mathbb{P}^1}(H,Q,\epsilon,\tm_i) \,=\, \sum_{k=0}^{\infty} \frac{1}{\prod_{\ell=1}^k (H+\tm_1 + \ell \epsilon)(H+\tm_2 + \ell \epsilon) } Q^{\frac{H}\epsilon+k} \ .
\end{equation}
Here, $H$ is the hyperplane divisor of $\mathbb{P}^1$ and the twisted masses $\tm_i$ correponds to the equivariant parameters of the $(\mathbb{C}^*)^2$-action canonically acting on the homogeneous coordinates of the projective line $\mathbb{P}^1$. Thus, in terms of the hyperplane class $H$ and the equivariant parameters $\tm_i$ the equivariant cohomology ring reads
\begin{equation}
   H^*_{(\mathbb{C}^*)^2}(\mathbb{P}^1,\mathbb{C}) \,=\, \left. \mathbb{C}[H,\tm_i] \middle/ (H+\tm_1)(H+\tm_2) \right. \ .
\end{equation}   
As observed in general in Section~\ref{sec:Giv}, we can check that the Givental $I$-function of $\mathbb{P}^1$ is indeed annihilated by the differential operator ideal $\boldsymbol{\mathcal{I}}_S$, i.e., 
\begin{equation}
    \mathcal{L}(Q,\epsilon,\tm_i)I_{\mathbb{P}^1}(H,Q,\epsilon,\tm_i) \,=\, 0 \ .
\end{equation}

A $U(1)$ gauged linear sigma model with $n+1$ chiral multiplets $\phi_i$, $i=1,\ldots,n+1$ that have twisted masses $\tm_i$, gauge charges $+1$, and R-charge $0$ is a natural generalization of the spectrum given in Table~\ref{tab:P1}. The gauge theory's classical target space is then given by the complex projective space $\mathbb{P}^n$, and the correlator relations yield the differential operator
\begin{equation} \label{eq:LPn}
  \mathcal{L}(Q,\epsilon,\tm_i) = (\epsilon \Theta +\tm_1)\cdots(\epsilon \Theta + \tm_{n+1})  - Q  \ ,
\end{equation}
which now annihilates the equivariant Givental $I$-function of $\mathbb{P}^n$ \cite{MR1653024,MR2276766}
\begin{equation}
  I_{\mathbb{P}^n}(H,Q,\epsilon,\tm_i) \,=\, \sum_{k=0}^{\infty} \frac{1}{\prod_{\ell=1}^k (H+\tm_1 + \ell \epsilon)\cdots(H+\tm_{n+1} + \ell \epsilon) } Q^{\frac{H}\epsilon+k} \ .
\end{equation}  
Here, $H$ and $\tm_i$ furnish the hyperplane class and the equivariant parameters of the equivariant cohomology ring $\left. \mathbb{C}[H,\tm_i] \middle/ (H+\tm_1)\cdots(H+\tm_{n+1}) \right.$ of the projective space $\mathbb{P}^n$.

\subsubsection{The quartic Fano threefold $\mathbb{P}^{4}[4]$} \label{sec:FP4[4]}
Let us now consider the gauged linear sigma model with Abelian gauge group $U(1)$ and the matter spectrum listed in Table~\ref{tab:P4[4]}.
\begin{table}[h]
\hfil
\hbox{
\vbox{
\offinterlineskip
\halign{\strut\vrule width1.2pt\hfil~#~\hfil\vrule width0.5pt&\hfil~#~\hfil\vrule width0.5pt&\hfil~#~\hfil\vrule width0.5pt&\hfil~#~\hfil\vrule width1.2pt\cr
\noalign{\hrule height 1.2pt}
Chiral multiplets & $U(1)$~charge & $U(1)_R$~charge & twisted masses\cr
\noalign{\hrule height 1.2pt}
$\phi_i$, $i=1,\ldots,5$ & $+1$ & $0$ & $\tm_i$ \cr
\noalign{\hrule height 0.5pt}
$P$ & $-4$ & $2$ & $\tm_P$ \cr
\noalign{\hrule height 1.2pt}
}}}
\hfil
\caption{Matter spectrum of the $U(1)$ gauged linear sigma model of the quartic Fano threefold $\mathbb{P}^4[4]$.} \label{tab:P4[4]}
\end{table}
The associated classical target space geometry is obtained in two steps: First, the symplectic quotient~\eqref{eq:SympQuotient} with a positive Fayet--Iliopoulos parameter yields the non-compact toric variety $\mathcal{O}(-4)_{\mathbb{P}^4}$. Second, the F-term constraint imposed by the superpotential results in the target space geometry being the family of quartic hypersurfaces $\mathbb{P}^4[4]$. This is a Fano threefold of index~one, second Betti number $b_2=1$, and degree four.

The generating south pole correlator relation stems from the first two rational functions~\eqref{eq:DefWandG}, which here read
\begin{equation} \label{eq:geez}
  g_0(w,\tm_i,\tm_P,\epsilon)\,=\,1 \ , \quad
  g_1(w,\tm_i,\tm_P,\epsilon) \,=\,  \frac{(w+\tm_1)\cdots(w+\tm_5)}{(4w-\tm_P)\cdots (4w -\tm_P-3\epsilon)} 
\end{equation}
and through the constraint~\eqref{eq:RelFinalCond1} yield the polynomials
\begin{equation}
   \alpha_0(w,\tm_i,\tm_P,\epsilon)\,=\,(w+\tm_1)\cdots(w+\tm_5) \ , 
   \quad \alpha_1(w,\tm_i,\tm_P,\epsilon) =-(4w-\tm_P)\cdots (4w -\tm_P-3\epsilon)\ .
\end{equation}
These immediately determine the south pole correlator relation \eqref{eq:RSDef} as
\begin{multline}
   R_S(Q,\tm_i,\tm_P,\epsilon,\kappa_{n,\,\cdot\,}) \,=\, 
  \left\langle \sigma_N^n  (\sigma_S+\tm_1)\cdots(\sigma_S+\tm_5) \right\rangle \\
   - Q  \left\langle \sigma_N^n (4\sigma_S-\tm_P+\epsilon)\cdots (4\sigma_S -\tm_P+4\epsilon)\right\rangle \ ,
\end{multline}
corresponding to the differential operator
\begin{equation}
    \mathcal{L}(Q,\epsilon,\tm_i,\tm_P)\,=\, (\epsilon\Theta+\tm_1)\cdots(\epsilon\Theta+\tm_5) 
    - Q (4\epsilon\Theta-\tm_P+\epsilon)\cdots (4\epsilon\Theta -\tm_P+4\epsilon) \ .
\end{equation}
For the studied quartic hypersurface threefold the Givental $I$-function reads
\begin{equation}
    I_{\mathbb{P}^4[4]}(H,Q,\epsilon,\tm_i,\tm_P)\,=\, 
    \sum_{k=0}^{\infty} \frac{\prod_{\ell=1}^{4k}(4 H - \tm_P + \ell \epsilon)}{\prod_{\ell=1}^k (H+\tm_1 + \ell \epsilon)\cdots(H+\tm_{5} + \ell \epsilon) } Q^{\frac{H}\epsilon+k} \ ,
\end{equation}
which is again a solution to the differential equation
\begin{equation}
   \mathcal{L}(Q,\epsilon,\tm_i,\tm_P) I_{\mathbb{P}^4[4]}(H,Q,\epsilon,\tm_i,\tm_P) \,=\, 0 \ .
\end{equation}
Here, the twisted masses $\tm_i$ and $\tm_P$ respectively are the equivariant parameters of the canonical $\mathbb{C}^*$-action on the base and on the fiber of the non-compact toric variety $\mathcal{O}(-4)_{\mathbb{P}^4}$.

Let us examine the texture of twisted masses in greater detail. From eq.~\eqref{eq:geez} we observe that the rational function $g_1(w,\tm_i,\tm_P,\epsilon)$ simplies if one of the twisted masses $\tm_i$ obeys $4\tm_i = - \tm_P$. With, for instance, $4\tm_5 = - \tm_P$ we arrive at
\begin{equation}
  g_0(w,\tm_i,\tm_P,\epsilon)\,=\,1 \ , \quad
  g_1(w,\tm_i,\tm_P,\epsilon) \,=\,  \frac{(w+\tm_1)\cdots(w+\tm_4)}{4(4w-\tm_P-\epsilon)\cdots (4w -\tm_P-3\epsilon)}  \ .
\end{equation}
Following the general discussion in Section~\ref{sec:Reduction}, these simplified rational functions define the reduced syzygy module $M_S^0$, which --- due to the observed simplification in the rational functions~\eqref{eq:geez} --- does not coincide with the limiting syzygy module $M_S^\text{lim}$. 

We now want to argue that this reduced syzygy module $M_S^0$ results in valid correlator relations. To this end we observe that for a positive Fayet--Iliopoulos parameter only the divisors attributed to the chiral fields $\phi_i$ contribute, while the chiral field $P$ does not contribute to the residue symbol~\eqref{eq:ResSymbol}. The set of contributing poles $\Theta(\xi>0)$ and non-contributing poles $\Omega(\xi>0)$ for all topological sectors $Q^k$ are readily determined to be
\begin{equation}
    \Theta(\xi>0,\tm_i) \,=\,\bigcup_{i=1}^4  \left\{\epsilon \mathbb{Z}_{\le 0} - \tm_i \right\}  \,\subset\, \mathbb{C} \ , \quad
    \Omega(\xi>0,\tm_P) \,=\,  \left\{ \tfrac14 \epsilon\mathbb{Z}_{>0} + \tfrac14 \tm_P \right\}  \,\subset\, \mathbb{C} \ ,
\end{equation}
with $\Theta(\xi>0,\tm_i) \cap \Omega(\xi>0,\tm_P) = \emptyset$. In the aforementioned limit $\tm_i \rightarrow -\tm_P/4$ the non-generic sets --- e.g., $\Theta(\xi>0,\tm_1,\tm_2,\tm_3,\tm_4,-\tm_P/4)$ and $\Omega(\xi>0,\tm_P)$ for $i=5$ --- still do not overlap and thus fulfill condition~\eqref{eq:OverlapCond}. Hence, there appears no dangerous overlap of poles, and we conclude that the reduced syzygy module $M_S^0$ is in the described limit of twisted masses indeed of physical relevance for correlator relations.\footnote{In the negative Fayet--Ilipoulos phase $\xi<0$ the role of the contributing and non-contributing chiral fields is exchanged such that we find $\Theta(\xi>0) = \Omega(\xi<0)$ and $\Omega(\xi>0) \equiv \Theta(\xi<0)$. In this phase the condition~\eqref{eq:OverlapCond} is fulfilled as well, resulting in the same conclusion for the syzygy module~$M_S^0$.} As a consequence, for $4\tm_5 = - \tm_P$ we find the reduced south pole correlator relation
\begin{multline}
   R_S(Q,\tm_1,\ldots,\tm_4,\tm_P,\epsilon,\kappa_{n,\,\cdot\,}) \,=\, 
   \left\langle \sigma_N^n  (\sigma_S+\tm_1)\cdots(\sigma_S+\tm_4) \right\rangle \\
   - 4\,Q\,\left\langle \sigma_N^n (4\sigma_S-\tm_P+\epsilon)\cdots (4\sigma_S -\tm_P+3\epsilon)\right\rangle \ ,
\end{multline}
which yields the reduced differential operator
\begin{multline}
    \mathcal{L}(Q,\epsilon,\tm_1,\ldots,\tm_4,\tm_P)\,=\, (\epsilon\Theta+\tm_1)\cdots(\epsilon\Theta+\tm_4) \\
    -4\,Q\,(4\epsilon\Theta-\tm_P+\epsilon)\cdots (4\epsilon\Theta -\tm_P+3\epsilon) \ .
\end{multline}
It generates the physical relevant differential ideal $\boldsymbol{\mathcal{I}}_S^0$ in the discussed limit of twisted masses. Note that this reduced differential operator becomes in particular relevant in the limit of vanishing twisted masses, $\tm_i=\tm_P=0$.

\subsubsection{The complex Grassmannian fourfold $\operatorname{Gr}(2,4)$} \label{sec:exGr24}
As our next example, we consider the gauged linear sigma model with the non-Abelian gauge group $U(2)$ and non-Abelian matter spectrum as displayed in Table~\ref{tab:Gr(2,4)}.
\begin{table}[h]
\hfil
\hbox{
\vbox{
\offinterlineskip
\halign{\strut\vrule width1.2pt height15pt#&\hfil~#~\hfil\vrule width0.5pt&\hfil~#~\hfil\vrule width0.5pt&\hfil~#~\hfil\vrule width0.5pt&\hfil~#~\hfil&#\vrule width1.2pt\cr
\noalign{\hrule height 1.2pt}
&\multispan4 non-Abelian gauge theory spectrum:\hfil&\cr
&Chiral multiplets & $U(2)$~Representation & $U(1)_R$~charge & twisted masses&\cr
\noalign{\hrule height 1.2pt}
&$\phi_i$, $i=1,\ldots,4$ & $\Box_{+1}$ & $0$ & $\tm_i$&\cr
\noalign{\hrule height 1.2pt}
\omit\vrule width0pt height2pt\cr
\noalign{\hrule height 1.2pt}
&\multispan4 Abelian Coulomb branch gauge theory spectrum:\hfil&\cr
&Chiral multiplets & $U(1)\times U(1)$~charge & $U(1)_R$~charge & twisted masses&\cr
\noalign{\hrule height 1.2pt}
&$\phi_i^{(1)}$, $i=1,\ldots,4$ & $(+1,0)$ & $0$ & $\tm_i$&\cr
\noalign{\hrule height 0.5pt}
&$\phi_i^{(2)}$, $i=1,\ldots,4$ & $(0,+1)$ & $0$ & $\tm_i$&\cr
\noalign{\hrule height 0.5pt}
&$W_\pm$ & $(\pm1,\mp1)$ & $2$ & $0$&\cr
\noalign{\hrule height 1.2pt}
}}}
\hfil
\caption{The top part of the table shows the chiral matter multiplets of the $U(2)$ gauged linear sigma model of the complex Grassmannian fourfold~$\operatorname{Gr}(2,4)$, where the $U(2)$ representation is specified in terms of the Young tableau of the non-Abelian subgroup $SU(2)$ together with the charge of the diagonal $U(1)$ subgroup as a subscript. The bottom part of the table lists the chiral spectrum in the Coulomb branch of the gauge theory, which comprises the decomposition of the non-Abelian matter multiplets into  representations of the unbroken Abelian subgroup $U(1)\times U(1)$ together with the $W_\pm$ bosons.} \label{tab:Gr(2,4)}
\end{table}
In the Coulomb branch the matter spectrum decomposes into representations of the Abelian subgroup $U(1)\times U(1)$ together with the $W_\pm$ multiplets of the broken gauge group $U(2)$, as listed in the second half of Table~\ref{tab:Gr(2,4)}.

For positive Fayet--Iliopoulos parameter $\xi$ the symplectic quotient yields the complex Grassmannian fourfold~$\operatorname{Gr}(2,4)$ as classical target space geometry of this gauge theory, i.e.,
\begin{equation}
   \left. \mathbb{C}^{4\times 2} \middle/ \mu^{-1}(\xi \cdot \boldsymbol{1})\right. \,\simeq\, \operatorname{Gr}(2,4) \ ,
\end{equation}
where $\mu: \mathbb{C}^{4\times 2} \to \mathfrak{u}(2)$ is the moment map into the Lie algebra $\mathfrak{u}(2)$ for the $U(2)$ action on the matter fields $\phi_i$ spanning the vector space $\mathbb{C}^{4\times 2}$. 
The Pl\"ucker embedding $\operatorname{Pl}: \operatorname{Gr}(2,4) \hookrightarrow \mathbb{P}(\Lambda^2\mathbb{C}^4)$ of the Grassmannian fourfold $\operatorname{Gr}(2,4)$ with its unique quadratic Pl\"ucker relation identifies this particular Grassmannian with a quadratic hypersuface in $\mathbb{P}^5$, i.e., 
\begin{equation} \label{eq:Gr24P52}
   \operatorname{Gr}(2,4) \,\simeq\, \mathbb{P}^5[2] \ .
\end{equation}   
The Fano fourfold $\operatorname{Gr}(2,4)$ has index four and degree two.

As discussed in Section~\ref{sec:nonAbCor} we first consider the Coulomb branch spectrum in order to arrive at the correlator relations of the non-Abelian gauge theory. The relevant polynomials~\eqref{eq:DefWandG} read 
\begin{multline}
  g_{p_1,p_2}(w_1,w_2,\tm_i,\epsilon) \,=\, 
   \prod_{i=1}^4 \left[\prod_{s_1=0}^{p_1-1}(w_1 +\tm_i - s_1\epsilon) \prod_{s_2=0}^{p_2-1} (w_2 +\tm_i - s_2\epsilon)\right] \\
   \times (-1)^{p_1-p_2}\frac{w_1 - w_2 -(p_1-p_2)\epsilon }{w_1 - w_2} \ .
\end{multline}
These polynomials lead to the syzygy polynomials $\alpha_{p_1,p_2}$, among which we find for $g_{0,0}$ and $g_{1,0}$ the solution
\begin{equation} \label{eq:syzGr24i}
  \alpha_{0,0} \,=\,  (w_1 +\tm_1)\cdots(w_1+\tm_4)(w_1 - w_2-\epsilon)\ , \quad \alpha_{1,0}\,=\,(w_1-w_2)  \ ,
\end{equation}
and for $g_{0,1}$, $g_{1,0}$, $g_{1,1}$, and $g_{0,2}$ the solution
\begin{equation} \label{eq:syzGr24ii}
\begin{aligned}
  \alpha_{0,1} \,&=\, (w_2 +\tm_1)\cdots(w_2+\tm_4) 
     - \epsilon \bigg[ 2\left(\sum_{1\le i < j<k \le 4} \tm_i \tm_j \tm_k\right)  + \left(\sum_{1\le i < j \le 4}\tm_i \tm_j \right) \left(w_1+3 w_2\right)\\
     & +\left(\sum_{i=1}^4 \tm_i\right)  \left(4w_2^2+w_1 w_2+w_1^2\right)+\left(5w_2^3+w_1^3+w_1^2 w_2+w_1 w_2^2\right)\bigg]  \\
    & \qquad + \epsilon^2 \left[ \left(w_1^2 + 2 w_1 w_2 + 9 w_2^2\right) + \left(\sum_{i=1}^4 \tm_i\right) \left( w_1 + 5 w_2 \right) +
      2 \left(\sum_{1\le i < j \le 4}\tm_i \tm_j \right) \right] \\
    & \qquad - \epsilon^3   \left[ \left(w_1 + 7 w_2 \right) + 2 \left(\sum_{i=1}^4\tm_i \right)  \right] + 2 \epsilon^4\ , \\
 \alpha_{1,0} \,&=\, -(w_2 +\tm_1)\cdots(w_2+\tm_4) \ ,\ \qquad
  \alpha_{1,1} \,=\, -1 \ ,\ \qquad 
  \alpha_{0,2} \,=\, 1 \ .
\end{aligned}  
\end{equation}

The Coulomb branch gauge theory with the Abelian gauge group $U(1)\times U(1)$ has two Fayet--Iliopoulos parameters $(\xi_1,\xi_2)$ corresponding to the parameters $(Q_1,Q_2)$,  where the non-Abelian topological sectors are labeled by $Q'\equiv Q_1 \equiv Q_2$.\footnote{Note that in this basis for the Fayet--Iliopoulos parameters the formal parameters $\vec Q$ used in Section~\ref{sec:nonAbCor} correspond to $\vec Q=(\sqrt{Q_1 Q_2},\sqrt{Q_1/Q_2})$.} The Weyl group $\mathcal{W}_G$ of $U(2)$ is $\mathbb{Z}_2$, which exchanges the two $U(1)$ factors of $U(1)\times U(1)$ and acts by permuting the $\sigma_{S_i}$, $i=1,2$, insertions in the Coulomb branch correlators.
The first syzygy~\eqref{eq:syzGr24i} together with its Weyl orbit thus determines the Coulomb branch south pole correlator relations
\begin{equation} \label{eq:RS12}
   R_S^{(i)}(\kappa_{\vec n,\,\cdot\,}) \,=\, \left\langle \vec\sigma_N^{\vec n} (\sigma_{S,i} +\tm_1)\cdots(\sigma_{S,i}+\tm_4) (\sigma_{S,i} - \sigma_{S,i+1}-\epsilon)\right\rangle
      + Q_i  \left\langle \vec\sigma_N^{\vec n}(\sigma_{S,i}-\sigma_{S,i+1}+\epsilon)\right\rangle \ ,
\end{equation}
for $i=1,2$ and with the identification $\sigma_{S,3}\equiv\sigma_{S,1}$. By restricting to the non-Abelian physical parameter $Q'$ and by projecting to the $\mathcal{W}_G$-invariant part with the Reynolds operator~\eqref{eq:ReynoldsOp2}, we obtain from both relations~\eqref{eq:RS12} the $\mathbb{Z}_2$ invariant correlator relation
\begin{multline} \label{eq:RGr24}
    R_S^{\mathbb{Z}_2}\,=\, \frac12 \left\langle f_N(\vec\sigma_N) (\sigma_{S,1} +\tm_1)\cdots(\sigma_{S,1}+\tm_4) (\sigma_{S,1}-\sigma_{S,2} - \epsilon) \right\rangle \\
    + \frac12 \left\langle f_N(\vec\sigma_N) (\sigma_{S,2} +\tm_1)\cdots(\sigma_{S,2}+\tm_4)(\sigma_{S,2}-\sigma_{S,1} - \epsilon) \right\rangle 
      + Q' \epsilon \left\langle f_N(\vec\sigma_N) \right\rangle \ .
\end{multline}
Analogously, the second syzygy~\eqref{eq:syzGr24ii} yields the $\mathcal{W}_G$-invariant south pole correlator relation
\begin{multline} \label{eq:TGr24}
    T_S^{\mathbb{Z}_2}\,=\, \left\langle f_N(\vec\sigma_N)(\sigma_{S,1}+\sigma_{S,2})(\sigma_{S,1}^2+\sigma_{S,2}^2) \right\rangle 
    +  \left(\sum_{i=1}^4\ \tm_i \right) \left\langle  f_N(\vec\sigma_N) (\sigma_{S,1}^2 + \sigma_{S,1}\sigma_{S,2} + \sigma_{S,2}^2) \right\rangle \\
    + \left(\sum_{1\le i < j \le 4}\tm_i \tm_j \right) \left\langle  f_N(\vec\sigma_N) (\sigma_{S,1} +  \sigma_{S,2}) \right\rangle
    + \left(\sum_{1\le i < j <k\le 4}\tm_i \tm_j \tm_k \right) \left\langle  f_N(\vec\sigma_N) \right\rangle \ .
\end{multline}

From the Weyl invariant correlator relations~\eqref{eq:RGr24} and \eqref{eq:TGr24} we can via the Luna--Richardson isomorphism~\eqref{eq:Restriction} construct the $G$-invariant non-Abelian south pole correlator relations. The $U(2)$-invariant polynomial ring $\mathbb{C}[\mathfrak{u}(2)]^{U(2)}$ is generated by the expressions $\operatorname{tr}(\adjs)$ and $\operatorname{tr}(\adjs^2)$, which map in the Coulomb branch to the symmetric polynomials $\sigma_1+\sigma_2$ and $\sigma_1^2 + \sigma_2^2$, respectively. Thus, obtaining the non-Abelian correlator relations amounts to replacing the symmetric functions in two variables in terms of the $U(2)$-invariant generators $\operatorname{tr}(\adjs)$ and $\operatorname{tr}(\adjs^2)$. After a few steps of algebra we arrive at the non-Abelian south pole relations
\begin{multline} \label{eq:RelR}
  R_S^{U(2)}\,=\, 
   \frac{1}{2} \left\langle f_N(\adjs_N) \left[\operatorname{tr}(\adjs_S)^3\operatorname{tr}(\adjs_S^2)-2\operatorname{tr}(\adjs_S) \operatorname{tr}(\adjs_S^2)^2
   -\epsilon\left(\frac{ \operatorname{tr}(\adjs_S)^4}{2}-\operatorname{tr}(\adjs_S)^2  \operatorname{tr}(\adjs_S^2)-\frac{ \operatorname{tr}(\adjs_S^2)^2}{2}\right)\right] \right\rangle \\
   + \left(\sum_{i=1}^4\frac{\tm_i}4\right) \left\langle f_N(\adjs_N) \left[ \operatorname{tr}(\adjs_S)^4
     -\operatorname{tr}(\adjs_S)^2 \operatorname{tr}(\adjs_S^2)-2 \operatorname{tr}(\adjs_S^2)^2+\epsilon\left(3 \operatorname{tr}(\adjs_S) \operatorname{tr}(\adjs_S^2)
   -\operatorname{tr}(\adjs_S)^3\right) \right] \right\rangle  \\
   +\left(\sum_{1\le i < j \le 4} \frac{\tm_i \tm_j}2 \right) \left\langle f_N(\adjs_N) \left[  \operatorname{tr}(\adjs_S)^3-2 \operatorname{tr}(\adjs_S) \operatorname{tr}(\adjs_S^2)+\epsilon \operatorname{tr}(\adjs_S^2) \right] \right\rangle \\
   +\left(\sum_{1\le i < j<k \le 4}\frac{\tm_i \tm_j \tm_k}2\right) \left\langle f_N(\adjs_N)  \left[  \operatorname{tr}(\adjs_S)^2-2 \operatorname{tr}(\adjs_S^2)+\epsilon \operatorname{tr}(\adjs_S) \right] \right\rangle 
   +\epsilon \left( \tm_1 \tm_2 \tm_3 \tm_4    - Q' \right)  \left\langle f_N(\adjs_N)\right\rangle  \ ,
\end{multline}
and
\begin{multline} \label{eq:RelT}
  T_S^{U(2)}\,=\, 
   \left\langle f_N(\adjs_N) \operatorname{tr}(\adjs_S) \operatorname{tr}(\adjs_S^2)\right\rangle 
   +\left(\sum_{i=1}^4\frac{\tm_i}2\right) \left\langle f_N(\adjs_N) \left[ \operatorname{tr}(\adjs_S)^2+\operatorname{tr}(\adjs_S^2) \right] \right\rangle \\
   +\left(\sum_{1\le i < j \le 4} {\tm_i \tm_j} \right) \left\langle f_N(\adjs_N) \operatorname{tr}(\adjs_S)\right\rangle 
   +\left(\sum_{1\le i < j<k \le 4}\tm_i \tm_j \tm_k \right) \left\langle f_N(\adjs_N) \right\rangle  \ .
\end{multline}

We now want to give a geometric interpretation of the derived correlator relations. For simplicity we consider the limit of vanishing twisted masses $\tm_i=0$. Let us first describe the cohomology elements of the Grassmannian $\operatorname{Gr}(2,4)$, generated by Schubert cycles $\sigma_{\nu}$ of Young tableaux with $\nu$ with at most two rows and two columns, see, e.g., ref.~\cite{MR1288523}.

There is a surjective ring homomorphism $\rho_{\operatorname{Gr}(2,4)}$ from the Schur polynomials $s_\nu$ to the Schubert cycles $\sigma_\nu$ with $\rho_{\operatorname{Gr}(2,4)}(\sigma_\nu) = s_\nu$. For the Grassmannian $\operatorname{Gr}(2,4)$ these are the Schur polynomials in two variables, generating the symmetric polynomial ring $\mathbb{C}[x_1,x_2]^{S_2}$ and obeying
\begin{equation}
  s_\nu \cdot s_\mu \,=\, s_{\nu\otimes\mu} \ ,
\end{equation}
in terms of the tensor product $\otimes$ of Young tableaux of the permutation group $S_2$. The kernel of the ring homomorphism $\rho_{\operatorname{Gr}(2,4)}$ is given by the two relations\footnote{For the computation of Young tableaux, see for instance ref.~\cite{MR1824028}.}
\begin{equation}
   s_{3} \,=\,x_1^3 + x_1^2x_2 + x_1 x_2^2+x_2^3 \,=\,0 \ , \qquad
   s_{4} \,=\,x_1^4 + x_1^3x_2 + x_1^2 x_2^2+x_1 x_2^3+x_2^4 \,=\, 0 \ ,
\end{equation}
and the cohomology ring becomes
\begin{equation} \label{eq:Gr24coh}
     H^*(\operatorname{Gr}(2,4),\mathbb{C}) 
     \,\simeq\, \left. \raisebox{0.4ex}{$\mathbb{C}[x_1,x_2]^{S_2}$} \middle/ \raisebox{-0.4ex}{$\left\langle s_{3}, s_{4} \right\rangle$} \right. 
     \,\simeq\, \left. \raisebox{0.4ex}{$\mathbb{C}[N_1,N_2]$} \middle/ \raisebox{-0.4ex}{$\left\langle N_1N_2, -\frac{N_1^4}4+N_1^2N_2+\frac{N_2^2}4 \right\rangle$} \right.\ .
\end{equation}
In the last step the ideal $\left\langle s_{3}, s_{4} \right\rangle$ of $\ker \rho$ of symmetric polynomials is expressed in terms of the Newton polynomials $N_\ell = x_1^\ell + x_2^\ell$, $\ell=1,2$, which in turn generate the symmetric polynomial ring $\mathbb{C}[x_1,x_2]^{S_2}$.

For complex Grassmannian varieties the deformation of the classical cohomology ring to the quantum cohomology ring is established in refs.~\cite{Intriligator:1991an,Vafa:1991uz,Witten:1993xi}. More generally, for Fano varieties Siebert and Tian show that if the ordinary cohomology ring is a polynomial ring with relations as in formula~\eqref{eq:Gr24coh}, then the quantum cohomology is captured by a $Q'$-dependent deformation of these relations \cite{MR1621570}. Applied to the Grassmannian $\operatorname{Gr}(2,4)$ this deformation yields the quantum cohomology ring \cite{MR1621570}
\begin{equation} \label{eq:qCohGr24}
    H_\star^*(\operatorname{Gr}(2,4),\mathbb{C}) 
    \,\simeq\, \left. \raisebox{0.4ex}{$\mathbb{C}[N_1,N_2][[Q']]$} \middle/ \raisebox{-0.4ex}{$\left\langle N_1N_2, -\frac{N_1^4}4+N_1^2N_2+\frac{N_2^2}4 + Q' \right\rangle$} \right.\ ,
\end{equation}

The presented formulation relates directly to the gauge theory correlators and its correlator relations, as discussed in the following. Due to Schur--Weyl reciprocity, we first note that the gauge invariant insertions are canonically identified with the Newton polynomials $N_r= x_1^r + x_2^r$ according to
\begin{equation} \label{eq:DicOp}
  \operatorname{tr}(\adjs^r) \, \longleftrightarrow \, N_r \ .
\end{equation}  
Let us know connect via this correspondence the correlator relations to the target space geometry. For vanishing twisted masses, the correlator relation~\eqref{eq:RelT} generalizes to 
\begin{equation} \label{eq:RelTprime}
\begin{aligned}
  {T}_S^{(k),U(2)}\,&=\, \left\langle f_N(\adjs_N) \operatorname{tr}(\adjs_S)^k \operatorname{tr}(\adjs_S^2)\right\rangle \ , \quad k=1,2,\ldots \ , \\
  {T'}_S^{U(2)}\,&=\, \left\langle f_N(\adjs_N) \operatorname{tr}(\adjs_S) \operatorname{tr}(\adjs_S^2)^2\right\rangle + 2 Q' \epsilon \left\langle f_N(\adjs_N) \right\rangle\ ,
\end{aligned}  
\end{equation}
which are obtained from the syzygy polynomials~\eqref{eq:syzGr24ii} after an overall multiplication with suitable powers of the $(w_1 + w_2)$ or $(w_1^2 + w_2^2)$. Combining these relations with the correlator relation~\eqref{eq:RelR}, we obtain the modified correlator relation
\begin{equation} \label{eq:RelRprime}
  {R'}_S^{U(2)}\,=\,  \left\langle f_N(\adjs_N) \left(-\frac{ \operatorname{tr}(\adjs_S)^4}{4}+\operatorname{tr}(\adjs_S)^2  \operatorname{tr}(\adjs_S^2)+\frac{ \operatorname{tr}(\adjs_S^2)^2}{4}\right) \right\rangle + Q' \left\langle f_N(\adjs_N) \right\rangle \ .
\end{equation}
Thus, applying the dictionary~\eqref{eq:DicOp}, we immediately see that the correlator relations ${T}_S^{(k),U(2)}$ and ${R'}_S^{U(2)}$ precisely realize the quantum cohomology relations in the defintion of the quantum cohomology ring~\eqref{eq:qCohGr24}. Furthermore, our results match the proposal for the realization of the quantum cohomology ring of Grassmannians in the context of the A-twisted gauged linear sigma model spelt out in the Appendix of ref.~\cite{Hori:2000kt}. This demonstrates that the correlators of the studied non-Abelian gauged linear sigma model compute quantum cohomology products of the Grassmannian fourfold $\operatorname{Gr}(2,4)$. Namely, we for instance explicitly find
\begin{equation} \label{eq:Gr24QProd}
\begin{aligned}
  \left\langle \operatorname{tr}(\adjs_S)^4 \right\rangle \,&=\, \int \sigma_1^4 \,=\, 2 \ , 
  &\left\langle \operatorname{tr}(\adjs_S^2)^2 \right\rangle \,&=\, \int (\sigma_2-\sigma_{1,1})^2 \,=\, 2   \ , \\
  \left\langle \operatorname{tr}(\adjs_S)^2\operatorname{tr}(\adjs_S^2) \right\rangle \,&=\, \int \sigma_1^2(\sigma_2-\sigma_{1,1}) \,=\, 0  \ , 
  &\left\langle \operatorname{tr}(\adjs_S)^8 \right\rangle \,&=\, \int \sigma_1^8 \,=\, 8 Q' \ , \\
  \left\langle \operatorname{tr}(\adjs_S^2)^4 \right\rangle \,&=\, \int (\sigma_2-\sigma_{1,1})^4 \,=\, -8 Q'   \ , 
  &\left\langle \operatorname{tr}(\adjs_S)^2\operatorname{tr}(\adjs_S^2)^3 \right\rangle \,&=\, \int \sigma_1^2(\sigma_2-\sigma_{1,1})^3 \,=\, 0 \ ,
\end{aligned}  
\end{equation}
where the Schubert cycles are multiplied with the $Q'$-dependent quantum product $\star$ and then integrated over the Grassmannian~$\operatorname{Gr}(2,4)$. Here, we have used the following relations among Newton and Schur polynomials
\begin{equation}
  N_1 \,=\, s_1 \ , \qquad N_2 \,=\, s_2 - s_{1,1} \ .
\end{equation}
The first three correlators compute the classical intersection numbers of the Grassmannian $\operatorname{Gr}(2,4)$, whereas the remaining correlators show the degree one contributions in some of the quantum products. Note that the quantum products~\eqref{eq:Gr24QProd} are in accord with the non-Abelian selection rule~\eqref{eq:sruleNonAb}. 

Finally, let us remark that --- upon multiplying the syzgy polynomials~\eqref{eq:syzGr24i} with the overall factor $(w_1+w_2)$ --- we arrive at a correlator relations of degree five in the adjoint insertion~$\adjs_S$. Removing the quadratic insertions~$\operatorname{tr}(\adjs_S^2)$ with the help of correlator relations of the type~\eqref{eq:RelTprime}, it is straight forward to then deduce the degree five relation 
\begin{equation}
    0 \,=\, \left\langle f_N(\adjs_N)  \operatorname{tr}(\adjs_S)^5 \right\rangle - 2 Q'  \left\langle f_N(\adjs_N)  (2 \operatorname{tr}(\adjs_S) + \epsilon) \right\rangle \ ,
 \end{equation}
which yields the (reduced) differential operator 
\begin{equation} 
     \mathcal{L}(Q',\epsilon) = \left(\epsilon\,\Theta\right)^5   - 2 Q' (2 \epsilon\, \Theta + \epsilon) \ .
\end{equation}
It annihilates the Givental $I$-function of quadratic hypersurfaces in $\mathbb{P}^5$ \cite{MR1653024,MR2276766}
\begin{equation}
    I_{\mathbb{P}^5[2]}(H,Q',\epsilon)\,=\, 
    \sum_{k=0}^{\infty} \frac{\prod_{\ell=1}^{2k}(2 H + \ell \epsilon)}{\prod_{\ell=1}^k (H+\ell \epsilon)^6} Q'^{\frac{H}\epsilon+k} \ ,
\end{equation}
where $H$ is the hyperplane class of $\mathbb{P}^5$. Remembering the geometric identification~\eqref{eq:Gr24P52}, the appearance of this Givental $I$-function as the solution to the determined differential operator also confirms the correspondence between gauge theory correlators and the quantum product of the complex Grassmannian~$\operatorname{Gr}(2,4)$.

\subsection{Calabi--Yau varieties as target spaces}
In this section we study examples of both Abelian and non-Abelian gauged linear sigma models with Calabi--Yau target spaces with focus on the interplay between the gauge theory correlators and the Picard--Fuchs differential equations of the Calabi--Yau geometries as developed in Section~\ref{sec:CYtarget}. We study gauged linear sigma model examples for both compact and non-compact target space examples, for Calabi--Yau threefold targets with one and two K\"ahler moduli, and Calabi--Yau fourfold target spaces with one K\"ahler modulus with minimal as well as non-minimal Picard--Fuchs operators.\footnote{For the notion of non-minimal Picard--Fuchs operators of Calabi--Yau fourfolds, see ref.~\cite{Gerhardus:2016iot}.}

\subsubsection{The quintic Calabi--Yau threefold}
The quintic hypersurface $\mathbb{P}^4[5]$ in the complex projective space $\mathbb{P}^4$ is the standard example of a compact Calabi--Yau threefold. The quintic arises as the target space of the gauged linear sigma model with the Abelian gauge group $U(1)$ and the matter spectrum listed in Table~\ref{tab:SpecQuintic}. 
\begin{table}[h]
\hfil
\hbox{
\vbox{
\offinterlineskip
\halign{\strut\vrule width1.2pt\hfil~#~\hfil\vrule width0.5pt&\hfil~#~\hfil\vrule width0.5pt&\hfil~#~\hfil\vrule width0.5pt&\hfil~#~\hfil\vrule width1.2pt\cr
\noalign{\hrule height 1.2pt}
Chiral multiplets & $U(1)$~charge & $U(1)_R$~charge & twisted masses\cr
\noalign{\hrule height 1.2pt}
$\phi_i$, $i=1,\ldots,5$ & $+1$ & $0$ & $\tm_i$ \cr
\noalign{\hrule height 0.5pt}
$P$ & $-5$ & $2$ & $\tm_P$ \cr
\noalign{\hrule height 1.2pt}
}}}
\hfil
\caption{Matter spectrum of the $U(1)$ gauged linear sigma model of the quintic Calabi--Yau threefold $\mathbb{P}^4[5]$.} \label{tab:SpecQuintic}
\end{table}

The derivation of the south pole correlator relations $R_S$ and the differential operators $\mathcal{L}$ parallels the discussion of the gauged linear sigma model of the Fano threefold $\mathbb{P}^4[4]$ presented in Section~\ref{sec:FP4[4]}. The analog computation eventually yields the differential operator 
\begin{equation} \label{eq:LQmass}
   \mathcal{L}(Q,\epsilon,\tm_i,\tm_P)\,=\, (\epsilon\Theta+\tm_1)\cdots(\epsilon\Theta+\tm_5) 
    + Q (5\epsilon\Theta-\tm_P+\epsilon)\cdots (5\epsilon\Theta -\tm_P+5\epsilon) \ .
\end{equation}

In the limit of vanishing twisted masses the defining polynomials~\eqref{eq:DefWandG} of the correlator relations simplify such that the corresponding limiting syzygy module $M^\text{lim}_S$ does not coincide with the reduced syzygy module $M_S^0$. By the same analysis as spelled out in detail for the Fano threefold $\mathbb{P}^4[4]$ in Section~\ref{sec:FP4[4]} we confirm that  $M_S^0$ encodes physically relevant correlator relations of the quintic threefold. As a consequence we find the reduced differential operator 
\begin{equation} \label{eq:LQ1}
   \mathcal{L}(Q,\epsilon)\,=\, \Theta^4 + 5\,Q\,(5\Theta+1)(5\Theta+2)(5\Theta+3)(5\Theta +4)\ ,
\end{equation}
which is the well-known Picard--Fuchs differential operator of the quintic Calabi--Yau threefold. We should emphasis that from the reduced syzygy module $M_S^0$ we directly obtain the order four Picard--Fuchs operator for the quintic threefold. Other methods --- for instance obtaining the GKZ system from the defining toric data of the quintic hypersurface --- often yield the order five differential operator, which is given by eq.~\eqref{eq:LQmass} in the limit of vanishing twisted masses. Arriving at the Picard--Fuchs operator of the desired minimal order is not a coincidence for specific examples of compact Calabi--Yau manifolds, but instead is a general feature of the presented approach.

These two differential operators  annihilate the Givental $I$-function of the quintic hypersurface  respectively for generic and vanishing twisted masses, namely $\mathcal{L} I_{\mathbb{P}^4[5]} = 0$ for \cite{MR1653024,MR2276766}\footnote{Note that the K\"ahler parameter $Q$ differs by a minus sign from refs.~\cite{MR1653024,MR2276766}.}
\begin{equation}
    I_{\mathbb{P}^4[5]}(H,Q,\epsilon,\tm_i,\tm_P)\,=\, 
    \sum_{k=0}^{\infty} \frac{\prod_{\ell=1}^{5k}(5 H - \tm_P + \ell \epsilon)}{\prod_{\ell=1}^k (H+\tm_1 + \ell \epsilon)\cdots(H+\tm_{5} + \ell \epsilon) } (-Q)^{\frac{H}\epsilon+k} \ .
\end{equation}

We now illustrate the computation of the Picard--Fuchs differential operator from gauged linear sigma model correlators. The quintic Calabi--Yau threefold has a single K\"ahler modulus. Hence, its Picard--Fuchs operator is given by the correlator formula~\eqref{eq:LCY3} in terms of the gauge theory correlators $\kappa_{0,3}$, $\kappa_{2,3}$, and $\kappa_{3,4}$, which are explicitly computed to be
\begin{equation} 
   \kappa_{0,3} \,=\, \frac{5}{1+5^{5}Q} \ , \quad
   \kappa_{2,3} \,=\, \frac{-6250 Q}{(1+5^{5}Q)^{2}} \epsilon^2 \ , \quad
   \kappa_{3,4} \,=\, \frac{100Q(-6+59375Q)}{(1+5^{5}Q)^{3}} \epsilon^4 \ .
\end{equation} 
Inserting these correlator into the formula~\eqref{eq:LCY3} yields the differential operator 
\begin{equation}
   \mathcal{L}\,=\,\frac{5^2 \epsilon^4}{(1+5^5 Q)^3}\left[ \Theta^4 +5\,Q\,(5\Theta+1)(5\Theta+2)(5\Theta+3)(5\Theta +4) \right] \ ,
\end{equation}
which (up to a negligible prefactor) agrees with the expected Picard--Fuchs operator~\eqref{eq:LQ1} of the quinitc Calabi--Yau threefold.

\subsubsection{The R\o{}dland Calabi--Yau threefold} \label{sec:Rod}
As our next example we consider the non-Abelian gauged linear sigma model studied by Hori and Tong \cite{Hori:2006dk}. This gauge theory realizes as its two geometric phases the two derived-equivalent families of Calabi--Yau threefold target space varieties \cite{MR2475813,Kuznetsov:2006arxiv}, first constructed by R\o{}dland~\cite{MR1775415}. The non-Abelian gauge group is $U(2)$ together with the charged chiral matter spectrum listed in Table~\ref{tab:SpecRod}. 
\begin{table}[h]
\hfil
\hbox{
\vbox{
\offinterlineskip
\halign{\strut\vrule width1.2pt height15pt#&\hfil~#~\hfil\vrule width0.5pt&\hfil~#~\hfil\vrule width0.5pt&\hfil~#~\hfil\vrule width0.5pt&\hfil~#~\hfil&#\vrule width1.2pt\cr
\noalign{\hrule height 1.2pt}
&\multispan4 non-Abelian gauge theory spectrum:\hfil&\cr
&Chiral multiplets & $U(2)$~Representation & $U(1)_R$~charge & twisted masses&\cr
\noalign{\hrule height 1.2pt}
&$\phi_i$, $i=1,\ldots,7$ & $\Box_{+1}$ & $0$ & $\tm_i$&\cr
\noalign{\hrule height 0.5pt}
&$P^j$, $j=1,\ldots,7$ & $\mathbf{1}_{-2}$ & $2$ & $\tm_P^j$&\cr
\noalign{\hrule height 1.2pt}
\omit\vrule width0pt height2pt\cr
\noalign{\hrule height 1.2pt}
&\multispan4 Abelian Coulomb branch gauge theory spectrum:\hfil&\cr
&Chiral multiplets & $U(1)\times U(1)$~charge & $U(1)_R$~charge & twisted masses&\cr
\noalign{\hrule height 1.2pt}
&$\phi_i^{(1)}$, $i=1,\ldots,7$ & $(+1,0)$ & $0$ & $\tm_i$&\cr
\noalign{\hrule height 0.5pt}
&$\phi_i^{(2)}$, $i=1,\ldots,7$ & $(0,+1)$ & $0$ & $\tm_i$&\cr
\noalign{\hrule height 0.5pt}
&$P^j$, $j=1,\ldots,7$ & $(-1,-1)$ & $2$ & $\tm_P^j$&\cr
\noalign{\hrule height 0.5pt}
&$W_\pm$ & $(\pm1,\mp1)$ & $2$ & $0$&\cr
\noalign{\hrule height 1.2pt}
}}}
\hfil
\caption{The table shows the chiral matter multiplets of the $U(2)$ Hori--Tong gauged linear sigma model for the R\o{}dland Calabi--Yau threefold target spaces. The $U(2)$ representation is specified in terms of the Young tableau of the non-Abelian subgroup $SU(2)$ together with the charge of the diagonal $U(1)$ subgroup as a subscript. Moreover, the table lists the chiral spectrum in the Coulomb branch of the gauge theory, where the Coulomb branch chiral fields fall into representations of $U(1)\times U(1)$.} \label{tab:SpecRod}
\end{table}
Furthermore, the table shows the decomposition of the non-Abelian spectrum into Abelian chiral multiplets in the Coulomb branch spectrum with unbroken gauge group $U(1)\times U(1)$. 

The geometric phases of this Hori--Tong gauged linear sigma model are analyzed in detail in ref.~\cite{Hori:2006dk}. For positive Fayet--Ilipoulos parameter $\xi$ we obtain via the symplectic quotient construction together with the F-term constraints the degree $42$ Calabi--Yau threefold target space~$X_{1^7}$. This Calabi--Yau threefold is a complete intersection in the Grassmannian $\operatorname{Gr}(2,7)$, which is given via the Pl\"ucker embedding $\operatorname{Pl}: \operatorname{Gr}(2,7) \hookrightarrow \mathbb{P}(\Lambda^2\mathbb{C}^7)$ as the intersection of $\operatorname{Gr}(2,7) \cap \mathbb{P}^{13}$, where the F-term constraints realize the intersecting projective subspace $\mathbb{P}^{13} \subset \mathbb{P}(\Lambda^2\mathbb{C}^7)$. For negative Fayet--Ilipoulos parameter $\xi$ the Hori--Tong gauged linear sigma model realizes a strong coupling phase, which never the less yields a derived equivalent Calabi--Yau threefold, namely the Pfaffian non-complete intersection Calabi--Yau variety of degree $14$ \cite{MR1775415}. Both Calabi--Yau manifolds have a single K\"ahler modulus as parametrized by the Fayet--Ilipoulos parameter $\xi$.

Similarly as for the complex Grassmannian fourfold~$\operatorname{Gr}(2,4)$ of Section~\ref{sec:exGr24}, it is now possible to deduce non-Abelian correlator relations from the Abelian Coulomb branch spectrum summarized at the bottom of Table~\ref{tab:SpecRod}, which then --- upon restricting to the linear gauge invariant insertions $\operatorname{tr}\adjs$ --- yield differential operators. However, instead of reiterating this straight forward but tedious computation, we in this example analyze the correlation functions, which then also allow us to derive the Picard--Fuchs differential operator as developed in Section~\ref{sec:CYtarget}. This example demonstrates that correlator formulas for Picard--Fuchs equations are applicable and particularly powerful for non-Abelian gauged linear sigma models, as other methods are often more intricate to implement.

Let us first connect the correlators of the Hori--Tong gauged linear sigma model to the geometry of the Calabi--Yau threefold~$X_{1^7}$. The correlators of respectively gauge invariant south and north pole insertions of the type~\eqref{eq:NonAbCorrSpecial} arise from insertions of the type $\operatorname{tr}(\adjs)$ and $\operatorname{tr}(\adjs^2)$. There are two distinct correlators with south pole insertions cubic in the adjoint field $\adjs$ (in the absence of north pole insertions), which are already computed in ref.~\cite{Closset:2015rna}
\begin{equation} \label{eq:X17cor1}
  \kappa_{0,3}\,=\,\left\langle \operatorname{tr}(\adjs_S)^3 \right\rangle \,=\, 
  \frac{14(3 +  Q)}{1 + 57 Q - 289 Q^2 - Q^3} \ , \quad
  \left\langle \operatorname{tr}(\adjs_S)\operatorname{tr}(\adjs_S^2) \right\rangle \,=\,
  \frac{14(1-9 Q)}{1 + 57 Q - 289 Q^2 - Q^3} \ .
\end{equation}
In order to geometrically interpret these correlators let us first interpret the gauge theory insertions $\operatorname{tr}(\adjs_S)$ and $\operatorname{tr}(\adjs_S^2)$ as elements of the twisted chiral ring of the conformal field theory at the infrared fixed point of the renormalization group flow. In the large volume limit $Q\to 0$ the twisted chiral ring becomes the cohomology ring $H^\text{ev}(X_{1^7},\mathbb{Z})$ \cite{Lerche:1989uy}, which (as a vector space) is generated by the integral generators
\begin{equation} \label{eq:X17gen}
  \left\langle 1, \iota^*\sigma_1, \tfrac{1}{42} \iota^*\sigma_1^2, \tfrac{1}{42} \iota^*\sigma_1^3 \right\rangle \,\simeq\, H^\text{ev}(X_{1^7},\mathbb{Z}) \ .
\end{equation}
These generators are induced from the embedding $\iota: X_{1^7} \hookrightarrow \operatorname{Gr}(2,7)$ into the ambient space. In particular, the Schubert cycle~$\sigma_1$ generates the cohomology group~$H^2(\operatorname{Gr}(2,7),\mathbb{Z})$, whereas the Schubert cycles $\sigma_{1,1}$ and $\sigma_2$ span the cohomology group $H^4(\operatorname{Gr}(2,7),\mathbb{Z})$. However, pulled back to the Calabi--Yau threefold $X_{1^7}$ the latter cycles become linearly dependent representatives in cohomology, i.e.,
\begin{equation} \label{eq:cohrelX17}
  \iota^* \sigma_{1,1} \sim \frac12 \iota^*\sigma_{2} \sim \frac{1}{3} \iota^*\sigma_1^2 \ .
\end{equation}
Repeating the arguments of Section~\ref{sec:exGr24}, we readily map the gauge invariant insertions $\operatorname{tr}(\adjs_S)$ and $\operatorname{tr}(\adjs_S^2)$ in the large volume limit $Q\to0$ to the the cohomology elements
\begin{equation}
  \operatorname{tr}(\adjs_S) \,\simeq\, \iota^*\sigma_1 \ , \quad
  \operatorname{tr}(\adjs_S^2) \,\simeq\, \iota^*\sigma_2 - \iota^*\sigma_{1,1} \ . 
\end{equation}
As consequence, at large volume the gauge theory correlators yield the classical intersection numbers of $X_{1^7}$, namely
\begin{equation}
\begin{aligned}
  \lim_{Q\to 0} \left\langle \operatorname{tr}(\adjs_S)^3 \right\rangle \,&=\, \int_{X_{1^7}} \iota^*\sigma_1^3 \,=\, 42 \ , \\
  \lim_{Q\to 0} \left\langle \operatorname{tr}(\adjs_S) \operatorname{tr}{(\adjs_S^2)} \right\rangle \,&=\, 
  \int_{X_{1^7}} \iota^*\sigma_1\cup(\iota^*\sigma_{1,1}-\iota^*\sigma_{2}) \,=\, 14 \ .
\end{aligned}  
\end{equation}  

For non-zero $Q$ the twisted chiral ring is replaced by the quantum cohomology ring $H_\star^\text{ev}(X_{1^7},\mathbb{Z})$, where the product in ordinary cohomology is deformed to the quantum product in quantum cohomology. However, neither the quantum cohomology groups $H_\star^\text{ev}(X_{1^7},\mathbb{Z})$ as vector spaces nor the (non-degenerate) intersection pairing $\langle \alpha, \beta \rangle = \int_{X_{1^7}} \alpha \cup \beta$ is modified at the quantum level; see, e.g., the review~\cite{Morrison:2000bt}. Since in the present example the cohomology group~$H_\star^\text{4}(X_{1^7},\mathbb{Z})$ is one dimensional, the gauge theory insertions $\operatorname{tr}(\adjs_S)^2$ and $\operatorname{tr}(\adjs_S^2)$ must --- even for non-vanishing $Q$ --- become linearly dependent in the infrared. Moreover, since the bilinear intersection product is $Q$-independent and non-degenerate, we infer from the correlators~\eqref{eq:X17cor1} the infrared quantum-deformed relation
\begin{equation} \label{eq:qrel}
   (1 - 9 Q) \operatorname{tr}(\adjs_S)^2 \,\sim_\text{IR} \, (3+Q) \operatorname{tr}{(\adjs_S^2)} \ ,
\end{equation}   
which reduces to the classical cohomology relation~\eqref{eq:cohrelX17} for $Q\to 0$. For generic values of $Q$ both gauge theory insertions $\operatorname{tr}(\adjs_S)^2$ and $\operatorname{tr}{(\adjs_S^2)}$ represent non-trivial (but linearly dependent) twisted chiral ring elements. However, for the special values $Q=-3$ or $Q=\frac19$ the gauge theory insertions $\operatorname{tr}(\adjs_S)^2$ or $\operatorname{tr}{(\adjs_S^2)}$ respectively flow to zero. This implies that the twisted chiral ring of the Calabi--Yau threefold $X_{1^7}$ for $Q\ne -3$ is generated by the gauge invariant insertions $\operatorname{tr}(\adjs_S)$, whereas for $Q=-3$ it is generated by the two gauge invariant insertions $\operatorname{tr}(\adjs_S)$ and $\operatorname{tr}(\adjs_S^2)$. In the latter situation the second generator is now required due to the non-generic quantum relation $\operatorname{tr}(\adjs_S)^2\sim_\text{IR} 0$, resulting from the degeneration $\lim_{Q\to -3}\kappa_{0,3}=0$ of the Yukawa coupling correlator~\eqref{eq:X17cor1}.

Finally, we want to extract the Picard--Fuchs differential equation of the quantum K\"ahler moduli space of the R\o{}dland Calabi--Yau threefolds by using the universal correlator formula~\eqref{eq:LCY3}. Apart from the Yukawa coupling correlator $\kappa_{0,3}$ already listed in eq.~\eqref{eq:X17cor1} this requires the correlators $\kappa_{2,3}$ and $\kappa_{3,4}$, which for the given example are readily calculated to be
\begin{equation}
\begin{aligned}
\kappa_{2,3} \,&=\,  \frac{28 Q (-51 + 787 Q + 75 Q^2 + Q^3)}{(1 + 57 Q - 
  289 Q^2 - Q^3)^2} \epsilon^2 \ ,\\
\kappa_{3,4} \,&=\, \frac{14 Q  (-15 + 3340 Q - 32415 Q^2 + 614760 Q^3 + 20747 Q^4 - 
   20 Q^5 + 3 Q^6)}{(1 + 57 Q - 289 Q^2 -Q^3)^3} \epsilon^4 \ .
\end{aligned}
\end{equation}
Upon explicitly plugging in the values of these correlators into formula~\eqref{eq:LCY3}, we find the Picard--Fuchs operator
\begin{multline}
\mathcal{L}  \,=\, \frac{196  \epsilon^4}{(1 + 57 Q - 289 Q^2 - Q^3)^3} \Big[ 
  (3 + Q)^2 (1 + 57 Q - 289 Q^2 - Q^3) \Theta^4  \\
  +4 Q (3+Q) (85 - 867 Q - 149 Q^2 - Q^3) \Theta^3 
 +2 Q (408 - 7597 Q - 2353 Q^2 - 239 Q^3 - 3 Q^4) \Theta^2 \\
  +2 Q (153 - 4773 Q - 675 Q^2 - 87 Q^3 - 2 Q^4) \Theta 
 +Q (45 - 2166 Q - 12 Q^2 - 26 Q^3 - Q^4) \Big]\ ,
\end{multline}
which is in agreement with the literature~\cite{MR1775415}.\footnote{Note that the Picard--Fuchs operator exhibits an apparent singularity at $Q=-3$. While this corresponds to a smooth point in moduli space with regular solutions, we observe that the Yukawa coupling $\kappa_{0,3}$ vanishes with the above discussed implications on the chiral ring. Such apparent singularities are a consequence of the universal form of the Picard--Fuchs operator \eqref{eq:LCY3v0}, which implies either that $\kappa_{0,3}$ vanishes (as in the given example) or that the discriminant locus has a spurious singular component. See also the discussion in ref.~\cite{Cynk:2017wo}.} This example illustrates that simply computing the three correlators~$\kappa_{0,3}$, $\kappa_{2,3}$, and $\kappa_{3,4}$ is a powerful approach to derive Picard--Fuchs operators of any Calabi--Yau threefolds with a single K\"ahler modulus --- even for projective non-complete intersection varieties or for complete intersections in non-toric varieties.

\subsubsection{The sextic Calabi--Yau fourfold}
As our next example we consider the sextic Calabi--Yau fourfold $\mathbb{P}^5[6]$ in the complex projective space $\mathbb{P}^5$. The spectrum of the associated $U(1)$ gauged linear sigma model is displayed in Table~\ref{tab:SpecSextic}. 
\begin{table}[h]
\hfil
\hbox{
\vbox{
\offinterlineskip
\halign{\strut\vrule width1.2pt\hfil~#~\hfil\vrule width0.5pt&\hfil~#~\hfil\vrule width0.5pt&\hfil~#~\hfil\vrule width0.5pt&\hfil~#~\hfil\vrule width1.2pt\cr
\noalign{\hrule height 1.2pt}
Chiral multiplets & $U(1)$~charge & $U(1)_R$~charge & twisted masses\cr
\noalign{\hrule height 1.2pt}
$\phi_i$, $i=1,\ldots,6$ & $+1$ & $0$ & $\tm_i$ \cr
\noalign{\hrule height 0.5pt}
$P$ & $-6$ & $2$ & $\tm_P$ \cr
\noalign{\hrule height 1.2pt}
}}}
\hfil
\caption{Matter spectrum of the $U(1)$ gauged linear sigma model of the sextic Calabi--Yau fourfold $\mathbb{P}^5[6]$.} \label{tab:SpecSextic}
\end{table}

The sextic Calabi--Yau fourfold has one K\"ahler parameter, and its quantum K\"ahler moduli space is governed by a Picard--Fuchs operator of order five.\footnote{As the sextic Calab--Yau fourfold is a hypersurface in projective space its Picard--Fuchs operator is necessarily of minimal order; c.f., the discussion in ref.~\cite{Gerhardus:2016iot}.} Therefore, we can reconstruct the Picard--Fuchs operator (for vanishing twisted masses) with formula~\eqref{eq:CY4N5cs}, which requires the knowledge of the correlators $\kappa_{0,4}$, $\kappa_{3,3}$, $\kappa_{4,4}$, which (for $\tm_i = \tm_P =0$) read
\begin{equation}
\begin{aligned} 
  \kappa_{0,4} \,&=\, \frac{6}{1-6^6 Q}  \ , \qquad
  \kappa_{3,3} \,=\, \frac{7776 Q (13 - 1026432 Q)}{( 1 - 6^6 Q)^3} \epsilon^2  \ , \\ 
  \kappa_{4,4} \,&=\, \frac{1728 Q (5 + 810648 Q + 60677807616 Q^2 + 4151263228821504 Q^3)}{( 1 - 6^6 Q)^5} \epsilon^4 \ .
\end{aligned}
\end{equation} 
Explicitly plugging in these correlators into eqs.~\eqref{eq:CY4N5cs}, we find the Picard--Fuchs operator 
\begin{equation} \label{eq:PFSextic}
   \mathcal{L}(Q) \,=\, \frac{6^5 \epsilon^5}{(1-6^6 Q)^6}\left[\Theta^5-6 Q (6\Theta+1)(6\Theta+2)(6\Theta+3)(6\Theta+4)(6\Theta+5) \right] \ ,
\end{equation}
which up to a dispensible overall factor is of the well-known expected form.

Alternatively --- as detailed for the example in Section~\ref{sec:FP4[4]} --- we obtain from the south pole correlator relation
\begin{multline}
   R_S(Q,\tm_i,\tm_P,\epsilon,\kappa_{n,\,\cdot\,}) \,=\, 
   \left\langle \sigma_N^n  (\sigma_S+\tm_1)\cdots(\sigma_S+\tm_6) \right\rangle \\
   - Q\,\left\langle \sigma_N^n (6\sigma_S-\tm_P+\epsilon)\cdots (6\sigma_S -\tm_P+6\epsilon)\right\rangle \ ,
\end{multline}
the differential operator of order six
\begin{equation}
    \mathcal{L}(Q,\epsilon,\tm_i,\tm_P)\,=\, (\epsilon\Theta+\tm_1)\cdots(\epsilon\Theta+\tm_6) 
    - Q (6\epsilon\Theta-\tm_P+\epsilon)\cdots (6\epsilon\Theta -\tm_P+6\epsilon) \ .
\end{equation}
For vanishing twisted masses this operator contains as a factor the Picard--Fuchs operator~\eqref{eq:PFSextic} of the order five, which in the absence of twisted masses arises from the reduced syzygy module $M_S^0$.

\subsubsection{Calabi--Yau subvarieties of the complex Grassmannian $\operatorname{Gr}(2,5)$}
To further illustrate the method of deriving Picard--Fuchs differential equations from gauge theory correlators, we study a list of complete intersection Calabi--Yau subvarieties of the complex Grassmannian $\operatorname{Gr}(2,5)$ of different dimensions. The Pl\"ucker embedding $\operatorname{Pl}: \operatorname{Gr}(2,5) \hookrightarrow \mathbb{P}(\Lambda^2 \mathbb{C}^5)$ realizes the complex Grassmannian $\operatorname{Gr}(2,5)$ as non-complete intersection variety of complex dimension six and first Chern class $c_1(\operatorname{Gr}(2,5)) = 5 \sigma_1$ in terms of the Schubert class~$\sigma_1$. Let us now consider the subvarieties 
\begin{equation}
  X_{n_1,\ldots,n_k} \,=\, \operatorname{Gr}(2,5) \cap \mathbb{P}(\Lambda^2 \mathbb{C}^5)[n_1,\ldots,n_k] \ , \qquad \dim_{\mathbb{C}}  X_{n_1,\ldots,n_k} \,=\, 6-k \ ,
\end{equation}
given as intersections of the Grassmannian with a generic complete intersection projective variety $\mathbb{P}(\Lambda^2 \mathbb{C}^5)[n_1,\ldots,n_k]$. 

Similarly as the R\o{}dland Calabi--Yau threefold of Section~\ref{sec:Rod}, we describe the defined varieties $X_{n_1,\ldots,n_k}$ as target spaces of non-Abelian gauged linear sigma models with gauge group $U(2)$ and the matter spectrum described in Table~\ref{tab:Gr25}.
\begin{table}[h]
\hfil
\hbox{
\vbox{
\offinterlineskip
\halign{\strut\vrule width1.2pt height15pt#&\hfil~#~\hfil\vrule width0.5pt&\hfil~#~\hfil\vrule width0.5pt&\hfil~#~\hfil\vrule width0.5pt&\hfil~#~\hfil&#\vrule width1.2pt\cr
\noalign{\hrule height 1.2pt}
&\multispan4 non-Abelian gauge theory spectrum:\hfil&\cr
&Chiral multiplets & $U(2)$~Representation & $U(1)_R$~charge & twisted masses&\cr
\noalign{\hrule height 1.2pt}
&$\phi_i$, $i=1,\ldots,5$ & $\Box_{+1}$ & $0$ & $\tm_i$&\cr
\noalign{\hrule height 0.5pt}
&$P^j$, $j=1,\ldots,k$ & $\mathbf{1}_{-2n_j}$ & $2$ & $\tm_P^j$&\cr
\noalign{\hrule height 1.2pt}
\omit\vrule width0pt height2pt\cr
\noalign{\hrule height 1.2pt}
&\multispan4 Abelian Coulomb branch gauge theory spectrum:\hfil&\cr
&Chiral multiplets & $U(1)\times U(1)$~charge & $U(1)_R$~charge & twisted masses&\cr
\noalign{\hrule height 1.2pt}
&$\phi_i^{(1)}$, $i=1,\ldots,5$ & $(+1,0)$ & $0$ & $\tm_i$&\cr
\noalign{\hrule height 0.5pt}
&$\phi_i^{(2)}$, $i=1,\ldots,5$ & $(0,+1)$ & $0$ & $\tm_i$&\cr
\noalign{\hrule height 0.5pt}
&$P^j$, $j=1,\ldots,k$ & $(-n_j,-n_j)$ & $2$ & $\tm_P^j$&\cr
\noalign{\hrule height 0.5pt}
&$W_\pm$ & $(\pm1,\mp1)$ & $2$ & $0$&\cr
\noalign{\hrule height 1.2pt}
}}}
\hfil
\caption{The table shows the chiral matter multiplets of the $U(2)$ gauged linear sigma model for the target spaces $X_{n_1,\ldots,n_k}\subset \operatorname{Gr}(2,5)$ in terms of representations of the gauge group $U(2)$. At the bottom we display the chiral spectrum in the Coulomb branch of the gauge theory, where the Coulomb branch fields fall into representations of $U(1)\times U(1)$.} \label{tab:Gr25}
\end{table}
Note that the axial anomaly of the $U(1)_R$-symmetry vanishes for the displayed spectrum provided that
\begin{equation}
  \sum_{i=1}^k n_i \,=\, 5 \ .
\end{equation}
This constraint implies the Calabi--Yau condition for the subvariety $X_{n_1,\ldots,n_k}$, as it implies a vanishing first Chern class of the gauge theory target space.  

\bigskip
Let us now turn to the specific analysis of Calabi--Yau subvarieties of $\operatorname{Gr}(2,5)$ in various dimensions, determining their Picard--Fuchs operators from the gauge theory perspective (in the limit of vanishing twisted masses $\tm_i \to 0, \tm_P^j \to 0$):

\paragraph{The elliptic curve $X_{1^5} \subset \operatorname{Gr}(2,5)$:}
The Picard--Fuchs operator of the quantum K\"ahler moduli space of the elliptic curve $X_{1^5}$ is according to eq.~\eqref{eq:LCY1} fully determined by the two correlation functions
\begin{equation} 
  \kappa_{0,1} \,=\, \frac{5}{1 - 11 Q - Q^2}  \ , \qquad
  \kappa_{1,2} \,=\, \frac{5 Q (3 - 32 Q - 14 Q^2 - Q^3)}{(1-11 Q - Q^2)^3}  \epsilon^2\ .
\end{equation} 
The resulting Picard--Fuchs operator reads 
\begin{equation}
  \mathcal{L}(Q) \,=\, \frac{5 \epsilon^2}{\left(1-11Q-Q^2\right)^2}\left[\Theta ^2-Q\left(11 \Theta ^2+11 \Theta +3\right)-Q^2(\Theta +1)^2\right] \ .
\end{equation}

\paragraph{Polarized $K3$ surface $X_{1^3,2}  \subset \operatorname{Gr}(2,5)$:}
The two-dimensional Calabi--Yau variety $X_{1^3,2}$ is a polarized K3 surface with single K\"ahler modulus. As a consequence its Picard--Fuchs operator is given by formula~\eqref{eq:K3Op}, which requires the gauge theory correlators 
\begin{equation} 
\kappa_{0,2} \,=\, \frac{10}{1 - 44 Q - 16 Q^2}  \ , \qquad
\kappa_{2,2} \,=\, \frac{40 Q(3 - 250 Q - 356 Q^2 - 112 Q^3)}{ (1 - 44 Q - 16 Q^2)^3}  \epsilon^2 \ . 
\end{equation}
Upon explicitly plugging in these correlators in eq.~\eqref{eq:K3Op}, we find for the Picard-Fuchs operator the expression 
\begin{multline}
  \mathcal{L}(Q)\,=\,\frac{8000 \epsilon^3}{(1 - 44 Q - 16 Q^2)^4} \\
  \times\left[
  \Theta ^3 -2 Q (2 \Theta +1) \left(11 \Theta ^2+11 \Theta +3\right) -4 Q^2 (\Theta +1) (2 \Theta +1) (2 \Theta +3)
  \right] \ . 
\end{multline}

\paragraph{Calabi--Yau threefold $X_{1^2,3}   \subset \operatorname{Gr}(2,5)$:}
The Calabi--Yau threefold $X_{1^2,3}$ has a single K\"ahler modulus, such that we can construct its Picard--Fuchs operator of order four from the correlators 
\begin{equation}
\begin{aligned}
\kappa_{0,3} \,&=\, \frac{15}{1 - 297 Q - 729 Q^2}  \ , \qquad 
\kappa_{2,3} \,=\, \frac{45 Q (49 + 351 Q)}{( 1 - 297 Q - 729 Q^2)^2} \epsilon^2   \ , \\ 
\kappa_{3,4} \,&=\,\frac{135 Q (2 + 1847 Q + 10125 Q^2 + 94041 Q^3)}{( 1 - 297 Q - 729 Q^2)^3}\epsilon^4 \ .
\end{aligned}
\end{equation} 
such that we arrive with eq.~\eqref{eq:LCY3} at the explicit result 
\begin{multline}
   \mathcal{L}(Q)  \,=\, \frac{225 \epsilon^4}{(1 - 297 Q - 729 Q^2)^3} 
   \left[ 
  \Theta ^4 -3 Q(3 \Theta +1) (3 \Theta +2) \left(11 \Theta ^2+11 \Theta +3\right)\right. \\
  \left.-9 Q^2(3 \Theta +1) (3 \Theta +2) (3 \Theta +4) (3 \Theta +5)
 \right] \ .
\end{multline}

\paragraph{Calabi--Yau threefold $X_{1,2^2}   \subset \operatorname{Gr}(2,5)$:}
Similarly as the previous example, we compute for the Calabi--Yau threefold $X_{1,2^2}$ with a single K\"ahler modulus the correlators 
\begin{equation}
\begin{aligned}
  \kappa_{0,3} \,&=\, \frac{20}{1-176 Q -256 Q^2}  \ , \\
  \kappa_{2,3} \,&=\, \frac{80Q (23 + 96 Q) }{(1-176 Q -256 Q^2)^2} \epsilon^2   \ , \\ 
  \kappa_{3,4} \,&=\, \frac{80Q(3 + 1624 Q + 5440 Q^2 + 27648 Q^3)}{(1-176 Q -256 Q^2)^3} \epsilon^4 \ .
\end{aligned}
\end{equation} 
in order to determine with eq.~\eqref{eq:LCY3} the Picard Fuchs operator 
\begin{multline}
   \mathcal{L}(Q)  \,=\, \frac{400 \epsilon^4}{(1 - 176 Q - 256 Q^2)^3} \\
   \times \left[
   \Theta ^4 -4 Q(2 \Theta +1)^2 \left(11 \Theta ^2+11 \Theta +3\right)-16 Q^2(2 \Theta +1)^2 (2 \Theta +3)^2 \right]  \ .
\end{multline}

\paragraph{Calabi--Yau fourfold: $X_{1,4} \subset \operatorname{Gr}(2,5)$:}
The Calabi--Yau fourfold target space $X_{1,4}$ has a single K\"ahler modulus and realizes a next-to-minimal order Picard--Fuchs system \cite{Honma:2013hma,Gerhardus:2016iot}, that is to say the Picard--Fuchs operator of its quantum K\"ahler moduli space is of order six. From the correlators
\begin{equation}
\begin{aligned} 
  \kappa_{0,4} \,&=\, \frac{40}{1 - 2816 Q - 65536 Q^2} \ ,  \\
  \kappa_{3,3} \,&=\, \frac{640 Q (81 - 347392 Q - 29818880 Q^2 - 570425344 Q^3)}{(1 - 2816 Q - 65536 Q^2)^3} \epsilon^2 \ , \\ 
  \kappa_{4,4} \,&=\,\frac{640 Q }{(1 - 2816 Q - 65536 Q^2)^5} (9 + 40096 Q + 134025216 Q^2 + 1397306163200 Q^3 \\ 
    &\qquad+  145439672238080 Q^4 + 8824474166099968 Q^5 + 267876216898322432 Q^6\\
    &\qquad + 3544332906740580352 Q^7)  \epsilon^4\ ,\\
  \kappa_{5,5} \,&=\, \frac{-5120 Q^2 }{(1 - 2816 Q - 65536 Q^2)^7}  (9971 + 105910560 Q + 1073969029120 Q^2 \\
    &+ 931195571404800 Q^3 +  691651807931269120 Q^4 + 100651036033033437184 Q^5 \\
    &+   4365384330728355921920 Q^6 + 117058417198543132426240 Q^7 \\
    &+ 5651505923432303191654400 Q^8 + 99919371819418105665290240 Q^9\\
    & + 2338062535134692823881744384 Q^{10}) \epsilon^6 \ ,
\end{aligned}
\end{equation}
we indeed confirm that $\kappa_{5,5}\neq \kappa_{5,5}^{N=5}$ with $\kappa_{5,5}^{N=5}$ as in \eqref{eq:k55N5CY4}, such that the operator can not be of order five. By further inserting them into the formula~\eqref{eq:CY4LN6} for next-to-minimal non-degenerate Picard--Fuchs operators we find 
\begin{multline}
  \mathcal{L}(Q)\,=\,\frac{\epsilon^6}{1 - 2816 Q - 65536 Q^2} \\
  \times \left[ (\Theta -1) \Theta ^5 -8 Q (2 \Theta +1) (4 \Theta +1) (4 \Theta +3) \left(11 \Theta ^2+11\Theta +3\right) \Theta  \right. \\
  \left.-64  Q^2 (2 \Theta +1) (2 \Theta +3) (4 \Theta +1) (4 \Theta +3) (4 \Theta+5) (4 \Theta +7)  \right] \ .
\end{multline}
This operator agrees with the result derived in refs.~\cite{Honma:2013hma,Gerhardus:2016iot}.

\paragraph{Calabi--Yau fourfold: $X_{2,3} \subset \operatorname{Gr}(2,5)$:}
The Calabi--Yau fourfold $X_{2,3}$ is similar to the previous example: It has a single K\"ahler modulus and a next-to-minimal order Picard--Fuchs system. By inserting the correlators
\begin{equation}
\begin{aligned} 
\kappa_{0,4} \,&=\, \frac{30}{1 - 1188 Q - 11664 Q^2}  \ ,  \\
  \kappa_{3,3} \,&=\,-\frac{360 Q (-49 + 86454 Q + 3105540 Q^2 + 24879312 Q^3)}{(1 - 1188 Q - 11664 Q^2)^3} \epsilon^2 \ , \\ 
  \kappa_{4,4} \,&=\,\frac{1080Q}{(1 - 1188 Q - 11664 Q^2)^5}  (2 + 2699 Q + 3039552 Q^2 + 21454542900 Q^3 \\
  &+ 951682497840 Q^4 + 24054397155072 Q^5 + 304084792134144 Q^6 + 1665027883348992 Q^7)\epsilon^4\ ,\\
  \kappa_{5,5} \,&=\,- \frac{77760 Q^2}{(1 - 1188 Q - 11664 Q^2)^7} (86 + 363305 Q + 1693192230 Q^2 + 624218132700 Q^3\\
  & + 214574221678770 Q^4 + 13324967851401792 Q^5 +  255596825554411680 Q^6\\
  & + 3077410414888817280 Q^7 + 
  56902089882303360000 Q^8 + 429222039529007646720 Q^9 \\
  &+ 3901536457986209734656 Q^{10}) \epsilon^6 \ .
\end{aligned}
\end{equation}
into eq.~\eqref{eq:CY4LN6} we find the order six Picard--Fuchs operator
\begin{multline}
  \mathcal{L}(Q)\,=\, \frac{\epsilon^6}{1 - 1188 Q - 11664 Q^2} \\
  \times \left[ (\Theta -1) \Theta ^5 -6 Q (2 \Theta +1) (3 \Theta +1) (3 \Theta +2) \left(11 \Theta ^2+11\Theta +3\right) \Theta \right. \\
  \left. -36  Q^2 (2 \Theta +1) (2 \Theta +3) (3 \Theta +1) (3 \Theta +2) (3 \Theta+4) (3 \Theta +5)\right] \ ,
\end{multline}
which is in agreement with the literature \cite{Honma:2013hma,Gerhardus:2016iot}.

\subsubsection{The local conifold Calabi--Yau threefold $\mathcal{O}(-1)\oplus\mathcal{O}(-1)\to\mathbb{P}^1$} \label{sec:Coni}
As our next example, we consider the Abelian $U(1)$ gauged linear sigma model with the matter content listed in Table~\ref{tab:SpecConi}.
\begin{table}[h]
\hfil
\hbox{
\vbox{
\offinterlineskip
\halign{\strut\vrule width1.2pt\hfil~#~\hfil\vrule width0.5pt&\hfil~#~\hfil\vrule width0.5pt&\hfil~#~\hfil\vrule width0.5pt&\hfil~#~\hfil\vrule width1.2pt\cr
\noalign{\hrule height 1.2pt}
Chiral multiplets & $U(1)$~charge & $U(1)_R$~charge & twisted masses\cr
\noalign{\hrule height 1.2pt}
$\phi_i$, $i=1,2$ & $+1$ & $0$ & $\tm_\phi^i$ \cr
\noalign{\hrule height 0.5pt}
$\psi_i$, $i=1,2$ & $-1$ & $0$ & $\tm_\psi^i$ \cr
\noalign{\hrule height 1.2pt}
}}}
\hfil
\caption{Matter spectrum of the $U(1)$ gauged linear sigma model with the non-compact conifold Calabi--Yau threefold $\mathcal{O}(-1)\oplus\mathcal{O}(-1)\to\mathbb{P}^1$ as its target space.} \label{tab:SpecConi}
\end{table}
Since the axial $U(1)_R$ anomaly cancels for this spectrum, we expect the analyzed gauged linear sigma model to have a Calabi--Yau target space. Note further that there is non superpotential, due to matter multiplets having $U(1)_R$ charge zero. The symplectic quotient~\eqref{eq:SympQuotient} thus describes a non-compact Calabi--Yau target space \eqref{eq:CYTot}, which for both positive and negative Fayet--Ilipoulos parameter $\xi$ realizes the non-compact conifold Calabi--Yau threefold $\mathcal{O}(-1)\oplus\mathcal{O}(-1)\to\mathbb{P}^1$.\footnote{Changing the Fayet--Ilipoulos parameter $\xi$ from $+\infty$ to $-\infty$ relates the two conifold phases by a flop transition.} 

We now determine the correlator relations from the spectrum in Table~\ref{tab:SpecConi}, which gives rise to rational functions
\begin{equation} \label{eq:Coni_gs}
  g_p(w,\tm_\phi^i,\tm_\psi^i,\epsilon) \,=\, 
  \prod_{s=0}^{p-1}\frac{(w + \tm_\phi^1 - \epsilon s)(w + \tm_\phi^2 - \epsilon s)}{(w - \tm_\psi^1 - (s+1)\epsilon )(w - \tm_\psi^2 - (s+1)\epsilon)} \ , \quad p=0,1,2,\ldots \ .
\end{equation}
The relation~\eqref{eq:RSDef} is solved by the syzygy polynomials 
\begin{equation}
  \alpha_0(w,\tm_\phi^i,\tm_\psi^i,\epsilon)\,=\,(w+\tm_\phi^1)(w+\tm_\phi^2)\ , \quad \alpha_1(w,\tm_\phi^i,\tm_\psi^i,\epsilon) \,=\, -(w -\tm_{\psi}^1-\epsilon)(w -\tm_{\psi}^2-\epsilon) \ ,
\end{equation}
that give rise to the south pole correlator relation
\begin{equation} \label{eq:correlConi}
  R_S(Q,\tm_\phi^i,\tm_\psi^i,\epsilon,\kappa_{n,\,\cdot\,}) \,=\,
  \left\langle \sigma_N^n (\sigma_S+\tm_\phi^1)(\sigma_S+\tm_\phi^2) \right\rangle
   - Q  \left\langle \sigma_N^n (\sigma_S-\tm_\psi^1)(\sigma_S-\tm_\psi^2) \right\rangle \ .
\end{equation}   
  
Let us now consider certain special limits of the twisted masses $\tm_\phi^i$ and $\tm_\psi^j$, for instance the limit $\tm_\phi^1 \, \to \, - \tm_\psi^1 - \epsilon$. The rational functions~\eqref{eq:Coni_gs} then simplify and the syzygy polynomials become
\begin{equation} \label{eq:Coni_nong}
\begin{aligned}
  \alpha_0(w,- \tm_\psi^1 - \epsilon,\tm_\phi^2,\tm_\psi^1,\tm_\psi^2,\epsilon)\,&=\,w+\tm_\phi^2\ , \\
  \alpha_1(w,- \tm_\psi^1 - \epsilon,\tm_\phi^2,\tm_\psi^1,\tm_\psi^2,\epsilon)\,&=\, -w +\tm_{\psi}^2+ \epsilon \ .
\end{aligned}  
\end{equation}
The corresponding limiting module $M_S^{\lim}$ of correlator relations therefore becomes a proper submodule of the non-generic syzygy module $M_S^0$ generated by the syzygy polynomials~\eqref{eq:Coni_nong}. However, as discussed in general in Section~\ref{sec:Reduction}, for this example the non-generic syzygy module~$M_S^0$ does not describe the south pole correlator relations of the gauge theory. Namely, in the phase with positive Fayet--Ilipoulos parameter $\xi >0$ the contributing poles (arising from the chiral fields $\phi_i$) and the non-contributing poles (attributed to the chiral fields $\psi_i$) respectively yield the pole sets
\begin{equation}
    \Theta(\xi>0,\tm_i) \,=\,\bigcup_{i=1}^2  \left\{\epsilon \mathbb{Z}_{\le 0} - \tm^i_\phi \right\}  \,\subset\, \mathbb{C} \ , \quad
    \Omega(\xi>0) \,=\,  \bigcup_{i=1}^2  \left\{\epsilon \mathbb{Z}_{\ge 0} + \tm^i_\psi \right\}  \,\subset\, \mathbb{C}  \ .
\end{equation}
While these sets do not intersect for generic values of the twisted masses $\tm_\phi^i$ and $\tm_\psi^i$, in the limit $\tm_\phi^1 \, \to \, - \tm_\psi^1 - \epsilon$ there is a non-empty intersection $\Theta(\xi>0) \cap \Omega(\xi>0) = \{ \tm_\psi^1, \tm_\psi^1 + \epsilon \}$. This indicates that the additional relations in the non-generic syzygy module $M_S^0$ do not realize physically valid correlator relations, which we also confirmed by explicit calculations.

From the correlator relation~\eqref{eq:correlConi} we easily read off the (equivariant) differential operator
\begin{equation}
    \mathcal{L}(Q,\epsilon,\tm_\phi^i,\tm_\psi^i)\,=\, 
    (\epsilon\Theta+\tm_\phi^1) (\epsilon\Theta+\tm_\phi^2) - Q (\epsilon\Theta-\tm_\psi^1) (\epsilon\Theta-\tm_\psi^2) \ .  
\end{equation}
Note that this operator does not further reduce in the limit $\tm_\phi^1 \, \to \, - \tm_\psi^1 - \epsilon$, precisely due to the discussed phenomenon of the non-generic syzygy module $M_S^0$ being unphysical in this example.

\subsubsection{The local Calabi--Yau threefold $\mathcal{O}(-3)_{\mathbb{P}^2}$}
As our second example of a gauge theory with a non-compact Calabi--Yau target space we study the Abelian $U(1)$ gauged linear sigma model with the chiral matter spectrum given in Table~\ref{tab:SpecLocalP2}.
\begin{table}[h]
\hfil
\hbox{
\vbox{
\offinterlineskip
\halign{\strut\vrule width1.2pt\hfil~#~\hfil\vrule width0.5pt&\hfil~#~\hfil\vrule width0.5pt&\hfil~#~\hfil\vrule width0.5pt&\hfil~#~\hfil\vrule width1.2pt\cr
\noalign{\hrule height 1.2pt}
Chiral multiplets & $U(1)$~charge & $U(1)_R$~charge & twisted masses\cr
\noalign{\hrule height 1.2pt}
$\phi_i$, $i=1,\ldots,3$ & $+1$ & $0$ & $\tm_i$ \cr
\noalign{\hrule height 0.5pt}
$P$ & $-3$ & $0$ & $\tm_P$ \cr
\noalign{\hrule height 1.2pt}
}}}
\hfil
\caption{Matter spectrum of the $U(1)$ gauged linear sigma model with the Calabi-Yau threefold $\mathcal{O}(-3)_{\mathbb{P}^2}$ as its target space.} \label{tab:SpecLocalP2}
\end{table}
Similar to the example in Section~\ref{sec:Coni}, the symplectic quotient~\eqref{eq:SympQuotient} with a positive Fayet--Ilipoulos parameter $\xi$ yields the non-compact Calabi--Yau threefold $\mathcal{O}(-3)_{\mathbb{P}^2}$.

As in the previous examples, the first step in deriving south pole correlator relations is to determine the rational functions 
\begin{equation} 
  g_p(w,\tm_i,\tm_p,\epsilon) \,=\, 
 (-1)^p \frac{\prod_{s=0}^{p-1}(w + \tm_1 - \epsilon s)(w + \tm_2 - \epsilon s)(w + \tm_3 - \epsilon s)}{\prod_{s=0}^{3p-1}(3w - \tm_p - (s+1)\epsilon )} \ , \quad p=0,1,2,\ldots \  .
\end{equation}
These define the syzygy module $M_S$ and we arrive at the south pole correlator relation 
\begin{multline}
   R_S(Q,\tm_i,\tm_P,\epsilon,\kappa_{n,\,\cdot\,}) \,=\,
  \left\langle \sigma_N^n (\sigma_S+\tm_1)(\sigma_S+\tm_2)(\sigma_S+\tm_2) \right\rangle \\
+  Q  \left\langle \sigma_N^n (3\sigma_S-\tm_P)(3\sigma_S-\tm_P+\epsilon)(3\sigma_S-\tm_P+2\epsilon) \right\rangle \ ,
\end{multline}
which gives rise to the differential operator
\begin{multline} \label{eq:LlocalP2}
   \mathcal{L}(Q,\epsilon,\tm_i,\tm_P)\,=\, 
    (\epsilon\Theta+\tm_1)(\epsilon\Theta+\tm_2)(\epsilon\Theta+\tm_3) \\
      + Q (3\epsilon\Theta-\tm_P) (3\epsilon\Theta-\tm_P+\epsilon)(3\epsilon\Theta-\tm_P+2\epsilon) \ .
\end{multline}

By analogous reasoning as for the example of the conifold target space we find that the limiting syzygy module $M_S^{\lim}$ entirely captures the correlator relations for non-generic twisted masses. The non-generic syzygy module $M_S^0$ becomes unphysical in particular limits, for instance when $\tm_1 \to -\frac{mp}{3} - \epsilon$. As consequence, the differential operator~\eqref{eq:LlocalP2} does also not further reduce for special values of the twisted masses.

\subsubsection{The two parameter Calabi--Yau threefold $\widehat{\mathbb{WP}}^{4}_{1,1,2,2,2}[8]$} \label{sec:WP11222}
Let us now examine the Abelian gauged linear sigma model with gauge group $U(1)\times U(1)$ and chiral matter space as presented in Table~\ref{tab:WP11222}.
\begin{table}[h]
\hfil
\hbox{
\vbox{
\offinterlineskip
\halign{\strut\vrule width1.2pt\hfil~#~\hfil\vrule width0.5pt&\hfil~#~\hfil\vrule width0.5pt&\hfil~#~\hfil\vrule width0.5pt&\hfil~#~\hfil\vrule width1.2pt\cr
\noalign{\hrule height 1.2pt}
Chiral multiplets & $U(1)\times U(1)$~charge & $U(1)_R$~charge & twisted masses\cr
\noalign{\hrule height 1.2pt}
$P$ & $(-4,0)$ & $2$ & $\tm_P$ \cr
\noalign{\hrule height 0.5pt}
$X_i$, $i=1,2,3$ & $(+1,0)$ & $0$ & $\tm_X^i$ \cr
\noalign{\hrule height 0.5pt}
$Y_\alpha$, $\alpha=1,2$ & $(0,+1)$ & $0$ & $\tm_Y^\alpha$ \cr
\noalign{\hrule height 0.5pt}
$Z$ & $(+1,-2)$ & $0$ & $\tm_Z$ \cr
\noalign{\hrule height 1.2pt}
}}}
\hfil
\caption{The chiral matter spectrum of the $U(1)\times U(1)$ gauged linear sigma model of the two parameter Calabi--Yau threefold $\widehat{\mathbb{WP}}^{4}_{1,1,2,2,2}[8]$.} \label{tab:WP11222}
\end{table}
For positively chosen Fayet--Ilipoulos parameters $(\xi_1,\xi_2)$ the symplectic quotient~\eqref{eq:SympQuotient} and the F-term constraints realize (for generic choices) the smooth Calabi--Yau threefold $\widehat{\mathbb{WP}}^{4}_{1,1,2,2,2}[8]$ with two K\"ahler moduli. It is a Calabi--Yau hypersurface in the toric variety obtained from a certain resolution of the singular weighted projective space $\mathbb{WP}^{4}_{1,1,2,2,2}$. From a mirror symmetry point of view this Calabi--Yau threefold and its two-dimensional quantum K\"ahler moduli space are analyzed in detail in ref.~\cite{Candelas:1993dm}. Here, we demonstrate the gauge theory approach to derive the order two and order three Picard--Fuchs differential operators governing the quantum K\"ahler moduli space.

From the spectrum in Table~\ref{tab:WP11222} we first determine the rational functions~\eqref{eq:RSDef}, which here read
\begin{multline}
   g_{p_1,p_2}(w_1,w_2,\tm_P,\tm_X^i,\tm_Y^\alpha,\tm_Z,\epsilon)\\
  \,=\, g_{p_1,p_2}^{Z}(w_1,w_2,\tm_Z,\epsilon) \cdot \frac{\prod_{i=1}^{3} \prod_{s=0}^{p_1-1}(w_1+\tm_X^i-\epsilon s) \prod_{\alpha=1}^{2} \prod_{s=0}^{p_2-1}(w_2+\tm_Y^\alpha-\epsilon s)}{\prod_{s=0}^{4p_1-1}(4w_1-\tm_{P}-\epsilon s)} 
\end{multline}
with 
\[
   g_{p_1,p_2}^{Z}(w_1,w_2,\tm_Z,\epsilon)= 
\begin{cases}
\prod\limits_{s=0}^{p_1-2p_2-1}(w_1-2w_2+\tm_{Z}-\epsilon s)& \text{if } p_1-2p_2 \geq 1\ ,\\
\prod\limits_{s=0}^{2p_2-p_1-1}(w_1-2w_2+\tm_{Z}+\epsilon (1+s))^{-1}              & \text{if } p_1-2p_2 \leq -1\ ,\\
1 & \text{if } p_1-2p_2 = 0 \ .
\end{cases}
\]
In the limit of vanishing twisted masses they simplify to\footnote{The criterion for factorizaton~\eqref{eq:OverlapCond} is fulfilled for $\tm_P=\tm_X^i=\tm_Y^\alpha=0$ and $\tm_Z$ generic. Already for this choice of twisted masses the displayed factorization occurs. After this factorization we can safely set $\tm_Z$ to zero as well.}
\begin{equation*}
\begin{aligned}
   g_{p_1,p_2}(w_1,w_2,\epsilon) \,&=\, \frac{w_1^2\prod_{s=1}^{p_1-1}(w_1-\epsilon s)^3 \prod_{s=0}^{p_2-1}(w_2-\epsilon s)^2}
   {4\prod_{s=1}^{4p_1-1}(4w_1-\epsilon s)} \ g_{p_1,p_2}^{Z}(w_1,w_2,0,\epsilon) \quad \text{for } p_1\geq1 \ ,\\
   g_{0,p_2}(w_1,w_2,\epsilon) \,&=\, g_{0,p_2}^{Z}(w_1,w_2,0,\epsilon)\prod_{s=0}^{p_2-1}(w_2-\epsilon s)^2 \ ,
\end{aligned}
\end{equation*}
and define the reduced syzygy module $M_S^0$, which has $M_S^{\lim}$ as a proper submodule. The analysis of the restricted cone structure reveals that $M_S^{0}$ realizes physically valid reduced correlator relations. In particular, we obtain the generating south pole correlator relations
\begin{equation}
\begin{aligned}
  R_S(\vec Q,\epsilon,\kappa_{\vec n,\,\cdot\,}) \,&=\,
     \left\langle \vec{\sigma}_N^{\vec n} \sigma_{S,2}^2 \right\rangle
     - Q_2 \left\langle \vec{\sigma}_N^{\vec n}\left(\sigma_{S,1}-2 \sigma_{S,2}\right)\left(\sigma_1-2 \sigma_2 - \epsilon\right)\right\rangle  \ , \\
  T_S(\vec Q,\epsilon,\kappa_{n,\,\cdot\,}) \,&=\,
   \left\langle \vec{\sigma}_N^{\vec n} \sigma_{S,1}^2\left(\sigma_{S,1} - 2 \sigma_{S,2} \right) \right\rangle
     - 8 Q_1 \left\langle \vec{\sigma}_N^{\vec n}\left(4\sigma_{S,1}+\epsilon\right)\left(2\sigma_{S,1}+\epsilon\right)\left(4\sigma_{S,1}+3\epsilon\right) \right\rangle \ ,  
\end{aligned}   
\end{equation}   
which respectively yield the Picard--Fuchs operators
\begin{equation} \label{eq:PFop11222}
\begin{aligned}
  \mathcal{L}_2(\vec{Q}) \,&=\, \Theta_2^2 - Q_2 ( \Theta_1 - 2 \Theta_2)(\Theta_1 - 2 \Theta_2 -1)  \ , \\
  \mathcal{L}_3(\vec{Q}) \,&=\, \Theta_1^2(\Theta_1 - 2 \Theta_2 ) - 8 Q_1 (4\Theta_1 +1)(2\Theta_1+1)(4\Theta_1 + 3) \ .
\end{aligned}
\end{equation}
These Picard--Fuchs operators are of order two and order three --- as required for a two-dimensional quantum K\"ahler moduli space of a Calabi--Yau threefold \cite{Hosono:1994ax} --- and they agree with the differential operators presented in the literature~\cite{Candelas:1993dm}. 

Alternatively, we can calculate both Picard--Fuchs operators~\eqref{eq:PFop11222} from the gauge theory correlators by employing formulas~\eqref{eq:3Fold2ParamL2SolveMainText} and \eqref{eq:CY32L3MainText}, which were derived for Calabi--Yau threefolds with two K\"ahler moduli. When using the differential relations of Table~\ref{tab:RepList} as well as the constraints in eq.~\eqref{eq:CY32ParamAddConstraint1} it suffices to compute the seven correlators
\begin{equation}
\begin{aligned}
  \kappa_{(0,0),(0,3)} \,&=\, \frac{4 Q_2 (1 + 4 Q_2 - 256 Q_1 (1 + 12 Q_2))}
    {\Delta_1^2 \Delta_2}\ , \qquad
   \kappa_{(0,0),(1,2)} \,=\,- \frac{8 (1 - 512 Q_1) Q_2}
   {\Delta_1 \Delta_2}\ , \\
   \kappa_{(0,0),(2,1)} \,&=\, \frac{4 - 1024 Q_1}
   {\Delta_2} \ , \qquad
   \kappa_{(0,0),(3,0)} \,=\, \frac{8}
   {\Delta_2} \ , \\
   \kappa_{(0,0),(0,5)} \,&=\, 
   -\frac{4 Q_2 \epsilon^2}{\Delta_1^4\Delta_2^3}
   \Big[3 + 95 Q_2 + 140 Q_2^2 - 16 Q_1 (240 + 9117 Q_2 + 14120 Q_2^2 + 400 Q_2^3) \\
    & +98304 Q_1^2 (20 + 894 Q_2 + 999 Q_2^2 - 1544 Q_2^3 + 560 Q_2^4) \\
    &- 2097152 Q_1^3 (240 + 12433 Q_2 + 2516 Q_2^2 - 87120 Q_2^3 +  32704 Q_2^4) \\
    &+ 2147483648 Q_1^4 (30 + 1779 Q_2 - 2071 Q_2^2 - 28220 Q_2^3 + 23088 Q_2^4 + 5568 Q_2^5) \\
    &- 206158430208 Q_1^5 (1 - 4 Q_2)^2 (16 + 1203 Q_2 + 6200 Q_2^2 + 2800 Q_2^3)\Big]
    \ , \\
   \kappa_{(0,0),(1,4)} \,&=\, 
   \frac{8 Q_2 \epsilon^2}
   {\Delta_1^3 \Delta_2^3}
   \Big[2 + 7 Q_2 - 32 Q_1 (100 + 313 Q_2 + 28 Q_2^2) \\
     &+ 24576 Q_1^2 (80 + 269 Q_2 - 576 Q_2^2 + 560 Q_2^3) \\
     & - 2097152 Q_1^3 (280 + 1213 Q_2 - 7224 Q_2^2 + 6928 Q_2^3) \\
     &+ 536870912 Q_1^4 (160 + 977 Q_2 - 9100 Q_2^2 + 9136 Q_2^3 + 
        5568 Q_2^4) \\
        &- 274877906944 Q_1^5 (1 - 4 Q_2)^2 (18 + 301 Q_2 + 308 Q_2^2) 
        \Big] \ , \\
   \kappa_{(0,0),(2,3)} \,&=\, -\frac{64 Q_1 Q_2 \epsilon^2}
   {\Delta_1 \Delta_2^3 }
   \Big[ 11 + 6144 Q_1 (6 - 35 Q_2) - 393216 Q_1^2 (83 - 420 Q_2) \\
    &+ 134217728 Q_1^3 (65 - 217 Q_2 - 348 Q_2^2) 
     - 4294967296 Q_1^4 (177 - 56 Q_2 - 2608 Q_2^2) \Big] \ ,
\end{aligned}
\end{equation}
given in terms of the discriminants
\begin{equation}
   \Delta_1 \,=\, 1- 4 Q_2 \ , \qquad \Delta_2 \,=\, 1-512 Q_1 + 65536 Q_1^2 ( 1 - 4 Q_2) \ .
\end{equation}
These correlators are in agreement with condition~\eqref{eq:CY32Assumption}, such that formulas~\eqref{eq:3Fold2ParamL2SolveMainText} and \eqref{eq:CY32L3MainText} are indeed applicable, and yield the differential operators 
\begin{equation} \label{eq:PFop11222ii}
\begin{aligned}
\mathcal{L}'_2(\vec{Q}) \,&=\, \frac{64 Q_2 \epsilon^2(1 - 256 Q_1 + 4 Q_2 - 3072 Q_1 Q_2)}{\Delta_1^4 \Delta_2^2} \cdot \mathcal{L}_2(\vec{Q}) \ , \\
\mathcal{L}'_3(\vec{Q}) \,&=\,  \frac{64 Q_2^2 \epsilon^3}{\Delta_1^4 \Delta_2^2} \cdot \big[ Q_2 \mathcal{L}_3(\vec{Q}) -(1-384 Q_1)\mathcal{L}_2(\vec{Q}) \\ & \hspace*{3cm} +(1-256 Q_1)\,\Theta_1\mathcal{L}_2(\vec{Q})+(2-1024 Q_1)\,\Theta_2 \mathcal{L}_2(\vec{Q})\big] \ .
\end{aligned}
\end{equation}
These operators generate the same Picard--Fuchs system as $\mathcal{L}_2(\vec{Q})$ and $\mathcal{L}_3(\vec{Q})$ given in eq.~\eqref{eq:PFop11222}.

\subsubsection{The two parameter Calabi--Yau threefold $\widehat{\mathbb{WP}}^{4}_{1,1,1,6,9}[18]$}
The discussion of the Abelian gauged linear sigma model with gauge group $U(1)\times U(1)$ and chiral matter spectrum given in Table~\ref{tab:WP11169} proceeds as for the example discussed in the previous Section~\ref{sec:WP11222}. 
\begin{table}[h]
\hfil
\hbox{
\vbox{
\offinterlineskip
\halign{\strut\vrule width1.2pt\hfil~#~\hfil\vrule width0.5pt&\hfil~#~\hfil\vrule width0.5pt&\hfil~#~\hfil\vrule width0.5pt&\hfil~#~\hfil\vrule width1.2pt\cr
\noalign{\hrule height 1.2pt}
Chiral multiplets & $U(1)\times U(1)$~charge & $U(1)_R$~charge & twisted masses\cr
\noalign{\hrule height 1.2pt}
$P$ & $(-6,0)$ & $2$ & $\tm_P$ \cr
\noalign{\hrule height 0.5pt}
$X$ & $(+3,0)$ & $0$ & $\tm_X$ \cr
\noalign{\hrule height 0.5pt}
$Y$ & $(+2,0)$ & $0$ & $\tm_Y$ \cr
\noalign{\hrule height 0.5pt}
$Z$ & $(+1,-3)$ & $0$ & $\tm_Z$ \cr
\noalign{\hrule height 0.5pt}
$\phi_i$, $i=1,2,3$ & $(0,+1)$ & $0$ & $\tm_i$ \cr
\noalign{\hrule height 1.2pt}
}}}
\hfil
\caption{The chiral matter spectrum of the $U(1)\times U(1)$ gauged linear sigma model of the two parameter Calabi--Yau threefold $\widehat{\mathbb{WP}}^{4}_{1,1,1,6,9}[18]$.} \label{tab:WP11169}
\end{table}
For positive Fayet--Ilipoulos parameters $(\xi_1,\xi_2)$ the symplectic quotient~\eqref{eq:SympQuotient} and the F-term constraints yield the smooth Calabi--Yau threefold $\widehat{\mathbb{WP}}^{4}_{1,1,1,6,9}[18]$ with two K\"ahler moduli. This threefold is a hypersurface in the resolution $\widehat{\mathbb{WP}}^{4}_{1,1,1,6,9}$ of the singular weighted projective space $\mathbb{WP}^{4}_{1,1,1,6,9}$. Its quantum K\"ahler moduli space is analyzed with mirror symmetry techniques in ref.~\cite{Candelas:1994hw}. 

By performing the same steps as in the previous example we find the generating south pole correlator relations
\begin{equation}
\begin{aligned}
  R_S(\vec Q,\epsilon,\kappa_{\vec n,\,\cdot\,}) \,&=\,
     \left\langle \vec{\sigma}_N^{\vec n} \sigma_{S,1}\left(\sigma_{S,1}- 3 \sigma_{S,2}\right)\right\rangle
     - 12 Q_1 \left\langle \vec{\sigma}_N^{\vec n}\left(6\sigma_{S,1}+\epsilon\right)\left(6\sigma_{S,1}+5\epsilon\right) \right\rangle  \ , \\
  T_S(\vec Q,\epsilon,\kappa_{n,\,\cdot\,}) \,&=\,
   \left\langle \vec{\sigma}_N^{\vec n} \sigma_{S,2}^3\ \right\rangle
     - Q_2 \left\langle \vec{\sigma}_N^{\vec n}\left(\sigma_{S,1}-3 \sigma_{S,2}\right)\left(\sigma_{S,1}-3 \sigma_{S,2} -\epsilon\right)\left(\sigma_{S,1}-3 \sigma_{S,2}-2\epsilon\right) \right\rangle \ .
\end{aligned}   
\end{equation}
These correspond to the order two and order three Picard--Fuchs operators 
\begin{equation} \label{eq:PFop11169}
\begin{aligned}
  \mathcal{L}_2(\vec{Q}) \,&=\, \Theta_1(\Theta_1-3\Theta_2) - 12 Q_1 (6 \Theta_1 + 1)(6\Theta_1 +5)  \ , \\
  \mathcal{L}_3(\vec{Q}) \,&=\, \Theta_2^3 - Q_2 (\Theta_1-3\Theta_2)(\Theta_1-3\Theta_2-1)(\Theta_1-3\Theta_2-2) \ ,
\end{aligned}
\end{equation}
which agree with the results of ref.~\cite{Candelas:1994hw}.

For the alternative approach of determining the Picard--Fuchs differential operators from gauge theory correlator we first calculate the correlators 
\begin{equation}
\begin{aligned}
  \kappa_{(0,0),(0,3)} \,&=\, \frac{9 (1 - 1296 Q_1 + 559872 Q_1^2) Q_2}
    {\Delta_1 \Delta_2}\ , \qquad
   \kappa_{(0,0),(1,2)} \,=\, \frac{(1-432Q_2)^2}
   {\Delta_2}\ , \\
   \kappa_{(0,0),(2,1)} \,&=\, \frac{3(1-432Q_1)}
   {\Delta_2} \ , \qquad
   \kappa_{(0,0),(3,0)} \,=\, \frac{9}
   {\Delta_2} \ , \\
   \kappa_{(0,0),(0,5)} \,&=\, 
   \frac{ Q_2 \epsilon^2}{\Delta_1^3\Delta_2^3}
   \Big[ (29 - 675 Q_2 - 12 Q_1 (9391 - 218565 Q_2 + 7290 Q_2^2)  \\
   &+5184 Q_1^2 (37549 - 873045 Q_2 + 72900 Q_2^2) \\
   &+ 
   33695156183620386816 Q_1^8 (1 + 27 Q_2)^2 (3127 - 312768 Q_2 + 
      1931850 Q_2^2) \\
      &+ 
   6718464 Q_1^3 (-28849 + 677565 Q_2 + 1318761 Q_2^2 + 6948099 Q_2^3) \\
   &- 
   967458816 Q_1^4 (-125105 + 3040740 Q_2 + 24874938 Q_2^2 + 
      124967367 Q_2^3) \\
      &+ 
   417942208512 Q_1^5 (-115709 + 3032235 Q_2 + 62974665 Q_2^2 + 
      311680305 Q_2^3) \\
      &+ 
   541653102231552 Q_1^6 (22237 - 664335 Q_2 - 26718579 Q_2^2 - 
      137977830 Q_2^3 + 156775095 Q_2^4) \\
      &- 
   77998046721343488 Q_1^7 (21889 - 791505 Q_2 - 50712156 Q_2^2 - 
      309219930 Q_2^3 + 1408318650 Q_2^4))\Big]
    \ , \\
   \kappa_{(0,0),(1,4)} \,&=\, 
   \frac{36 Q_1 Q_2 \epsilon^2}
   {\Delta_1 \Delta_2^3}
   \Big[-5 + 19440 Q_1 + 559872 Q_1^2 (293 + 9531 Q_2) \\
   &+ 
  522427760640 Q_1^4 (671 + 19089 Q_2) - 
  403107840 Q_1^3 (979 + 28620 Q_2) \\
  &+ 
  225688792596480 Q_1^5 (-679 - 17469 Q_2 + 43011 Q_2^2) \\
  &+ 
  8423789045905096704 Q_1^7 (-343 - 1431 Q_2 + 211410 Q_2^2) \\
  &- 
  32499186133893120 Q_1^6 (-1025 - 18900 Q_2 + 256608 Q_2^2))
        \Big] \ , \\
   \kappa_{(0,0),(2,3)} \,&=\, \frac{60466176 Q_1^3 (-1 + 432 Q_1) Q_2 \epsilon^2}
   { \Delta_2^3 }
   \Big[ (-1 + 432 Q_1)^3 (353 + 28944 Q_1) \\
   &+ 
 2176782336 Q_1^3 (-295 + 28944 Q_1) Q_2 \Big] \ ,
\end{aligned}
\end{equation}
given in terms of the discriminants
\begin{equation}
   \Delta_1 \,=\, 1+27 Q_2 \ , \qquad \Delta_2 \,=\, (1 - 1296 Q_1 + 559872 Q_1^2 - 80621568 Q_1^3 (1 + 27 Q_2))\ .
\end{equation}
These correlators obey the condition~\eqref{eq:CY32Assumption}, hence formulas~\eqref{eq:3Fold2ParamL2SolveMainText} and \eqref{eq:CY32L3MainText} are applicable. Together with the differential relations of Table~\ref{tab:RepList} as well as the constraints in eq.~\eqref{eq:CY32ParamAddConstraint1} we arrive at the operators 
\begin{equation}
\begin{aligned}
  \mathcal{L}'_2(\vec{Q}) \,&=\, 
    \frac{-9 \epsilon^2 (1 - 1296 Q_1 + 559872 Q_1^2) Q_2}{\Delta_1^2 \Delta_2^2}\cdot \mathcal{L}'_2(\vec{Q})   \ , \\
\mathcal{L}'_3(\vec{Q}) \,&=\,  \frac{(-1+432 Q_1) \epsilon^3}{\Delta_1^2 \Delta_2^2} \cdot \big[ (1 - 432 Q_1)^2 \mathcal{L}_3(\vec{Q}) -3 (1 - 720 Q_1) Q_2\mathcal{L}_2(\vec{Q}) \\ & \hspace*{3cm} +(1 - 432 Q_1) Q_2\, \Theta_1\mathcal{L}_2(\vec{Q})- 6 (1 - 648 Q_1) Q_2\,\Theta_2 \mathcal{L}_2(\vec{Q})\big] \ ,
\end{aligned}
\end{equation}
which generate same Picard--Fuchs system as $\mathcal{L}_2(\vec{Q})$ and $\mathcal{L}_3(\vec{Q})$ given in eq.~\eqref{eq:PFop11169}

\subsection{Projective varieties with ample canonical class}
Finally, let us present examples of gauged linear sigma models with the ample canonical bundle property, as introduced in Section~\ref{sec:Acorr}. Simple examples are given by hypersurfaces of degree $d$ in projective spaces $\mathbb{P}^n$ with $d>n+1$ or --- specifically in complex dimension one --- by genus $g$ curves with $g>2$. As deduced in Section~\ref{sec:Acorr} from the selection rules, the correlators in such gauge theories are polynomials in inverse powers of the parameters $\vec Q$, and we can again derive correlator relations for correlation functions in such theories.

\paragraph{The genus three curve $\mathbb{P}^2[4]$:}
Let us consider the Abelian $U(1)$ gauged linear sigma model with the chiral matter spectrum as listed in Table~\ref{tab:Genus3}.
\begin{table}[h]
\hfil
\hbox{
\vbox{
\offinterlineskip
\halign{\strut\vrule width1.2pt\hfil~#~\hfil\vrule width0.5pt&\hfil~#~\hfil\vrule width0.5pt&\hfil~#~\hfil\vrule width0.5pt&\hfil~#~\hfil\vrule width1.2pt\cr
\noalign{\hrule height 1.2pt}
Chiral multiplets & $U(1)$~charge & $U(1)_R$~charge & twisted masses\cr
\noalign{\hrule height 1.2pt}
$\phi_i$, $i=1,\ldots,3$ & $+1$ & $0$ & $\tm_i$ \cr
\noalign{\hrule height 0.5pt}
$P$ & $-4$ & $2$ & $\tm_P$ \cr
\noalign{\hrule height 1.2pt}
}}}
\hfil
\caption{Chiral matter spectrum of the $U(1)$ gauged linear sigma model for the genus three curve~$\mathbb{P}^2[4]$.} \label{tab:Genus3}
\end{table}
For a positive Fayet--Iliopoulos parameter $\xi$ its classical target space geometry is $\mathbb{P}^2[4]$, which generically is a smooth curve of genus three. The south pole correlator relations are deduced from the rational functions~\eqref{eq:DefWandG}, which for $p=0$ and $p=1$ are given by
\begin{equation}
  g_0(w,\tm_i,\tm_P,\epsilon)\,=\, 1 \ , \quad
  g_1(w,\tm_i,\tm_P,\epsilon) \,=\,  \frac{(w+\tm_1)(w+\tm_2)(w+\tm_3)}{(4w-\tm_P)\cdots (4w -\tm_P-3\epsilon)} \ .
\end{equation}
Their corresponding syzygy polynomials read
\begin{equation}
\begin{aligned}
   \alpha_0(w,\tm_i,\tm_P,\epsilon)\,&=\,(w+\tm_1)(w+\tm_2)(w+\tm_3)\ , \\ 
   \alpha_1(w,\tm_i,\tm_P,\epsilon) \,&=\, -(4w-\tm_P)(4w-\tm_P-\epsilon)(4w-\tm_P-2\epsilon)(4w -\tm_P-3\epsilon)\ ,
\end{aligned}
\end{equation}
which with eq.~\eqref{eq:RSDef} result in the south pole correlator relation
\begin{multline}
   R_S(Q,\tm_i,\tm_P,\epsilon,\kappa_{n,\,\cdot\,}) \,=\, 
  \left\langle \sigma_N^n  (\sigma_S+\tm_1)\cdots(\sigma_S+\tm_3) \right\rangle \\
   - Q  \left\langle \sigma_N^n (4\sigma_S-\tm_P+\epsilon)\cdots (4\sigma_S -\tm_P+4\epsilon)\right\rangle \ .
\end{multline}
In the limit of vanishing twisted masses condition~\eqref{eq:OverlapCond} is fulfilled, and we obtain the non-generic south pole correlator relation  
\begin{equation} 
   R_S(Q,\epsilon,\kappa_{n,\,\cdot\,}) \,=\, 
  \left\langle \sigma_N^n\sigma_S^2 \right\rangle 
   - 8 Q  \left\langle \sigma_N^n (2\sigma_S+\epsilon) (4\sigma_S+\epsilon)  (4\sigma_S +3\epsilon)\right\rangle \ .
\end{equation}
This relation can explicitly be confirmed for the correlators calculated in the $\xi<0$ phase,
\begin{equation}
   \kappa_{0,0}\,=\, 0 \ , \quad 
   \kappa_{0,1}\,=\, 0 \ , \quad 
   \kappa_{0,2}\,=\, \frac{-v}{64} \ , \quad 
   \kappa_{0,3}\,=\, \frac{-v^2}{16384}+\frac{3 v \epsilon}{128} \ ,
\end{equation}
where the variable $v = Q^{-1}$.

\paragraph{The surface $\mathbb{P}^3[5]$ of ample canonical bundle:}
The Abelian $U(1)$ gauged linear sigma model with the chiral matter spectrum listed in Table~\ref{tab:SurP3[5]}
\begin{table}[th]
\hfil
\hbox{
\vbox{
\offinterlineskip
\halign{\strut\vrule width1.2pt\hfil~#~\hfil\vrule width0.5pt&\hfil~#~\hfil\vrule width0.5pt&\hfil~#~\hfil\vrule width0.5pt&\hfil~#~\hfil\vrule width1.2pt\cr
\noalign{\hrule height 1.2pt}
Chiral multiplets & $U(1)$~charge & $U(1)_R$~charge & twisted masses\cr
\noalign{\hrule height 1.2pt}
$\phi_i$, $i=1,\ldots,4$ & $+1$ & $0$ & $\tm_i$ \cr
\noalign{\hrule height 0.5pt}
$P$ & $-5$ & $2$ & $\tm_P$ \cr
\noalign{\hrule height 1.2pt}
}}}
\hfil
\caption{Chiral matter spectrum of the $U(1)$ gauged linear sigma model for the surface~$\mathbb{P}^3[5]$ with ample canonical bundle.} \label{tab:SurP3[5]}
\end{table}
yields for a positive Fayet--Iliopoulos parameter $\xi$ the classical target space geometry $\mathbb{P}^3[5]$, which is a surface with ample canonical bundle. By an analysis similar as for the previous example we determine the generic south pole correlator relation 
\begin{multline}
   R_S(Q,\tm_i,\tm_P,\epsilon,\kappa_{n,\,\cdot\,}) \,=\, 
  \left\langle \sigma_N^n  (\sigma_S+\tm_1)\cdots(\sigma_S+\tm_4) \right\rangle \\
   + Q  \left\langle \sigma_N^n (5\sigma_S-\tm_P+\epsilon)\cdots (5\sigma_S -\tm_P+5\epsilon)\right\rangle \ .
\end{multline}
Since condition~\eqref{eq:OverlapCond} is satisfied in the limit of twisted masses, we further find the non-generic south pole correlator relation
\begin{equation}
   R_S(Q,\epsilon,\kappa_{n,\,\cdot\,}) \,=\, 
  \left\langle \sigma_N^n\sigma_S^3 \right\rangle 
   + 5 Q  \left\langle \sigma_N^n (5\sigma_S+\epsilon) (5\sigma_S+2\epsilon)  (5\sigma_S +3\epsilon)(5\sigma_S +4\epsilon)\right\rangle \ .
\end{equation}

\section{Conclusions} \label{sec:Conc}
In this work we studied the structure and the geometric aspects of certain gauge theory correlators of two-dimensional $\mathcal{N}=(2,2)$ supersymmetric gauged linear sigma model, as defined and computed in ref.~\cite{Closset:2015rna} via localization techniques on the two sphere $S^2$ in a certain (off-shell) supergravity background. We established universal and non-trivial correlator relations among such correlators, which in a Hilbert space interpretation gave rise to a differential ideal annihilating the ground state of the gauge theory. Using only the gauge theory spectrum as an input, we offered a straightforward procedure to directly compute these correlator relations and their associated differential operators. The resulting differential equations conformed with a particular example of differential operators already derived in ref.~\cite{Closset:2015rna} and with quantum A-periods considered by Hori and Vafa \cite{Hori:2000kt}. This is expected from the close relationship between the performed gauge theory computation in this work and the A-twisted gauged linear sigma model employed in the context of mirror symmetry in ref.~\cite{Hori:2000kt}.

By matching the considered correlation functions to a quadratic pairing of the Givental $I$-function --- as proposed in refs.~\cite{Ueda:2016wfa,Kim:2016uq} --- we argued that the Givental $I$-function is also in the kernel of the derived differential ideal. This observation and the connection to the Hori--Vafa A-periods linked the analyzed gauge theory correlators to the quantum cohomology rings of the gauge theory target spaces. 

For two-dimensional $\mathcal{N}=(2,2)$ supersymmetric gauged linear sigma model with an IR limit to $\mathcal{N}=(2,2)$ superconformal field theories, we found that the computed differential operators were (in a geometric phase) the Picard--Fuchs operators of the associated Calabi--Yau target space. Furthermore, for particular classes of Calabi--Yau geometries --- namely for elliptic curves, for polarized K3 surfaces with a single K\"ahler modulus, for Calabi--Yau threefolds with up to two K\"ahler moduli, and for Calabi--Yau fourfolds with a single K\"ahler modulus with both order five and order six Picard--Fuchs operators --- we derived universal formulas for the coefficients of their Picard--Fuchs operators in terms of gauge theory correlators.

We exemplified our results and exhibited the connection to the quantum cohomology ring for various target space geometries, arising from both Abelian and non-Abelian gauged linear sigma models. For instance, we showed in an example how the gauge theory relations were in one-to-one correspondence with the quantum cohomology ring of complex Grassmannian target space Fano varieties. We demonstrated that for Calabi--Yau threefolds arising from non-Abelian gauged linear sigma models there are non-trivial IR quantum relations among the gauge theory operators, as dictated by the chiral ring structure of the conformal field theories. Moreover, we illustrated that the derived universal Picard--Fuchs operators for classes of Calabi--Yau geometries were in agreement with those studied in the explicit examples.

Let us finally point out some applications, open questions, and future research directions. We believe that the presented methods offer a powerful tool to explicitly compute the differential equations governing the quantum cohomology of the target space geometries, both for complete intersections in toric varieties and in more general GIT quotients. Furthermore, comparing correlator relations among dual $\mathcal{N}=(2,2)$ gauge theories --- as considered for instance in refs.~\cite{Hori:2006dk,Hori:2011pd,Closset:2017vvl} --- can provide a non-trivial duality check. The derived universal forms of Picard--Fuchs operators in terms of correlators together with the exhibited differential-algebraic relations among their coefficient functions may serve as a starting point to classify differential operators for classes of Calabi--Yau geometries --- as already performed for Calabi--Yau threefolds with a single K\"ahler modulus by Almkvist, van Enckevort, van Straten, and Zudilin \cite{AESZ}. It would be interesting to examine if a similar universal structure of differential operators can also be determined for other target space geometries beyond the Calabi--Yau case. Finally, for non-Abelian $\mathcal{N}=(2,2)$ gauge theories we have focused on the derivation of differential operators arising from gauge invariant operators that are linear in the twisted chiral fields. We still would like to determine differential operators also involving non-linear gauge invariant twisted chiral fields and to work out their geometric meaning on the level of the Givental $I$-function.

\section*{Acknowledgements}
We would like to thank
Cyril Closset,
Stefano Cremonesi,
Ken Intriligator,
Bumsig Kim,
Albrecht Klemm,
Peter Mayr,
Ronen Plesser,
Thorsten Schimannek
and
Cumrum Vafa
for discussions and correspondences. A.G. and U.N. are supported by the graduate school BCGS, and A.G. is supported by the Studienstiftung des deutschen Volkes.

\bigskip
\bigskip

\appendix

\section{Calabi--Yau target spaces} \label{app:CY}
In this part of the appendix we cover some rather technical aspects of the discussion associated to Section~\ref{sec:CYtarget}. First, we present the derivation of the formulas for the two generating Picard--Fuchs operators in the case of two parameter Calabi--Yau threefolds. Second, we consider one-parameter Calabi--Yau fourfolds with order six Picard--Fuchs operators. 

\subsection{Two parameter Calabi--Yau threefolds}\label{App:3Fold2Param}
We here give a detailed derivation of the formulas~\eqref{eq:3Fold2ParamL2SolveMainText} and \eqref{eq:CY32L3MainText}, which express the two generating Picard--Fuchs operators associated to Calabi--Yau threefolds with two K\"ahler parameters in terms of A-twisted correlators. In doing so, we will use the notation introduced in subsection~\ref{sec:CY32Param}.

\paragraph{The order $2$ operator:}
We first consider the order two operator $\mathcal{L}^{(2)}$ separately. The natural candidate of a matrix $M(I,J)$, from which we would try to determine $c^{(2)}$, is $M(I^{(2)},I^{(2)})$. Due to the selection rule, however, the first column  of this matrix is zero. The condition $M(I^{(2)},I^{(2)}) \cdot c^{(2)} = 0$ is therefore independent of $c^{(2)}_{(0,0)}$ and can not determine $\mathcal{L}^{(2)}$ entirely.

This leads us to consider more general matrices of the type $M^{(2)}_{a,b}=M(J^{(2)}_{a,b},I^{(2)})$, where $J^{(2)}_{a,b} =  \{(0, 1), (1, 0), (0, 2), (1, 1), (2, 0), (a, b)\}$ with $a,b \geq 0$ chosen such that the correlator $\kappa_{(a,b),(0,0)} =(-1)^{1+a+b} \kappa_{(0,0),(a,b)}$ and hence the first column of $M^{(2)}_{a,b}$ is non-zero at generic points in moduli space. As we saw above, there indeed exists $0\leq a^\prime \leq 3$ for which $\kappa_{(0,0),(a^\prime,3-a^\prime)}$ is guaranteed to not vanish identically. We can thus choose to work with $M^{(2)}_{a^\prime,3-a^\prime}$. For definiteness, we here assume $\kappa_{(0,0),(0,3)}$ to not vanish at generic points and consider the $6\times 6$ matrix $M_1 = M^{(2)}_{0,3}$,
\begin{equation}
M_1 =
\left(
\begin{array}{cccccc}
 0 & 0 & 0 & -\kappa _{(0,0),(0,3)} & -\kappa _{(0,0),(1,2)} & -\kappa _{(0,0),(2,1)} \\
 0 & 0 & 0 & -\kappa _{(0,0),(1,2)} & -\kappa _{(0,0),(2,1)} & -\kappa _{(0,0),(3,0)} \\
 0 & \kappa _{(0,0),(0,3)} & \kappa _{(0,0),(1,2)} & 0 & \kappa _{(0,2),(1,1)} & \kappa _{(0,2),(2,0)} \\
 0 & \kappa _{(0,0),(1,2)} & \kappa _{(0,0),(2,1)} & -\kappa _{(0,2),(1,1)} & 0 & \kappa _{(1,1),(2,0)} \\
 0 & \kappa _{(0,0),(2,1)} & \kappa _{(0,0),(3,0)} & -\kappa _{(0,2),(2,0)} & -\kappa _{(1,1),(2,0)} & 0 \\
 -\kappa _{(0,0),(0,3)} & \kappa _{(0,1),(0,3)} & -\kappa _{(0,3),(1,0)} & -\kappa _{(0,2),(0,3)} & -\kappa _{(0,3),(1,1)} & -\kappa _{(0,3),(2,0)}
\end{array}
\right)\ .
\end{equation}
The kernel of this matrix is generically one-dimensional --- with or without using the derivative rule --- and its up to scaling unique kernel element is given by
\begin{equation}\label{eq:3Fold2ParamL2Solve}
\begin{aligned}
 c^{(2)}_{(0,0)} = &+\kappa_{(0,0),(1,2)}\Big(\kappa _{(0,0),(3,0)} \kappa _{(0,2),(0,3)}+\kappa _{(0,2),(2,0)} \kappa _{(0,3),(1,0)}-\kappa _{(0,1),(0,3)} \kappa _{(1,1),(2,0)}\Big) \\
 &+\kappa _{(0,0),(2,1)}
   \Big(\kappa _{(0,1),(0,3)} \kappa _{(0,2),(2,0)}-\kappa _{(0,2),(1,1)} \kappa _{(0,3),(1,0)}+\kappa _{(0,0),(1,2)} \kappa _{(0,3),(1,1)} \\
 &+\kappa _{(0,0),(0,3)} \kappa _{(0,3),(2,0)}\Big)-\kappa _{(0,3),(2,0)} \kappa _{(0,0),(1,2)}^2-\kappa _{(0,0),(2,1)}^2 \kappa _{(0,2),(0,3)}\\ 
  &-\kappa _{(0,0),(3,0)} \kappa _{(0,1),(0,3)} \kappa _{(0,2),(1,1)}-\kappa _{(0,0),(0,3)} \kappa _{(0,0),(3,0)} \kappa _{(0,3),(1,1)} \\
 &-\kappa _{(0,0),(0,3)}
   \kappa _{(0,3),(1,0)} \kappa _{(1,1),(2,0)}\\
 c^{(2)}_{(0,1)} = &-\kappa _{(0,0),(0,3)} \Big(\kappa _{(0,0),(3,0)} \kappa _{(0,2),(1,1)}-\kappa _{(0,0),(2,1)} \kappa _{(0,2),(2,0)}+\kappa _{(0,0),(1,2)} \kappa _{(1,1),(2,0)}\Big) \\
 c^{(2)}_{(1,0)} = &+ \kappa _{(0,0),(0,3)} \Big(\kappa _{(0,0),(2,1)} \kappa _{(0,2),(1,1)}-\kappa _{(0,0),(1,2)} \kappa _{(0,2),(2,0)}+\kappa _{(0,0),(0,3)} \kappa _{(1,1),(2,0)}\Big) \\
  c^{(2)}_{(0,2)} = &+\kappa _{(0,0),(0,3)} \Big(\kappa _{(0,0),(2,1)}^2-\kappa _{(0,0),(1,2)} \kappa _{(0,0),(3,0)}\Big)\\
 c^{(2)}_{(1,1)} = &+\kappa _{(0,0),(0,3)} \Big(\kappa _{(0,0),(0,3)} \kappa _{(0,0),(3,0)}-\kappa _{(0,0),(1,2)} \kappa _{(0,0),(2,1)}\Big) \\
 c^{(2)}_{(2,0)} = &+\kappa _{(0,0),(0,3)} \Big(\kappa _{(0,0),(1,2)}^2-\kappa _{(0,0),(0,3)} \kappa _{(0,0),(2,1)}\Big)\ .
\end{aligned}
\end{equation}
These coefficient functions determine $\mathcal{L}^{(2)}$ according to eq.~\eqref{eq:LExpandCY32}. With the formulas listed in Table~\ref{tab:RepList} this expresses $\mathcal{L}^{(2)}$ in terms of only seven correlators, namely $\kappa_{(0,0),(a,3-a)}$ with $0 \leq a \leq 3$ and $\kappa_{(0,0),(b,5-b)}$ with $0 \leq b \leq 2$.\

The above discussion is valid only as long as the kernel of $M_1$ is not of dimension two or higher. Upon using the derivative relations and recalling the assumption that $\kappa _{(0,0),(0,3)}\neq 0$, this requirement is equivalent to
\begin{equation}\label{eq:3Fold2ParamNonGeneric}
\begin{aligned}
\kappa _{(0,0),(0,3)}\kappa _{(0,0),(2,1)} &\neq\kappa _{(0,0),(1,2)}^2 \\ \text{or} \quad \kappa _{(0,0),(3,0)}\kappa _{(0,0),(1,2)} &\neq \kappa_{(0,0),(2,1)}^2 \\ \text{or} \quad \kappa _{(0,0),(3,0)}\kappa _{(0,0),(0,3)} &\neq \kappa _{(0,0),(1,2)} \kappa_{(0,0),(2,1)}
\end{aligned}
\end{equation}
This is in turn equivalent to at least one of the $c^{(2)}_{a,2-a}$ for $a\leq 0 \leq 2$ given in eq.~\eqref{eq:3Fold2ParamL2Solve} being non-zero. Hence, $\mathcal{L}^{(2)}$ is in fact of order two.

Further, consider cases in which $\kappa _{(0,0),(0,3)}$ vanishes identically. There then has to be $0\leq \tilde{a} \leq 2$ with $\kappa_{(0,0),(\tilde{a},3-\tilde{a})} \neq 0$, and we could base our discussion on $M^{(2)}_{\tilde{a},3-\tilde{a}}$ instead. Such an analysis would be completely equivalent, its validity would again require eq.~\eqref{eq:3Fold2ParamNonGeneric}. In fact, one can always arrange for $\kappa _{(0,0),(0,3)}\neq 0$ by appropriately rotating the generators of the $U(1)^2$ factor in the gauge group.

\paragraph{The order $3$ operator:}
In order to determine the second generating operator, $\mathcal{L}^{(3)}$, we consider the antisymmetric $10\times 10$ matrix $M_2 =M(I^{(3)},I^{(3)})$. This appears to be the easiest choice of a matrix $M(I,J)$, for which the kernel condition $M(I,J) \cdot c^{(3)} = 0$ can potentially determine $c^{(3)}$ entirely.

By employing the basic properties~\eqref{eq:RelDerCY32Param}-\eqref{eq:SelecRuleCY32Param}, $M_2$ can be expressed in terms of only ten correlators, namely $\kappa_{(0,0),(a,b-a)}$ with $b=3,5$ and $0\leq a \leq b$. Even after doing so, $M_2$ remains to be of full rank. Recall, however, that the three vectors corresponding to $\mathcal{L}^{(2)}$, $\Theta_1\mathcal{L}^{(2)}$ and $\Theta_2\mathcal{L}^{(2)}$ have to be in the kernel of $M_2$ as well. This requirement imposes additional constraints on the correlators, which lead to the kernel of $M_2$ being at least three-dimensional. Due to its antisymmetry the rank of $M_2$ has to be even, we thus find its kernel to be at least four-dimensional. Note further that $M_1$ is submatrix of $M_2$, which by the assumptions made in the above paragraph is of rank five (at generic points in moduli space). The rank of $M_2$ is therefore found to be six, and the fourth kernel element needs to be $c^{(3)}$. Let us now carry out this procedure explicitly.

First, we demand that the vector corresponding to $\mathcal{L}^{(2)}$ is in the kernel of $M_2$. This imposes the three additional constraints $A_1 = A_2 = A_3 = 0$, where
{\begin{footnotesize}
\begin{equation}
\begin{aligned}
 A_1 = & +\kappa _{(0,3),(2,0)} \kappa _{(0,0),(1,2)}^3-\big(\kappa _{(0,0),(3,0)} \kappa _{(0,2),(0,3)}+\kappa _{(0,2),(2,0)} \kappa _{(0,3),(1,0)}+\kappa _{(0,0),(2,1)} \kappa _{(0,3),(1,1)} \\
 &-\kappa _{(0,1),(0,3)}\kappa _{(1,1),(2,0)}+\kappa _{(0,0),(0,3)} \kappa _{(1,2),(2,0)}\big) \kappa _{(0,0),(1,2)}^2+\big\{\kappa _{(0,2),(0,3)} \kappa _{(0,0),(2,1)}^2 \\
 &+\big[-\kappa _{(0,1),(0,3)} \kappa
   _{(0,2),(2,0)}+\kappa _{(0,2),(1,1)} \kappa _{(0,3),(1,0)}+\kappa _{(0,0),(0,3)} \big(\kappa _{(1,1),(1,2)}-\kappa _{(0,3),(2,0)}\big)\big] \kappa _{(0,0),(2,1)}\\
   &+\kappa _{(0,0),(3,0)} \big[\kappa
   _{(0,1),(0,3)} \kappa _{(0,2),(1,1)}+\kappa _{(0,0),(0,3)} \big(\kappa _{(0,2),(1,2)}+\kappa _{(0,3),(1,1)}\big)\big]\\
   &-\kappa _{(0,0),(0,3)} \big[\kappa _{(0,2),(2,0)} \kappa
   _{(1,0),(1,2)}+\big(\kappa _{(0,1),(1,2)}-\kappa _{(0,3),(1,0)}\big) \kappa _{(1,1),(2,0)}\big]\big\} \kappa_{(0,0),(1,2)}\\
   &+\kappa _{(0,0),(0,3)} \big[-\kappa _{(0,2),(1,2)} \kappa
   _{(0,0),(2,1)}^2+\big(\kappa _{(0,1),(1,2)} \kappa _{(0,2),(2,0)}+\kappa _{(0,2),(1,1)} \kappa _{(1,0),(1,2)}\\
   &+\kappa _{(0,0),(0,3)} \kappa _{(1,2),(2,0)}\big) \kappa _{(0,0),(2,1)}-\kappa _{(0,0),(3,0)}
   \big(\kappa _{(0,1),(1,2)} \kappa _{(0,2),(1,1)}+\kappa _{(0,0),(0,3)} \kappa _{(1,1),(1,2)}\big)\\
   &+\kappa _{(0,0),(0,3)} \kappa _{(1,0),(1,2)} \kappa _{(1,1),(2,0)}\big] \ ,\\[0.2em]
A_2 = & +\kappa _{(0,2),(0,3)} \kappa _{(0,0),(2,1)}^3-\big[\kappa _{(0,1),(0,3)} \kappa _{(0,2),(2,0)}-\kappa _{(0,2),(1,1)} \kappa _{(0,3),(1,0)}+\kappa _{(0,0),(1,2)} \kappa _{(0,3),(1,1)}\\
&+\kappa _{(0,0),(0,3)}
   \big(\kappa _{(0,2),(2,1)}+\kappa _{(0,3),(2,0)}\big)\big] \kappa _{(0,0),(2,1)}^2+\big[\kappa _{(0,3),(2,0)} \kappa _{(0,0),(1,2)}^2\\
   &+\big(-\kappa _{(0,0),(3,0)} \kappa _{(0,2),(0,3)}-\kappa
   _{(0,2),(2,0)} \kappa _{(0,3),(1,0)}+\kappa _{(0,1),(0,3)} \kappa _{(1,1),(2,0)}\\
   &+\kappa _{(0,0),(0,3)} \kappa _{(1,1),(2,1)}\big) \kappa _{(0,0),(1,2)}+\kappa _{(0,0),(3,0)} \big(\kappa _{(0,1),(0,3)}
   \kappa _{(0,2),(1,1)}+\kappa _{(0,0),(0,3)} \kappa _{(0,3),(1,1)}\big)\\
   &+\kappa _{(0,0),(0,3)} \big(\kappa _{(0,1),(2,1)} \kappa _{(0,2),(2,0)}+\kappa _{(0,2),(1,1)} \kappa _{(1,0),(2,1)}+\kappa
   _{(0,3),(1,0)} \kappa _{(1,1),(2,0)}\\
   &+\kappa _{(0,0),(0,3)} \kappa _{(2,0),(2,1)}\big)\big] \kappa _{(0,0),(2,1)}-\kappa _{(0,0),(0,3)} \big[\kappa _{(2,0),(2,1)} \kappa _{(0,0),(1,2)}^2+\big(\kappa
   _{(0,2),(2,0)} \kappa _{(1,0),(2,1)}\\
   &+\kappa _{(0,1),(2,1)} \kappa _{(1,1),(2,0)}\big) \kappa _{(0,0),(1,2)}-\kappa _{(0,0),(0,3)} \kappa _{(1,0),(2,1)} \kappa _{(1,1),(2,0}\\
   &+\kappa _{(0,0),(3,0)}
   \big(\kappa _{(0,1),(2,1)} \kappa _{(0,2),(1,1)}-\kappa _{(0,0),(1,2)} \kappa _{(0,2),(2,1)}+\kappa _{(0,0),(0,3)} \kappa _{(1,1),(2,1)}\big)\big]\ ,\\[0.2em]
A_3 = &-\kappa _{(0,0),(0,3)} \big(\kappa _{(0,0),(2,1)}^2-\kappa _{(0,0),(1,2)} \kappa _{(0,0),(3,0)}\big) \kappa _{(0,2),(3,0)}\\
&+\kappa _{(0,0),(0,3)} \kappa _{(1,0),(3,0)} \big(\kappa _{(0,0),(2,1)} \kappa
   _{(0,2),(1,1)}-\kappa _{(0,0),(1,2)} \kappa _{(0,2),(2,0)}+\kappa _{(0,0),(0,3)} \kappa _{(1,1),(2,0)}\big)\\
   &-\kappa _{(0,0),(0,3)} \kappa _{(0,1),(3,0)} \big(\kappa _{(0,0),(3,0)} \kappa
   _{(0,2),(1,1)}-\kappa _{(0,0),(2,1)} \kappa _{(0,2),(2,0)}+\kappa _{(0,0),(1,2)} \kappa _{(1,1),(2,0)}\big)\\
   &-\kappa _{(0,0),(3,0)} \big[-\kappa _{(0,3),(2,0)} \kappa _{(0,0),(1,2)}^2+\big(\kappa
   _{(0,0),(3,0)} \kappa _{(0,2),(0,3)}+\kappa _{(0,2),(2,0)} \kappa _{(0,3),(1,0)}\\
   &-\kappa _{(0,1),(0,3)} \kappa _{(1,1),(2,0)}\big) \kappa _{(0,0),(1,2)}-\kappa _{(0,0),(2,1)}^2 \kappa _{(0,2),(0,3)}-\kappa
   _{(0,0),(3,0)} \kappa _{(0,1),(0,3)} \kappa _{(0,2),(1,1)}\\
   &-\kappa _{(0,0),(0,3)} \kappa _{(0,0),(3,0)} \kappa _{(0,3),(1,1)}+\kappa _{(0,0),(2,1)} \big(\kappa _{(0,1),(0,3)} \kappa _{(0,2),(2,0)}-\kappa
   _{(0,2),(1,1)} \kappa _{(0,3),(1,0)}\\
   &+\kappa _{(0,0),(1,2)} \kappa _{(0,3),(1,1)}+\kappa _{(0,0),(0,3)} \kappa _{(0,3),(2,0)}\big)-\kappa _{(0,0),(0,3)} \kappa _{(0,3),(1,0)} \kappa
   _{(1,1),(2,0)}\big]\\
   &-\kappa _{(0,0),(0,3)} \big(\kappa _{(0,0),(0,3)} \kappa _{(0,0),(3,0)}-\kappa _{(0,0),(1,2)} \kappa _{(0,0),(2,1)}\big) \kappa _{(1,1),(3,0)}\\
   &+\kappa _{(0,0),(0,3)} \big(\kappa
   _{(0,0),(0,3)} \kappa _{(0,0),(2,1)}-\kappa _{(0,0),(1,2)}^2\big) \kappa _{(2,0),(3,0)}\ .
\end{aligned}\label{eq:CY32ParamAddConstraint1}
\end{equation}
\end{footnotesize}}
Upon insertion of the formulas listed in Table~\ref{tab:RepList} these expressions become linear in the independent five-point correlators $\kappa _{(0,0),(a,5-a)}$ with $0\leq a \leq 5$. We then make the simplifying assumption that
\begin{equation}\label{eq:CY32ParamAssumption}
\kappa _{(0,0),(0,3)}\kappa _{(0,0),(2,1)} \neq \kappa _{(0,0),(1,2)}^2  \ ,
\end{equation}
in case of which $A_1 = A_2 = A_3 = 0$ can be solved for $\kappa _{(0,0),(a,5-a)}$ with $a=3,4,5$. Note that assumption~\eqref{eq:CY32ParamAssumption} implies the previous assumption~\eqref{eq:3Fold2ParamNonGeneric}.

Second, we further demand that the vectors corresponding to $\Theta_1\mathcal{L}^{(2)}$ and $\Theta_2\mathcal{L}^{(2)}$ are in the kernel of $M_2$ as well. Given $A_i = 0$ and $\Theta_j A_i = 0$ (with $i=1,2,3$ and $j=1,2$), this imposes two additional, independent constraints. We choose these to be $B_1 = B_2 = 0$ with
\begin{equation}\label{eq:CY32ParamAddConstraint2}
\begin{aligned}
B_1 = &+ \kappa _{(0,0),(0,3)} \big[-\kappa _{(1,3),(2,0)} \kappa _{(0,0),(1,2)}^2+\kappa _{(0,2),(2,0)} \kappa _{(1,0),(1,3)} \kappa _{(0,0),(1,2)}\\
&+\kappa _{(0,1),(1,3)} \kappa _{(1,1),(2,0)} \kappa
   _{(0,0),(1,2)}+\kappa _{(0,0),(2,1)}^2 \kappa _{(0,2),(1,3)}+\kappa _{(0,0),(3,0)} \big(\kappa _{(0,1),(1,3)} \kappa _{(0,2),(1,1)}\\
&-\kappa _{(0,0),(1,2)} \kappa _{(0,2),(1,3)}+\kappa _{(0,0),(0,3)} \kappa
   _{(1,1),(1,3)}\big)-\kappa _{(0,0),(0,3)} \kappa _{(1,0),(1,3)} \kappa _{(1,1),(2,0)}\\
&-\kappa _{(0,0),(2,1)} \big(\kappa _{(0,1),(1,3)} \kappa _{(0,2),(2,0)}+\kappa _{(0,2),(1,1)} \kappa
   _{(1,0),(1,3)}+\kappa _{(0,0),(1,2)} \kappa _{(1,1),(1,3)}\\
&-\kappa _{(0,0),(0,3)} \kappa _{(1,3),(2,0)}\big)\big]-\kappa _{(0,0),(1,3)} \big[\kappa _{(0,3),(2,0)} \kappa _{(0,0),(1,2)}^2-\big(\kappa
   _{(0,0),(3,0)} \kappa _{(0,2),(0,3)}\\
&+\kappa _{(0,2),(2,0)} \kappa _{(0,3),(1,0)}-\kappa _{(0,1),(0,3)} \kappa _{(1,1),(2,0)}\big) \kappa _{(0,0),(1,2)}+\kappa _{(0,0),(2,1)}^2 \kappa _{(0,2),(0,3)}\\
&+\kappa
   _{(0,0),(3,0)} \kappa _{(0,1),(0,3)} \kappa _{(0,2),(1,1)}+\kappa _{(0,0),(0,3)} \kappa _{(0,0),(3,0)} \kappa _{(0,3),(1,1)}-\kappa _{(0,0),(2,1)} \big(\kappa _{(0,1),(0,3)} \kappa _{(0,2),(2,0)}\\
&-\kappa
   _{(0,2),(1,1)} \kappa _{(0,3),(1,0)}+\kappa _{(0,0),(1,2)} \kappa _{(0,3),(1,1)}+\kappa _{(0,0),(0,3)} \kappa _{(0,3),(2,0)}\big)\\
&+\kappa _{(0,0),(0,3)} \kappa _{(0,3),(1,0)} \kappa _{(1,1),(2,0)}\big] \\[0.2em]
B_2 = &+ \kappa _{(0,0),(0,3)} \big[-\kappa _{(0,4),(2,0)} \kappa _{(0,0),(1,2)}^2+\big(-\kappa _{(0,0),(3,0)} \kappa _{(0,2),(0,4)}+\kappa _{(0,2),(2,0)} \kappa _{(0,4),(1,0)}\\
&+\kappa _{(0,1),(0,4)} \kappa
   _{(1,1),(2,0)}\big) \kappa _{(0,0),(1,2)}+\kappa _{(0,0),(2,1)}^2 \kappa _{(0,2),(0,4)}+\kappa _{(0,0),(3,0)} \kappa _{(0,1),(0,4)} \kappa _{(0,2),(1,1)}\\
&-\kappa _{(0,0),(0,3)} \kappa _{(0,0),(3,0)} \kappa
   _{(0,4),(1,1)}+\kappa _{(0,0),(2,1)} \big(-\kappa _{(0,1),(0,4)} \kappa _{(0,2),(2,0)}-\kappa _{(0,2),(1,1)} \kappa _{(0,4),(1,0)}\\
&+\kappa _{(0,0),(1,2)} \kappa _{(0,4),(1,1)}+\kappa _{(0,0),(0,3)} \kappa
   _{(0,4),(2,0)}\big)-\kappa _{(0,0),(0,3)} \kappa _{(0,4),(1,0)} \kappa _{(1,1),(2,0)}\big]\\
&-\kappa _{(0,0),(0,4)} \big[\kappa _{(0,3),(2,0)} \kappa _{(0,0),(1,2)}^2-\big(\kappa _{(0,0),(3,0)} \kappa
   _{(0,2),(0,3)}+\kappa _{(0,2),(2,0)} \kappa _{(0,3),(1,0)}\\
&-\kappa _{(0,1),(0,3)} \kappa _{(1,1),(2,0)}\big) \kappa _{(0,0),(1,2)}+\kappa _{(0,0),(2,1)}^2 \kappa _{(0,2),(0,3)}+\kappa _{(0,0),(3,0)} \kappa
   _{(0,1),(0,3)} \kappa _{(0,2),(1,1)}\\
&+\kappa _{(0,0),(0,3)} \kappa _{(0,0),(3,0)} \kappa _{(0,3),(1,1)}-\kappa _{(0,0),(2,1)} \big(\kappa _{(0,1),(0,3)} \kappa _{(0,2),(2,0)}-\kappa _{(0,2),(1,1)} \kappa
   _{(0,3),(1,0)}\\
&+\kappa _{(0,0),(1,2)} \kappa _{(0,3),(1,1)}+\kappa _{(0,0),(0,3)} \kappa _{(0,3),(2,0)}\big)+\kappa _{(0,0),(0,3)} \kappa _{(0,3),(1,0)} \kappa _{(1,1),(2,0)}\big] \ .
\end{aligned}
\end{equation}
Since these expressions involve six-point correlators, insertion of the formulas listed in Table~\ref{tab:RepList} turns them into first order (linear partial) diffential equations for the remaining five-point correlators $\kappa _{(0,0),(a,5-a)}$ with $a=0,1,2$. The combination of all five additional constraints, $A_1 = A_2 = A_3 =0$ and $B_1 = B_2= 0$, finally implies that the vector $c^{(3)}$ given by
\begin{equation}\label{eq:CY32ParamC3}
\begin{aligned}
c^{(3)}_{(0,0)} = 
&+\kappa _{(0,3),(1,2)} \kappa _{(0,0),(1,2)}^2-\kappa _{(0,2),(1,2)} \kappa _{(0,3),(1,0)} \kappa _{(0,0),(1,2)}-\kappa _{(0,2),(0,3)} \kappa _{(1,0),(1,2)} \kappa _{(0,0),(1,2)}\\
&+\kappa _{(0,1),(0,3)} \kappa
   _{(1,1),(1,2)} \kappa _{(0,0),(1,2)}+\kappa _{(0,1),(1,2)} \left(\kappa _{(0,2),(1,1)} \kappa _{(0,3),(1,0)}-\kappa _{(0,0),(1,2)} \kappa _{(0,3),(1,1)}\right)\\
&+\kappa _{(0,0),(2,1)} \left(\kappa
   _{(0,1),(1,2)} \kappa _{(0,2),(0,3)}-\kappa _{(0,1),(0,3)} \kappa _{(0,2),(1,2)}-\kappa _{(0,0),(0,3)} \kappa _{(0,3),(1,2)}\right)\\
&+\kappa _{(0,1),(0,3)} \kappa _{(0,2),(1,1)} \kappa _{(1,0),(1,2)}+\kappa
   _{(0,0),(0,3)} \kappa _{(0,3),(1,1)} \kappa _{(1,0),(1,2)}\\
&+\kappa _{(0,0),(0,3)} \kappa _{(0,3),(1,0)} \kappa _{(1,1),(1,2)}\\
c^{(3)}_{(0,1)}=&+\kappa_{(0,0),(1,2)}\left(\kappa _{(0,0),(2,1)} \kappa _{(0,2),(0,3)}+\kappa _{(0,2),(1,1)} \kappa _{(0,3),(1,0)}+\kappa _{(0,0),(0,3)} \kappa _{(1,1),(1,2)}\right) \\
&+\kappa _{(0,0),(0,3)} \left(\kappa _{(0,2),(1,1)} \kappa _{(1,0),(1,2)}-\kappa _{(0,0),(2,1)} \kappa _{(0,2),(1,2)}\right)-\kappa _{(0,3),(1,1)} \kappa _{(0,0),(1,2)}^2\\
c^{(3)}_{(1,0)}=&+\kappa_{(0,0),(1,2)}\left(\kappa _{(0,1),(0,3)} \kappa _{(0,2),(1,1)}+\kappa _{(0,0),(0,3)} \left(\kappa _{(0,2),(1,2)}+\kappa _{(0,3),(1,1)}\right)\right)\\
&-\kappa _{(0,0),(0,3)} \left(\kappa _{(0,1),(1,2)} \kappa _{(0,2),(1,1)}+\kappa _{(0,0),(0,3)} \kappa _{(1,1),(1,2)}\right)-\kappa _{(0,2),(0,3)} \kappa _{(0,0),(1,2)}^2\\
c^{(3)}_{(0,2)}=&+\kappa _{(0,3),(1,0)} \kappa _{(0,0),(1,2)}^2+\kappa _{(0,0),(1,2)}\left(\kappa _{(0,0),(2,1)} \kappa _{(0,1),(0,3)}+\kappa _{(0,0),(0,3)} \kappa _{(1,0),(1,2)}\right)\\
&-\kappa _{(0,0),(0,3)} \kappa
   _{(0,0),(2,1)} \kappa _{(0,1),(1,2)}\\
c^{(3)}_{(1,1)}=&-\kappa _{(1,0),(1,2)} \kappa _{(0,0),(0,3)}^2+\kappa _{(0,0),(1,2)}\kappa _{(0,0),(0,3)} \left(\kappa _{(0,1),(1,2)}-\kappa _{(0,3),(1,0)}\right) \\
&-\kappa _{(0,0),(1,2)}^2 \kappa _{(0,1),(0,3)}\\
c^{(3)}_{(2,0)}= &\,\, 0\\
c^{(3)}_{(0,3)}=&+\kappa _{(0,0),(1,2)}\left(\kappa _{(0,0),(1,2)}^2-\kappa _{(0,0),(0,3)}\kappa _{(0,0),(2,1)}\right)\\
c^{(3)}_{(1,2)}=&+\kappa _{(0,0),(0,3)} \left(\kappa _{(0,0),(0,3)} \kappa _{(0,0),(2,1)}-\kappa _{(0,0),(1,2)}^2\right)\\
c^{(3)}_{(2,1)} = &\,\,0\\
c^{(3)}_{(3,0)}= &\,\,0
\end{aligned}
\end{equation}
is in the kernel of $M_2$. Due to assumption~\eqref{eq:CY32ParamAssumption}, the corresponding operator is indeed of order three and can not be a linear combination of $\mathcal{L}^{(2)}$ and its derivatives. Therefore, these coefficient functions indeed constitute the order three operator $\mathcal{L}^{(3)}$.

Together with the formulas listed in Table~\ref{tab:RepList} this expresses $\mathcal{L}^{(3)}$ in terms of the same seven correlators required to determine $\mathcal{L}^{(2)}$, namely $\kappa_{(0,0),(a,3-a)}$ with $0 \leq a \leq 3$ and $\kappa_{(0,0),(b,5-b)}$ with $0 \leq b \leq 2$.  Among these seven correlators there are the two additional constraints $B_1= B_2 = 0$, which we believe to futher uniquely fix one of the three remaining five-point correlators.

Lastly, consider cases in which assumption~\eqref{eq:CY32ParamAssumption} fails but assumption~\eqref{eq:3Fold2ParamNonGeneric} holds. Then, the second or the third inequality in~\eqref{eq:3Fold2ParamNonGeneric} needs to be true, and one could follow similar steps to determine different formulas for $\mathcal{L}^{(3)}$ that are valid in these cases. By an appropriate rotation of the generators of $U(1)^2$ one can, in fact, always arrange for assumption~\eqref{eq:3Fold2ParamNonGeneric} while preserving the initial assumption $\kappa_{(0,0),(0,3)}\neq 0$.

\begin{table}[htbp]
{\begin{footnotesize}
\begin{tabular}{ll}
$\kappa_{(0, 1),(0, 3)}  = \kappa_{(0, 0),(0, 3)}^{(0,1)} \ , $ &
$\kappa_{(0, 1),(1, 2)}  = \tfrac{1}{2}\kappa_{(0, 0),(0, 3)}^{(1,0)} +\tfrac{1}{2} \kappa_{(0, 0),(1, 2)}^{(0,1)}\ ,$ \\[0.6em]
$\kappa_{(0, 1),(2, 1)}  = \kappa_{(0, 0),(1, 2)}^{(1,0)} \ , $ &
$\kappa_{(0, 1),(3, 0)}  = \tfrac{3}{2}\kappa_{(0, 0),(2, 1)}^{(1,0)} - \tfrac{1}{2}\kappa_{(0, 0),(3,0)}^{(0,1)} \ ,$ \\[0.6em]
$\kappa_{(0, 2),(1, 1)}  = \tfrac{1}{2}\kappa_{(0, 0),(0, 3)}^{(1,0)} - \tfrac{1}{2}\kappa_{(0, 0),(1, 2)}^{(0,1)} \ , $ &
$\kappa_{(0, 2),(2, 0)}  = \kappa_{(0, 0),(1, 2)}^{(1,0)} - \kappa_{(0, 0),(2, 1)}^{(0,1)}\ , $ \\[0.6em]
$\kappa_{(0, 3),(1, 0)}  = \tfrac{1}{2}\kappa_{(0, 0),(0, 3)}^{(1,0)} - \tfrac{3}{2} \kappa_{(0, 0),(1, 2)}^{(0,1)} \ , $ &
$\kappa_{(1, 0),(1, 2)}  = \kappa_{(0, 0),(2, 1)}^{(0,1)}\ , $ \\[0.6em]
$\kappa_{(1, 0),(2, 1)}  = \tfrac{1}{2}\kappa_{(0, 0),(2, 1)}^{(1,0)} + \tfrac{1}{2} \kappa_{(0, 0),(3, 0)}^{(0,1)} \ , $ &
$\kappa_{(1, 0),(3, 0)}  = \kappa_{(0, 0),(3, 0)}^{(1,0)}\ , $ \\[0.6em]
$\kappa_{(1, 1),(2, 0)}  = \tfrac{1}{2}\kappa_{(0, 0),(2, 1)}^{(1,0)} - \tfrac{1}{2}\kappa_{(0, 0),(3, 0)}^{(0,1)} \ , \qquad \qquad $ &
$\kappa_{(0, 2),(0, 3)}  = -3 \kappa_{(0, 0),(0, 3)}^{(0,2)} + \kappa_{(0, 0),(0, 5)}\ , $ \\[0.6em]
\multicolumn{2}{l}{$\kappa_{(0, 2),(1, 2)}  = -\kappa_{(0, 0),(0, 3)}^{(1,1)} - 2 \kappa_{(0, 0),(1, 2)}^{(0,2)} + \kappa_{(0, 0),(1, 4)}\ , $} \\[0.6em]
\multicolumn{2}{l}{$\kappa_{(0, 2),(2, 1)}  = -2 \kappa_{(0, 0),(1, 2)}^{(1,1)} - \kappa_{(0, 0),(2, 1)}^{(0,2)} + \kappa_{(0, 0),(2, 3)}\ , $} \\[0.6em]
\multicolumn{2}{l}{$\kappa_{(0, 2),(3, 0)}  = -3 \kappa_{(0, 0),(2, 1)}^{(1,1)} + \kappa_{(0, 0),(3, 2)}\ , $ } \\[0.6em]
\multicolumn{2}{l}{$\kappa_{(0, 3),(1, 1)}  = - \tfrac{3}{2} \kappa_{(0, 0),(0, 3)}^{(1,1)} - \tfrac{3}{2}\kappa_{(0, 0),(1, 2)}^{(0,2)} + \kappa_{(0, 0),(1, 4)} \ ,$} \\[0.6em]
\multicolumn{2}{l}{$\kappa_{(0, 3),(2, 0)}  = -3 \kappa_{(0, 0),(1, 2)}^{(1,1)} + \kappa_{(0, 0),(2, 3)}\ , $ } \\[0.6em]
\multicolumn{2}{l}{$\kappa_{(1, 1),(1, 2)}  = -\tfrac{1}{2}\kappa_{(0, 0),(0, 3)}^{(2,0)} - \tfrac{3}{2} \kappa_{(0, 0),(1, 2)}^{(1,1)} - \kappa_{(0, 0),(2,1)}^{(0,2)} + \kappa_{(0, 0),(2, 3)} \ ,$} \\[0.6em]
\multicolumn{2}{l}{$\kappa_{(1, 1),(2, 1)}  = -\kappa_{(0, 0),(1, 2)}^{(2,0)} - \tfrac{3}{2}\kappa_{(0, 0),(2, 1)}^{(1,1)} -\tfrac{1}{2}\kappa_{(0, 0),(3,0)}^{(0,2)} + \kappa_{(0, 0),(3, 2)}\ ,$} \\[0.6em]
\multicolumn{2}{l}{$\kappa_{(1, 1),(3, 0)}  = -\tfrac{3}{2}\kappa_{(0, 0),(2, 1)}^{(2,0)} - \tfrac{3}{2}\kappa_{(0, 0),(3, 0)}^{(1,1)} + \kappa_{(0, 0),(4, 1)} \ ,$} \\[0.6em]
\multicolumn{2}{l}{$\kappa_{(1, 2),(2, 0)}  = -\kappa_{(0, 0),(1, 2)}^{(2,0)} - 2 \kappa_{(0, 0),(2, 1)}^{(1,1)} + \kappa_{(0, 0),(3, 2)} \ ,$}\\[0.6em]
\multicolumn{2}{l}{$\kappa_{(2, 0),(2, 1)}  = -2 \kappa_{(0, 0),(2, 1)}^{(2,0)} - \kappa_{(0, 0),(3, 0)}^{(1,1)} + \kappa_{(0, 0),(4, 1)}\ , $} \\[0.6em]
\multicolumn{2}{l}{$\kappa_{(2, 0),(3, 0)}  = -3 \kappa_{(0, 0),(3, 0)}^{(2,0)} + \kappa_{(0, 0),(5, 0)}\ ,$} \\[0.6em]
\multicolumn{2}{l}{$\kappa_{(0, 3),(1, 2)}  = -\kappa_{(0, 0),(0, 3)}^{(1,2)} + \tfrac{1}{2}\kappa_{(0, 0),(0, 5)}^{(1,0)} + \kappa_{(0, 0),(1, 2)}^{(0,3)} - \kappa_{(0, 0),(1, 4)}^{(0,1)} \ ,$}\\[0.6em]
\multicolumn{2}{l}{$\kappa_{(0, 3),(2, 1)}  = -\kappa_{(0, 0),(0, 3)}^{(2,1)} + \kappa_{(0, 0),(1, 4)}^{(1,0)} + \kappa_{(0, 0),(2, 1)}^{(0,3)} - \kappa_{(0, 0),(2, 3)}^{(0,1)}\ , $} \\[0.6em]
\multicolumn{2}{l}{$\kappa_{(0, 3),(3, 0)}  = -\tfrac{1}{4}\kappa_{(0, 0),(0, 3)}^{(3,0)} - \tfrac{9}{4} \kappa_{(0, 0),(1, 2)}^{(2,1)} + \tfrac{9}{4}\kappa_{(0, 0),(2, 1)}^{(1,2)} + \tfrac{3}{2}\kappa_{(0, 0),(2, 3)}^{(1,0)} + \tfrac{1}{4}\kappa_{(0, 0),(3,0)}^{(0,3)} - \tfrac{3}{2} \kappa_{(0, 0),(3, 2)}^{(0,1)} \ ,$}\\[0.6em]
\multicolumn{2}{l}{$\kappa_{(1, 2),(2, 1)}  = -\tfrac{1}{4}\kappa_{(0, 0),(0, 3)}^{(3,0)} - \tfrac{1}{4}\kappa_{(0, 0),(1, 2)}^{(2,1)} + \tfrac{1}{4}\kappa_{(0, 0),(2,1)}^{(1,2)} + \tfrac{1}{2}\kappa_{(0, 0),(2, 3)}^{(1,0)} + \tfrac{1}{4}\kappa_{(0,0),(3, 0)}^{(0,3)} - \tfrac{1}{2}\kappa_{(0, 0),(3, 2)}^{(0,1)} \ ,$} \\[0.6em]
\multicolumn{2}{l}{$\kappa_{(1, 2),(3, 0)}  = -\kappa_{(0, 0),(1, 2)}^{(3,0)} + \kappa_{(0, 0),(3, 0)}^{(1,2)} + \kappa_{(0, 0),(3, 2)}^{(1,0)} - \kappa_{(0, 0),(4, 1)}^{(0,1)}\ , $}\\[0.6em]
\multicolumn{2}{l}{$\kappa_{(2, 1),(3, 0)}  = -\kappa_{(0, 0),(2, 1)}^{(3,0)} + \kappa_{(0, 0),(3, 0)}^{(2,1)} + \tfrac{1}{2}\kappa_{(0, 0),(4, 1)}^{(1,0)} - \tfrac{1}{2}\kappa_{(0, 0),(5, 0)}^{(0,1)} \ . $}
\end{tabular}
\end{footnotesize}}
\caption{\label{tab:RepList} Formulas to express relevant four-, five- and six-point correlators in terms of $\kappa_{(0,0),(a,b-a)}$ with $b=3,5$ and $0\leq a \leq b$. We abbreviate $\kappa_{\vec{n},\vec{m}}^{(p,q)} = (\epsilon\Theta_1)^p (\epsilon \Theta_2)^q\kappa_{\vec{n},\vec{m}}$.}
\end{table}

\paragraph{Implications of the basic correlator properties:}
The basic properties~\eqref{eq:RelDerCY32Param}-\eqref{eq:SelecRuleCY32Param} allow to express all correlators in terms of those $\kappa_{\vec{n},\vec{m}}$ with $\vec{n} = 0$ and $|\vec{m}|_1\in 2\mathbb{N}+1$. For the case of three-point correlators we have made this explicit in eq.~\eqref{eq:3Fold2Param3PointRel}. In Table~\ref{tab:RepList} we list the corresponding formulas for those four-, five- and six-point correlators that appear in some of the above formulas.

\subsection{One parameter Calabi--Yau fourfolds with order six Picard--Fuchs operator}\label{App:4FoldN6}
As mentioned in Section~\ref{sec:CY4}, the order $N$ of the generating Picard--Fuchs associated to one parameter Calabi--Yau fourfolds is not universally fixed. While the minimal case $N=5$ was discussed in Section~\ref{sec:CY4}, we here consider the next to minimal case $N=6$. We refrain from disucssing cases with $N\geq 7$. 

We therefore define the index set $I^\prime =\{0, 1, 2, 3, 4, 5\}$ and consider the vector of coefficient functions $c^{(6)} = c(I^\prime)$, which determines the order six Picard--Fuchs operator $\mathcal{L}^{(6)}$ as
\begin{equation}\label{eq:LCY4ExpandN6}
\mathcal{L}^{(6)} \, = \, \sum_{k=0}^5 c^{(6)}_k \left(\epsilon \Theta\right)^k \ .
\end{equation} 
By assumption there is such $c^{(6)}$ and it needs to be in the kernel of the $7 \times 7$ matrix $M^{(6)}$, where
\begin{equation}
M^{(6)} = M(I^\prime,I^\prime) = \left(
\begin{array}{ccccccc}
 0 & 0 & 0 & 0 & \kappa _{0,4} & \kappa _{0,5} & \kappa _{0,6} \\
 0 & 0 & 0 & -\kappa _{0,4} & -\kappa _{1,4} & -\kappa _{1,5} & -\kappa _{1,6} \\
 0 & 0 & \kappa _{0,4} & \kappa _{2,3} & \kappa _{2,4} & \kappa _{2,5} & \kappa _{2,6} \\
 0 & -\kappa _{0,4} & \kappa _{2,3} & -\kappa _{3,3} & -\kappa _{3,4} & -\kappa _{3,5} & -\kappa _{3,6} \\
 \kappa _{0,4} & -\kappa _{1,4} & \kappa _{2,4} & -\kappa _{3,4} & \kappa _{4,4} & \kappa _{4,5} & \kappa _{4,6} \\
 \kappa _{0,5} & -\kappa _{1,5} & \kappa _{2,5} & -\kappa _{3,5} & \kappa _{4,5} & -\kappa _{5,5} & -\kappa _{5,6} \\
 \kappa _{0,6} & -\kappa _{1,6} & \kappa _{2,6} & -\kappa _{3,6} & \kappa _{4,6} & -\kappa _{5,6} & \kappa _{6,6} \\
\end{array}
\right) \ .
\end{equation}
Since $M^{(5)}$ is a submatrix of $M^{(6)}$, we find $\operatorname{rank} M^{(6)}\geq 5$ and $\operatorname{dim}\operatorname{ker} M^{(6)} \leq 2$ for generic points in moduli space. An explicit calculation shows
\begin{equation}\label{eq:CY4DetM6}
\begin{aligned}
\frac{\operatorname{det} M^{(6)}}{-\kappa_{0,4}^{5}} =
&+\left(\kappa_{6,6}+\alpha_1\right)\left(\kappa_{5,5}-\kappa_{5,5}^{N=5}\right) +\alpha_2\left(\kappa_{5,5}-\kappa_{5,5}^{N=5}\right)^2+\frac{1}{4}\left(\epsilon\Theta \,\kappa_{5,5} - \epsilon\Theta \, \kappa_{5,5}^{N=5}\right)^2 \ ,
\end{aligned}
\end{equation}
where $\alpha_1$ and $\alpha_2$ are non-zero for $\kappa_{5,5} = \kappa_{5,5}^{N=5}$. The kernel condition $\operatorname{det} M^{(6)} = 0$ is thus non-trivial and equivalent to
\begin{equation}\label{eq:CY4N6TwoSituations}
\begin{aligned}
&\kappa_{5,5} = \kappa_{5,5}^{N=5} \qquad &&\text{(degenerate situation)} \\
\text{or}\quad \big(& \,\kappa_{5,5} \neq \kappa_{5,5}^{N=5}\quad \text{and} \quad \kappa_{6,6} = \kappa_{6,6}^{N=6}\, \big)\qquad &&\text{(non-degenerate situation)} \ ,
\end{aligned}
\end{equation}
with
\begin{equation}\label{eq:CY4k66N6}
\begin{aligned}
\kappa_{6,6}^{N=6} = -\alpha_1 -\alpha_2\left(\kappa_{5,5}-\kappa_{5,5}^{N=5}\right)-  \frac{\left(\epsilon\Theta \,\kappa_{5,5} - \epsilon\Theta \, \kappa_{5,5}^{N=5}\right)^2}{4\left(\kappa_{5,5}-\kappa_{5,5}^{N=5}\right)}\ .
\end{aligned}
\end{equation}
We have evidence that the $\textit{degenerate situation}$ defined by eq.~\eqref{eq:CY4N6TwoSituations}, namely $\kappa_{5,5} = \kappa_{5,5}^{N=5}$, is in conflict with the assumption of there not being a Picard--Fuchs operator of order five. Therefore, we focus on the \textit{non-degenerate situation}, namely $\kappa_{5,5} \neq \kappa_{5,5}^{N=5}$ and $\kappa_{6,6} = \kappa_{6,6}^{N=6}$.

Recall that $\kappa_{5,5} \neq \kappa_{5,5}^{N=5}$ implies $\operatorname{det}M^{(5)} \neq 0$, such that the $6\times 6$ submatrix $M^{(5)}$ is of full rank. Upon imposing $\kappa_{6,6} = \kappa_{6,6}^{N=6}$ and $\kappa_{5,5} \neq \kappa_{5,5}^{N=5}$ the kernel of $M^{(6)}$ is thus exactly one-dimensional. An explicit calculation shows that it corresponds to the operator
\begin{equation}\label{eq:CY4LN6}
\mathcal{L}^{(6)}=\left(\epsilon \Theta +\beta_1\right)\left(\kappa_{0,4}^{-5}\,\mathcal{L}^{(5)}\right) + \beta_2 \left(\kappa_{5,5}-\kappa_{5,5}^{N=5}\right)
\end{equation}
where 
\begin{equation}
\beta_1 = -\tfrac{1}{2} \epsilon \Theta\, \operatorname{ln}\left(\kappa_{5,5}-\kappa_{5,5}^{\text{fixed}}\right) \ , \qquad \beta_2 = - \kappa_{0,4}^{-1} \ ,
\end{equation}
and $\mathcal{L}^{(5)}$ is the order five operator constituted by the coefficient functions in eq.~\eqref{eq:CY4N5cs}. Note that eq.~\eqref{eq:CY4LN6} expresses $\mathcal{L}^{(6)}$ entirely in terms of the four correlators $\kappa_{0,4} =\kappa_{2,2}$, $\kappa_{3,3}$, $\kappa_{4,4}$ and $\kappa_{5,5}$. Hence, there are two differential-algebraic relations among its coefficient functions.

We finally note that for eq.~\eqref{eq:CY4LN6} to transform covariantly under a rescaling $\mathcal{L}^{(5)} \mapsto g \cdot \mathcal{L}^{(5)}$, the coefficients $\beta_1$ and $\beta_2$ need to transform as $\beta_1 \mapsto\beta_1^{\prime}= \beta_1 - \epsilon\Theta\,\operatorname{ln} g$ and $\beta_2 \mapsto \beta_2^{\prime}= g\cdot \beta_2$. While $\beta_2^{\prime}$ cannot be set to zero, we can achieve $\beta_1^{\prime}=0$ by $g = (\kappa_{5,5}-\kappa_{5,5}^{N=5})^{-1/2}$.

\bigskip
\bibliographystyle{amsmod}
\bibliography{GJN}

\ifx\undefined\bysame
\newcommand{\bysame}{\leavevmode\hbox to3em{\hrulefill}\,}
\fi
\begin{thebibliography}{10}

\bibitem{Witten:1993yc}
E.~Witten, {\em {Phases of N=2 theories in two-dimensions}}, Nucl. Phys. {\bf
  B403} (1993) 159--222, [AMS/IP Stud. Adv. Math.1,143(1996)], {\tt
  arXiv:hep-th/9301042} {\tt [hep-th]}.

\bibitem{Morrison:1994fr}
D.~R. Morrison and M.~R. Plesser, {\em {Summing the instantons: Quantum
  cohomology and mirror symmetry in toric varieties}}, Nucl. Phys. {\bf B440}
  (1995) 279--354, {\tt arXiv:hep-th/9412236} {\tt [hep-th]}.

\bibitem{Morrison:1995yh}
D.~R. Morrison and M.~R. Plesser, {\em {Towards mirror symmetry as duality for
  two-dimensional abelian gauge theories}}, Nucl. Phys. Proc. Suppl. {\bf 46}
  (1996) 177--186, {\tt arXiv:hep-th/9508107} {\tt [hep-th]}.

\bibitem{Hori:2000kt}
K.~Hori and C.~Vafa, {\em {Mirror symmetry}}, {\tt arXiv:hep-th/0002222} {\tt
  [hep-th]}.

\bibitem{Closset:2015rna}
C.~Closset, S.~Cremonesi, and D.~S. Park, {\em {The equivariant A-twist and
  gauged linear sigma models on the two-sphere}}, JHEP {\bf 06} (2015) 076,
  {\tt arXiv:1504.06308} {\tt [hep-th]}.

\bibitem{Benini:2015noa}
F.~Benini and A.~Zaffaroni, {\em {A topologically twisted index for
  three-dimensional supersymmetric theories}}, JHEP {\bf 07} (2015) 127, {\tt
  arXiv:1504.03698} {\tt [hep-th]}.

\bibitem{Festuccia:2011ws}
G.~Festuccia and N.~Seiberg, {\em {Rigid Supersymmetric Theories in Curved
  Superspace}}, JHEP {\bf 06} (2011) 114, {\tt arXiv:1105.0689} {\tt [hep-th]}.

\bibitem{Pestun:2007rz}
V.~Pestun, {\em {Localization of gauge theory on a four-sphere and
  supersymmetric Wilson loops}}, Commun. Math. Phys. {\bf 313} (2012) 71--129,
  {\tt arXiv:0712.2824} {\tt [hep-th]}.

\bibitem{Ueda:2016wfa}
K.~Ueda and Y.~Yoshida, {\em {Equivariant A-twisted GLSM and Gromov--Witten
  invariants of CY 3-folds in Grassmannians}}, JHEP {\bf 09} (2017) 128, {\tt
  arXiv:1602.02487} {\tt [hep-th]}.

\bibitem{Kim:2016uq}
B.~Kim, J.~Oh, K.~Ueda, and Y.~Yoshida, {\em {Residue mirror symmetry for
  Grassmannians}}, {\tt arXiv:1607.08317} {\tt [math.AG]}.

\bibitem{Bonelli:2013mma}
G.~Bonelli, A.~Sciarappa, A.~Tanzini, and P.~Vasko, {\em {Vortex partition
  functions, wall crossing and equivariant Gromov-Witten invariants}}, Commun.
  Math. Phys. {\bf 333} (2015) 717--760, {\tt arXiv:1307.5997} {\tt [hep-th]}.

\bibitem{Candelas:1990rm}
P.~Candelas, X.~C. De~La~Ossa, P.~S. Green, and L.~Parkes, {\em {A Pair of
  Calabi-Yau manifolds as an exactly soluble superconformal theory}}, Nucl.
  Phys. {\bf B359} (1991) 21--74, [AMS/IP Stud. Adv. Math.9,31(1998)].

\bibitem{Batyrev:1994hm}
V.~V. Batyrev, {\em {Dual polyhedra and mirror symmetry for Calabi-Yau
  hypersurfaces in toric varieties}}, J. Alg. Geom. {\bf 3} (1994) 493--545,
  {\tt arXiv:alg-geom/9310003} {\tt [alg-geom]}.

\bibitem{Batyrev:1994pg}
V.~V. Batyrev and L.~A. Borisov, {\em {On Calabi-Yau complete intersections in
  toric varieties}}, {\tt arXiv:alg-geom/9412017} {\tt [alg-geom]}.

\bibitem{Jockers:2012dk}
H.~Jockers, V.~Kumar, J.~M. Lapan, D.~R. Morrison, and M.~Romo, {\em
  {Two-Sphere Partition Functions and Gromov-Witten Invariants}}, Commun. Math.
  Phys. {\bf 325} (2014) 1139--1170, {\tt arXiv:1208.6244} {\tt [hep-th]}.

\bibitem{Gomis:2012wy}
J.~Gomis and S.~Lee, {\em {Exact Kahler Potential from Gauge Theory and Mirror
  Symmetry}}, JHEP {\bf 04} (2013) 019, {\tt arXiv:1210.6022} {\tt [hep-th]}.

\bibitem{Gomis:2015yaa}
J.~Gomis, P.-S. Hsin, Z.~Komargodski, A.~Schwimmer, N.~Seiberg, and S.~Theisen,
  {\em {Anomalies, Conformal Manifolds, and Spheres}}, JHEP {\bf 03} (2016)
  022, {\tt arXiv:1509.08511} {\tt [hep-th]}.

\bibitem{Honda:2013uca}
D.~Honda and T.~Okuda, {\em {Exact results for boundaries and domain walls in
  2d supersymmetric theories}}, JHEP {\bf 09} (2015) 140, {\tt arXiv:1308.2217}
  {\tt [hep-th]}.

\bibitem{Sugishita:2013jca}
S.~Sugishita and S.~Terashima, {\em {Exact Results in Supersymmetric Field
  Theories on Manifolds with Boundaries}}, JHEP {\bf 11} (2013) 021, {\tt
  arXiv:1308.1973} {\tt [hep-th]}.

\bibitem{Hori:2013ika}
K.~Hori and M.~Romo, {\em {Exact Results In Two-Dimensional (2,2)
  Supersymmetric Gauge Theories With Boundary}}, 2013, {\tt arXiv:1308.2438}
  {\tt [hep-th]}.

\bibitem{Hosono:1994ax}
S.~Hosono, A.~Klemm, S.~Theisen, and S.-T. Yau, {\em {Mirror symmetry, mirror
  map and applications to complete intersection Calabi-Yau spaces}}, Nucl.
  Phys. {\bf B433} (1995) 501--554, [AMS/IP Stud. Adv. Math.1,545(1996)], {\tt
  arXiv:hep-th/9406055} {\tt [hep-th]}.

\bibitem{MR2282972}
G.~Almkvist and W.~Zudilin, {\em Differential equations, mirror maps and zeta
  values}, Mirror symmetry. {V}, AMS/IP Stud. Adv. Math., vol.~38, Amer. Math.
  Soc., Providence, RI, 2006, pp.~481--515, {\tt arXiv:math.NT/0402386} {\tt
  [math.NT]}.

\bibitem{MR2282974}
C.~van Enckevort and D.~van Straten, {\em Monodromy calculations of fourth
  order equations of {C}alabi-{Y}au type}, Mirror symmetry. {V}, AMS/IP Stud.
  Adv. Math., vol.~38, Amer. Math. Soc., Providence, RI, 2006, pp.~539--559,
  {\tt arXiv:math/0412539} {\tt [math.AG]}.

\bibitem{Strominger:1990pd}
A.~Strominger, {\em {S}pecial {G}eometry}, Commun. Math. Phys. {\bf 133} (1990)
  163--180.

\bibitem{Cecotti:1991me}
S.~Cecotti and C.~Vafa, {\em {Topological antitopological fusion}}, Nucl. Phys.
  {\bf B367} (1991) 359--461.

\bibitem{Aspinwall:2017loy}
P.~S. Aspinwall, M.~R. Plesser, and K.~Wang, {\em {Mirror Symmetry and
  Discriminants}}, {\tt arXiv:1702.04661} {\tt [hep-th]}.

\bibitem{MR1288523}
P.~Griffiths and J.~Harris, {\em Principles of algebraic geometry}, {John Wiley
  \& Sons}, Inc., New York, 1994.

\bibitem{Gerhardus:2015sla}
A.~Gerhardus and H.~Jockers, {\em {Dual pairs of gauged linear sigma models and
  derived equivalences of Calabi--Yau threefolds}}, J. Geom. Phys. {\bf 114}
  (2017) 223--259, {\tt arXiv:1505.00099} {\tt [hep-th]}.

\bibitem{Beem:2012mb}
C.~Beem, T.~Dimofte, and S.~Pasquetti, {\em {Holomorphic Blocks in Three
  Dimensions}}, JHEP {\bf 12} (2014) 177, {\tt arXiv:1211.1986} {\tt [hep-th]}.

\bibitem{Goddard:1976qe}
P.~Goddard, J.~Nuyts, and D.~I. Olive, {\em {Gauge Theories and Magnetic
  Charge}}, Nucl. Phys. {\bf B125} (1977) 1--28.

\bibitem{MR0255525}
J.~A. Dieudonn{\'e} and J.~B. Carrell, {\em Invariant theory, old and new},
  Advances in Math. {\bf 4} (1970) 1--80 (1970).

\bibitem{MR2004511}
I.~Dolgachev, {\em Lectures on invariant theory}, London Mathematical Society
  Lecture Note Series, vol. 296, Cambridge University Press, Cambridge, 2003.

\bibitem{MR0179268}
M.~Nagata, {\em Invariants of a group in an affine ring}, J. Math. Kyoto Univ.
  {\bf 3} (1964) 369--377.

\bibitem{MR544240}
D.~Luna and R.~W. Richardson, {\em A generalization of the {C}hevalley
  restriction theorem}, Duke Math. J. {\bf 46} (1979) 487--496.

\bibitem{MR2276766}
T.~Coates and A.~Givental, {\em Quantum {R}iemann-{R}och, {L}efschetz and
  {S}erre}, Ann. of Math. (2) {\bf 165} (2007) 15--53, {\tt arXiv:math/0110142}
  {\tt [math.AG]}.

\bibitem{KlubergStern:1974rs}
H.~Kluberg-Stern and J.~B. Zuber, {\em {Ward Identities and Some Clues to the
  Renormalization of Gauge Invariant Operators}}, Phys. Rev. {\bf D12} (1975)
  467--481.

\bibitem{Henneaux:1992ig}
M.~Henneaux and C.~Teitelboim, {\em {Quantization of gauge systems}}, Princeton
  University Press, 1992.

\bibitem{Closset:2014pda}
C.~Closset and S.~Cremonesi, {\em {Comments on $ \mathcal{N} $ = (2, 2)
  supersymmetry on two-manifolds}}, JHEP {\bf 07} (2014) 075, {\tt
  arXiv:1404.2636} {\tt [hep-th]}.

\bibitem{Gerhardus:2016iot}
A.~Gerhardus and H.~Jockers, {\em {Quantum periods of Calabi--Yau fourfolds}},
  Nucl. Phys. {\bf B913} (2016) 425--474, {\tt arXiv:1604.05325} {\tt
  [hep-th]}.

\bibitem{MR1653024}
A.~Givental, {\em A mirror theorem for toric complete intersections},
  Topological field theory, primitive forms and related topics ({K}yoto, 1996),
  Progr. Math., vol. 160, Birkh{\"a}user Boston, Boston, MA, 1998,
  pp.~141--175, {\tt arXiv:alg-geom/9701016} {\tt [alg-geom]}.

\bibitem{MR3126932}
I.~t. Ciocan-Fontanine, B.~Kim, and D.~Maulik, {\em Stable quasimaps to {GIT}
  quotients}, J. Geom. Phys. {\bf 75} (2014) 17--47, {\tt arXiv:1106.3724} {\tt
  [math.AG]}.

\bibitem{MR1824028}
B.~E. Sagan, {\em The symmetric group}, second ed., Graduate Texts in
  Mathematics, vol. 203, Springer-Verlag, New York, 2001, Representations,
  combinatorial algorithms, and symmetric functions.

\bibitem{Intriligator:1991an}
K.~A. Intriligator, {\em {Fusion residues}}, Mod. Phys. Lett. {\bf A6} (1991)
  3543--3556, {\tt arXiv:hep-th/9108005} {\tt [hep-th]}.

\bibitem{Vafa:1991uz}
C.~Vafa, {\em {Topological mirrors and quantum rings}}, Mirror Symmetry. {I},
  AMS/IP Stud. Adv. Math., vol.~9, Amer. Math. Soc., Providence, RI, 1998,
  pp.~97--120, {\tt arXiv:hep-th/9111017} {\tt [hep-th]}.

\bibitem{Witten:1993xi}
E.~Witten, {\em {The Verlinde algebra and the cohomology of the Grassmannian}},
  1993, {\tt arXiv:hep-th/9312104} {\tt [hep-th]}.

\bibitem{MR1621570}
B.~Siebert and G.~Tian, {\em On quantum cohomology rings of {F}ano manifolds
  and a formula of {V}afa and {I}ntriligator}, Asian J. Math. {\bf 1} (1997)
  679--695, {\tt arXiv:alg-geom/9403010} {\tt [math.AG]}.

\bibitem{Hori:2006dk}
K.~Hori and D.~Tong, {\em {Aspects of Non-Abelian Gauge Dynamics in
  Two-Dimensional N=(2,2) Theories}}, JHEP {\bf 05} (2007) 079, {\tt
  arXiv:hep-th/0609032} {\tt [hep-th]}.

\bibitem{MR2475813}
L.~Borisov and A.~C{\u{a}}ld{\u{a}}raru, {\em The {P}faffian-{G}rassmannian
  derived equivalence}, J. Algebraic Geom. {\bf 18} (2009) 201--222, {\tt
  arXiv:math/0608404} {\tt [math.AG]}.

\bibitem{Kuznetsov:2006arxiv}
A.~Kuznetsov, {\em {Homological projective duality for Grassmannians of
  lines}}, {\tt arXiv:math/0610957} {\tt [math.AG]}.

\bibitem{MR1775415}
E.~A. R{\o}dland, {\em The {P}faffian {C}alabi-{Y}au, its mirror, and their
  link to the {G}rassmannian {$G(2,7)$}}, Compositio Math. {\bf 122} (2000)
  135--149, {\tt arXiv:math/9801092} {\tt [math.AG]}.

\bibitem{Lerche:1989uy}
W.~Lerche, C.~Vafa, and N.~P. Warner, {\em {Chiral Rings in N=2 Superconformal
  Theories}}, Nucl. Phys. {\bf B324} (1989) 427--474.

\bibitem{Morrison:2000bt}
D.~R. Morrison, {\em {Geometric aspects of mirror symmetry}}, {\tt
  arXiv:math/0007090} {\tt [math-ag]}.

\bibitem{Cynk:2017wo}
S.~Cynk and D.~van Straten, {\em {Picard-Fuchs operators for octic arrangements
  I (The case of orphans)}}, {\tt arXiv:1709.09752} {\tt [math.AG]}.

\bibitem{Honma:2013hma}
Y.~Honma and M.~Manabe, {\em {Exact K\"ahler Potential for Calabi-Yau
  Fourfolds}}, JHEP {\bf 05} (2013) 102, {\tt arXiv:1302.3760} {\tt [hep-th]}.

\bibitem{Candelas:1993dm}
P.~Candelas, X.~De~La~Ossa, A.~Font, S.~H. Katz, and D.~R. Morrison, {\em
  {Mirror symmetry for two parameter models. 1.}}, Nucl. Phys. {\bf B416}
  (1994) 481--538, [AMS/IP Stud. Adv. Math.1,483(1996)], {\tt
  arXiv:hep-th/9308083} {\tt [hep-th]}.

\bibitem{Candelas:1994hw}
P.~Candelas, A.~Font, S.~H. Katz, and D.~R. Morrison, {\em {Mirror symmetry for
  two parameter models. 2.}}, Nucl. Phys. {\bf B429} (1994) 626--674, {\tt
  arXiv:hep-th/9403187} {\tt [hep-th]}.

\bibitem{Hori:2011pd}
K.~Hori, {\em {Duality In Two-Dimensional (2,2) Supersymmetric Non-Abelian
  Gauge Theories}}, JHEP {\bf 10} (2013) 121, {\tt arXiv:1104.2853} {\tt
  [hep-th]}.

\bibitem{Closset:2017vvl}
C.~Closset, N.~Mekareeya, and D.~S. Park, {\em {A-twisted correlators and Hori
  dualities}}, JHEP {\bf 08} (2017) 101, {\tt arXiv:1705.04137} {\tt [hep-th]}.

\bibitem{AESZ}
G.~Almkvist, C.~van Enckevort, D.~van Straten, and W.~Zudilin, {\em {Tables of
  Calabi--Yau equations}}, {\tt arXiv:math/0507430} {\tt [math.AG]}.

\end{thebibliography}
\end{document}